\documentclass[aps,prx,reprint,superscriptaddress,nofootinbib,longbibliography]{revtex4-2}

\usepackage[utf8]{inputenc}
\usepackage[T1]{fontenc}
\usepackage{lmodern}
\usepackage{microtype}
\usepackage{mathtools}

\usepackage{placeins}
\usepackage{xspace}
\usepackage{braket}
\usepackage{physics}
\usepackage{amsmath}
\usepackage{amsfonts}
\usepackage{graphicx}
\usepackage{hyperref}
\hypersetup{
  colorlinks=true,
  citecolor=blue,
  linkcolor=blue,
  urlcolor=blue
}

\usepackage{mdframed}
\usepackage{comment}
\usepackage{qcircuit}
\usepackage{multirow}
\usepackage{adjustbox}
\usepackage{color}

\begin{document}

\title{Machine Learning in the Search for New Fundamental Physics}

\author{Georgia Karagiorgi}
\email{georgia@nevis.columbia.edu}
\affiliation{Department of Physics, Columbia University, New York, NY 10027, USA}

\author{Gregor Kasieczka}
\email{gregor.kasieczka@uni-hamburg.de}
\affiliation{Institut f\"{u}r Experimentalphysik, Universit\"{a}t Hamburg, 22761 Hamburg, Germany}

\author{Scott Kravitz}
\email{swkravitz@lbl.gov}
\affiliation{Physics Division, Lawrence Berkeley National Laboratory, Berkeley, CA 94720, USA}

\author{Benjamin Nachman}
\email{bpnachman@lbl.gov}
\affiliation{Physics Division, Lawrence Berkeley National Laboratory, Berkeley, CA 94720, USA}
\affiliation{Berkeley Institute for Data Science, University of California, Berkeley, CA 94720, USA}

\author{David Shih}
\email{shih@physics.rutgers.edu}
\affiliation{NHETC, Department of Physics and Astronomy, Rutgers University, Piscataway, NJ 08854, USA}

\begin{abstract}
Machine learning plays a crucial role in enhancing and accelerating the search for new fundamental physics.  
We review the state of machine learning methods and applications for new physics searches in the context of terrestrial high energy physics experiments, including the Large Hadron Collider, rare event searches, and neutrino experiments.
While machine learning has a long history in these fields, the deep learning revolution (early 2010s) has yielded a qualitative shift in terms of the scope and ambition of research.
These modern machine learning developments are the focus of the present review.
\end{abstract}

\date{\today}
\maketitle

\section{Introduction}

High Energy Physics (HEP) is entering a new data-driven era.  For many decades, the Standard Model (SM) of particle physics has provided clear theoretical guidance to experiments, resulting in an extensive search program that culminated in the discovery of the Higgs boson~\cite{Aad:2012tfa,Chatrchyan:2012ufa}.  But while the SM is now complete, there are key experimental observations that compel the community to expand the search efforts for new particles and forces of nature beyond the SM (BSM).  For example, the existence of dark matter and dark energy is well-established~\cite{Zyla:2020zbs}, as is the mass of neutrinos~\cite{Fukuda:1998mi,Ahmad:2002jz}, and the baryon-anti-baryon asymmetry in the universe~\cite{Canetti:2012zc}---yet none of these observations are explained by the SM.   Additionally, aesthetic problems plague the SM, including the unexplained weak-scale mass of the Higgs boson, the existence of three generations of fermions, and the minuteness of the neutron dipole moment~\cite{Abel:2020gbr}.  Current and near-future HEP experiments have the potential to shed light on all of these fundamental challenges by creating new particles in the laboratory, or by observing interactions of new particles with normal matter or with other new particles.

This great potential for discovery comes with significant data challenges.
New particle interactions are expected to be rare, and their signature could be only subtly different from the SM.  This means that researchers must collect and sift through an immense amount of complex data in order to isolate potential BSM physics.  Machine Learning (ML) offers a powerful solution to this challenge.  Deep learning techniques\footnote{While the terms \textit{Artificial Intelligence}, \textit{Machine Learning}, and \textit{Statistical Learning} are often used to represent different concepts, there is no consensus and many consider them to be synonyms with differences arising mostly from sociology.  \textit{Deep Learning} is used here to mean Modern Machine Learning, with deep neural networks and other advanced tools that contain (much) more than tens of thousands of tunable parameters.} are well-suited for analyzing large amounts of data in many dimensions to find subtle patterns.  Multivariate analysis has been commonplace in HEP for decades (see e.g.\ the thousands of citations to Ref.~\cite{Hocker:2007ht}), but the latest tools will qualitatively extend the sensitivity to \textit{hypervariate} analysis whereby the entire phase space of available experimental information can be analyzed holistically.  

In tandem with the growing data volume, a related challenge is the increasing need for efficient (in terms of computational time, power, and resource utilization) and accurate data processing for high-throughput applications. Efforts to that end include the development and acceleration of deep learning-based processing algorithms on power-efficient hardware platforms.

In addition to the growing data challenge, there is also the compounding challenge in simulating expectations for what experiments may observe.  HEP experiments rely heavily on simulations for all aspects of research, from experimental design all the way to data analysis. Built on a thorough understanding of the SM and the fundamental laws of nature, these simulations are extremely comprehensive and sophisticated, but they are still only an approximation to nature.  It is therefore often necessary to combine simulations with information directly from data to improve simulation accuracy. The corresponding ML models must be robust against inaccuracies and be able to integrate uncertainties.

BSM physics can also be precisely simulated, and most searches are developed, optimized and interpreted in the context of a specific BSM model.  It is not possible to test every possible model in a single search, and even if this could be done, there would still be blind spots in the search program. Here, again, ML and in particular unsupervised ML methods are starting to provide a complementary search strategy that is liberated from model dependence compared with traditional searches.


%

This review focuses on the applications of modern ML to the search for new fundamental physics\footnote{For ML applications to SM physics, see e.g.~Ref.~\cite{Radovic:2018dip}.  A \textit{Living Review} of ML for particle physics that is continually updated with the latest methods and results can be found at Ref.~\cite{Feickert:2021ajf}.  In what follows, we will highlight the first paper(s) on various subjects; a complete list can be found in the Living Review.}.
To keep the scope manageable, we focus on terrestrial experiments that directly probe BSM effects and that aim to elucidate the particle nature of BSM.  This includes production of new, massive particles at the Large Hadron Collider (LHC), rare event searches---such as direct detection of dark matter---in the laboratory, and searches for weakly interacting particles like sterile neutrinos.  
These three science area \textit{frontiers} (energy, rare event, neutrino) share much in common, starting with their fundamental units of data: all three frontiers are {\it event} based, where each event represents a (nearly) independent and identically distributed draw from an ensemble of physics processes (typically, particle interactions with other particles, or matter in a detector). Although the structure and format of recorded events can differ between frontiers, 
a similar data pipeline
is followed for all experiments. %

The first step of the data pipeline is to record potentially interesting events with real-time or online algorithms %
(data acquisition).  Then, various pattern recognition algorithms are used to reconstruct and calibrate the properties of recorded events (data reconstruction).  Finally, certain events are selected based on their properties and used for statistical analyses (final data analysis).  A parallel track also runs on simulation, where synthetic events are generated and then passed through these same steps.  Machine learning is actively being integrated into each of the three steps, with significant innovation from HEP domain scientists to design custom solutions to our unique challenges. 

Below, we will highlight advances from ML in all aspects of the data pipeline, with an orientation towards new particle searches.  First, we will give a general overview of ML methods for new physics searches, many of which are still in the proof-of-concept stage.  Then, we will turn to frontier-specific highlights, describing ML methods that are actively being applied to data. While many ML-based reconstruction innovations are now being integrated into experimental workflows, data acquisition and final data analysis tools are not yet widely deployed, owing to the newness of the methods being developed, and the time it takes to turn them from proofs-of-concept into complete analyses.

\section{Machine Learning Foundations for Fundamental Physics}
\label{sec:overview}
Machine Learning involves up to five components: input features $x\in\mathbb{X}=\mathbb{R}^N$, targets $y\in\mathbb{Y}=\mathbb{R}^M$, a model $f:\mathbb{R}^N\rightarrow\mathbb{R}^M$ with tunable parameters, a loss functional $L[f]:\mathbb{X}^\mathbb{Y}\rightarrow \mathbb{R}$, and an optimization strategy.  For many applications, $N=\mathcal{O}(1)$ as the features are a fixed set built from the full phase space based on physical intuition. In this case, Boosted Decision Trees (BDTs) (an example of ``shallow'' ML) have long been the  tools of choice for machine learning in HEP. Meanwhile, deep learning methods, i.e.\ neural networks (NNs), are now able to readily process high-dimensional feature spaces where $N$ can be many thousand or more.  Additionally, $N$ may not be fixed for each data point.  For example, LHC proton-proton collision or neutrino/dark matter interaction events naturally produce a variable number of particles.

\subsection{Data Representations}

The choice of how to represent the data (e.g.\ four vectors of every particle in the event, energies of calorimeter deposits, etc.) can play an important role in the design (architecture) of the deep learning algorithm, and can have a significant impact on its performance. 

The most flexible and general architecture is that of the fully connected or dense neural network (DNN), where the features $x$ are simply flattened to a column vector, and fed to the DNN which is specified by the number of hidden layers, the number of nodes in each hidden layers, etc. Often the humble DNN is sufficiently powerful for many applications in HEP. But
fundamental physics events often respect various symmetries, which can be built in to machine learning architectures to reduce the number of parameters and increase performance.  

If $x$ can be represented as a fixed-size tensor with translational invariance across indices, then $f$ is usually a convolutional neural network (CNN)~\cite{fukushima1982neocognitron,lecun1989handwritten}.  
For example, it is often natural to represent fundamental physics events as images, with the pixel intensity given by the amount of energy deposited in a given detector region (e.g.,~a calorimeter cell).   If a particular physics signature can register anywhere in the detector, then translational invariance is  a good symmetry.  The first applications of CNNs to particle physics~\cite{deOliveira:2015xxd,Baldi:2016fql} showed that CNNs can still be useful even if translational symmetry is broken from pre-processing (image centering).  Studies of homogeneous detectors like those common in neutrino physics were the first to exploit the translational invariance of these methods~\cite{Aurisano:2016jvx}.

Other architectures are well-suited when $x$ has structure other than a translationally invariant tensor.  If $x$ is a sequence, then recurrent neural networks (RNNs) are particularly effective~\cite{rumelhart1986learning,LSTM}.  While CNNs share weights across space, RNNs share weights across time (location in the sequence).  As sequences can vary in length, RNNs were the first tools for processing variable-dimensional features in high energy physics (for jet flavor tagging~\cite{Guest:2016iqz}).  When $x$ is a hierarchy of sequences (called a tree), then recursive neural networks can be used in a similar fashion to RNNs.  Tree structures arise naturally in high energy jet clustering and were first studied in that context to identify Lorentz-boosted $W$ bosons decaying into hadrons~\cite{Louppe:2017ipp}.

One challenge with image, sequence, and tree representations of data is that they require a spatial or temporal order to the dimensions of $x$. While this is often natural, in many cases, there is no inherent order.  For example, the particles of the same type produced in collider and fixed-target experiments are indistinguishable due to quantum mechanics.  It is possible to impose an order (e.g.~sort by energy), but this does not necessarily reflect the underlying physics processes or associations.  There are now multiple architectures that can process variable-length and permutation-invariant\footnote{A neural network $f$ is \textit{invariant} under the operation of a group $G$ if $f(g(x))=f(x)$ for all $g\in G$ while the network is \textit{equivariant} if $f(g(x))=g(f(x))$.  See, e.g.~Ref.~\cite{Dolan:2020qkr,serviansky2020set2graph,Bogatskiy:2020tje,Shimmin:2021pkm} for examples of equivariant networks in high energy physics.} sets.  One possibility is deep sets~\cite{DBLP:conf/nips/ZaheerKRPSS17}, first adapted to particle physics as the Particle Flow Network~\cite{Komiske:2018cqr}.  In this framework, neural networks are decomposed into two parts: a network that embeds each of the $N/n$ components of $x_i$ into a \textit{latent space} $\Phi:\mathbb{R}^n\rightarrow\mathbb{R}^k$ and a second network $F$ that processes the sum over latent space vectors: $f(x)=F(\sum \Phi(x_i))$.  A second permutation invariant architecture is the graph neural network (first applied in particle physics in Refs.~\cite{Henrion:DLPS2017,choma2018graph,Qu:2019gqs} and reviewed in Ref.~\cite{1808887}), which makes use of locality in the passage of information between nodes of the graph and layers of the network. 

\subsection{Machine Learning Tasks}

Machine learning tasks are categorized by the learning target $y$.  When $y$ is discrete and finite, the task is called \textit{classification} and the typical loss functional is the cross entropy: $L[f]=\sum_i\sum_j \mathbb{I}[y_i=i]\log(f_j(x_i))$ for classes $j$, indicator function $\mathbb{I}$, and $\sum_j f_j=1$.  For continuous (or discrete and infinite) $y$, the task is called \textit{regression} and a common loss functional is the mean squared error: $L[f]=\sum_i (f(x_i)-y_i)^2$.  Other loss functionals correspond to different learning targets (mean, median, mode, etc.); see Ref.~\cite{Cheong:2019upg} for details in the context of high energy physics.  Both of the classification and regression tasks as described above are called \textit{supervised}, since each data point $x_i$ comes with a label $y_i$.  

All other cases are called \textit{less-than-supervised}.  {\it Unsupervised} learning proceeds without any labels.  Such approaches are typically designed to learn implicitly or explicitly the data probability density $p(x)$.  This can be useful for various tasks including generative modeling and anomaly detection.  Three standard approaches to unsupervised deep learning include Generative Adversarial Networks (GANs)~\cite{Goodfellow:2014:GAN:2969033.2969125,Creswell2018}, (Variational) Autoencoders (VAEs)~\cite{kingma2014autoencoding,Kingma2019}, and Normalizing Flows (NFs)~\cite{10.5555/3045118.3045281,Kobyzev2020}, first studied in HEP in Refs.~\cite{deOliveira:2017pjk,Mustafa_2019},~\cite{ATL-SOFT-PUB-2018-001,Hajer:2018kqm,Farina:2018fyg,Heimel:2018mkt}, and~\cite{Albergo:2019eim,Gao:2020zvv,Gao:2020vdv,Bothmann:2020ywa,Nachman:2020lpy}, respectively.  Each\footnote{In the case of autoencoders, this is only true for variational architectures.} of these methods learns to map a random variable $Z\in\mathbb{R}^k$ with a known probability density to the data.  For GANs, a second network $h$ is simultaneously trained to distinguish $f(Z)$ from $X$; when that network performs poorly, then $f$ is a good model of the data.  For VAEs, $f$ is the decoder of a two-network encoder-decoder setup.  The data are mapped from the data space into the latent space via the encoder and then back to the data space via the decoder, trying to preserve the data distribution and statistical properties of the latent space.  A normalizing flow is a series of invertible functions $f_i$ with tractable Jacobians in order to change $Z$ into $X$: $p_f(f(z))=p(z)\prod_i |\partial f_{i+1}^{-1}/f_{i}|$ for $f_0=z$.  NFs are optimized by maximizing the likelihood of the data: $L[f]=\log p_f$.

Due in part to quantum mechanics, it is often impossible to know the labels of individual examples from real data.  However, simulations (via Monte Carlo methods) play a key role in the development and execution of searches for new particles and it is usually possible to have per-instance labels by examining the Monte Carlo truth record.  In addition to purely unsupervised and supervised techniques, there is a spectrum of methods with varying levels of supervision, some of which use a mix of information from data and from simulation.  \textit{Semi-supervised} methods\footnote{Such a categorisation is not unique, see e.g.~\cite{10.1093/nsr/nwx106} for an alternative way of defining weak supervision. We follow the established usage in applications of machine learning for particle physics.} have labels available for some, but not all data.  \textit{Weakly-supervised} methods have labels for all data, but each label is noisy.  These approaches were first studied in HEP in Refs.~\cite{Dery:2017fap,Metodiev:2017vrx,Cohen:2017exh, Komiske:2018oaa}.  For example, if a set of events is known to be composed of two classes and the class proportions are $p_0$ and $p_1=1-p_0$, then randomly assigning the label $0$ with probability $p_0$ would constitute noisy labels.  This is the assumption that underlies Learning from Label Proportions~\cite{Dery:2017fap}.  Learning may still be possible even if the proportions are not known as is the case in Classification Without Labels~\cite{Metodiev:2017vrx}. 

\section{Machine Learning Searches for New Physics} %
\label{sec:modeldep}

In this section, we review the current state of the art in methods for direct new physics searches at particle detectors. Many of the ideas presented here are still at the proof-of-concept stage; in Sec.~\ref{sec:frontiers} we describe the status of methods that have been applied to actual data.

A search for new physics requires two essential components. The first is a method for achieving signal sensitivity by selecting events that preferentially contain new physics, removing as many of the (generally far more numerous) SM background events as possible. %
The second part is the careful and precise estimation of the SM background.  Events in data that have passed a selection are compared to a prediction of the SM background for the same selection. If the number of events or their distribution in some observable is inconsistent between data and SM background prediction, there is evidence for new phenomena.  If the data are consistent with a particular BSM theory, then that model's parameters can be estimated; otherwise, limits are placed on the new particle parameters.

\begin{figure}[h!]
\centering
\includegraphics[width=0.4\textwidth]{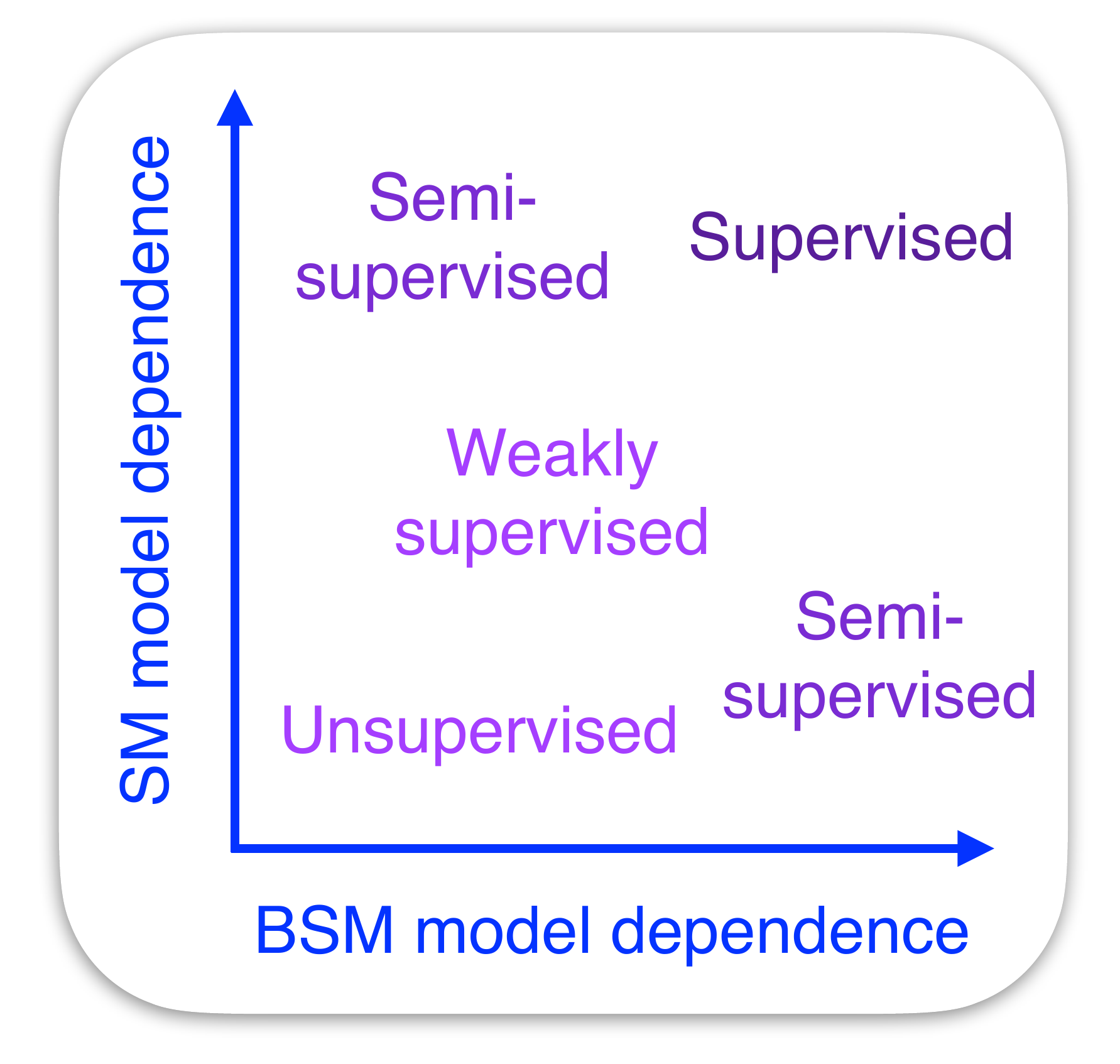}\\\includegraphics[width=0.4\textwidth]{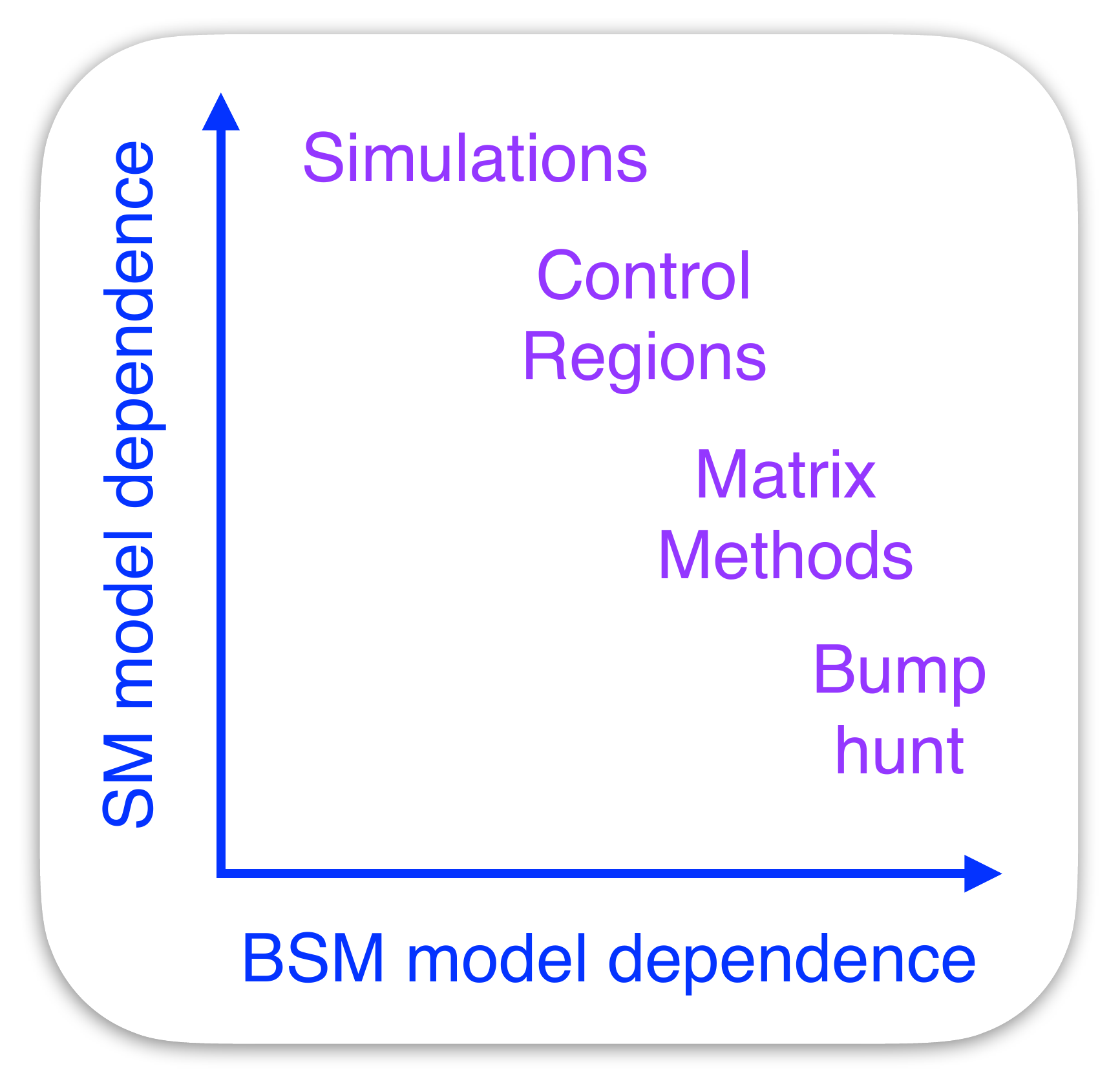}
\caption{Schematic diagrams for the landscape of model dependence for achieving signal sensitivity (top) and for calibrating the SM background (bottom).  \textit{Supervision} refers to the type of label information available during training.  \textit{Supervised} searches use simulation (labeled by construction) for the signal and the Standard Model background.  \textit{Semi-supervised} searches use data (unlabeled by construction) for either the background or the signal-sensitive sample.  \textit{Weakly supervised} searches have labels for every example, but the labels are noisy.  Finally, \textit{unsupervised} methods do not use any label information.  In rare cases with relatively simple processes, SM simulations can be used directly to estimate the background.  Most of the time, a combination of data and simulations is used to estimate the SM background.  The \textit{control region method} uses an auxiliary measurement in a signal-poor region to constrain the simulation.  Various \textit{matrix methods} (such as the ABCD method) use two independent features that are both signal-sensitive to predict the background.  \textit{Bump hunts} assume that the signal is localized in one dimension (often an invariant mass) where a sideband fit can be used to predict the background in the resonant region.  Figure adapted from Ref.~\cite{Nachman:2020lpy}.}
\label{fig:schematiclandscape}
\end{figure}

The role of SM and BSM modeling (often via simulations) in both achieving signal sensitivity and estimating the SM background can be used to categorize different search strategies, as illustrated in Fig.~\ref{fig:schematiclandscape}. At present, the majority of searches achieve signal sensitivity in a simulation-based and model-specific way. These searches begin by positing the existence of new particles with particular physics-theorized parameters (masses, couplings, etc.).  Given such a model, simulations can be used to predict what the new phenomena would look like in a detector.  Combined with simulations of the SM, and often augmented with data-driven approaches for background estimation, a search strategy can be devised which would reject SM processes and enhance the presence of new particle processes in a given set of data. Increasingly, searches use modern machine learning to train supervised classifiers for this purpose, and employ them in all stages of a data pipeline to enhance sensitivity to predictable signatures of new physics.  Some aspects of these searches are described in Sec.~\ref{sec:fullysupervised}.  

Meanwhile, a growing number of methods that make use of less-than-supervised ML techniques are being proposed for more model-agnostic and simulation-independent new physics searches.\footnote{A series of non-machine-learning semi-supervised searches have been conducted over the last decades at  D0~\cite{sleuth,Abbott:2000fb,Abbott:2000gx,Abbott:2001ke}, H1~\cite{Aaron:2008aa,Aktas:2004pz}, ALEPH~\cite{Cranmer:2005zn}, CDF~\cite{Aaltonen:2007dg,Aaltonen:2007ab,Aaltonen:2008vt}, CMS~\cite{CMS-PAS-EXO-14-016,CMS-PAS-EXO-10-021,CMS:2020ohc,Sirunyan:2020jwk}, and ATLAS~\cite{Aaboud:2018ufy,ATLAS-CONF-2014-006,ATLAS-CONF-2012-107}.  All of these searches share essentially the same approach: they compared (many) histograms of data to histograms of SM simulations and looked for discrepancies.} For example, various methods have been proposed that use the data itself to enhance the signal sensitivity. Data are unlabeled by construction and so any approach of this type is necessarily less-than-supervised. These and other ideas will be described in Sec.~\ref{sec:lessthan}.

\subsection{Signal Model-driven / Fully Supervised Searches}
\label{sec:fullysupervised}

Using simulations of the SM and new particles, 
fully supervised classifiers are trained to distinguish pure signal from pure background.  The resulting classifiers are then applied to data and used to enhance the presence of a potential signal.  There are generally two ways to train these supervised classifiers.  One approach is to train using all information in an event while another approach is to train using particular objects.  For example, a new heavy particle $X$ produced at a collider decaying into a pair of Higgs bosons with $m_X\gg m_h$ will result in two collimated sprays of particles (called \textit{jets}), one for each Higgs boson.  One option is that a classifier could be trained using the full event to exploit the properties of the $X$ production and decay.  A second option is that a dedicated \textit{tagger} for classifying Lorentz-boosted Higgs boson jets from generic quark and gluon jets could be constructed.  Particle taggers can then be combined with other tools to form an event-level classifier.  The sections below discuss signal sensitivity (object tagging in Sec.~\ref{sec:supervisedsignal} and event-level classification in Sec.~\ref{sec:supervisedsignalevent} ) and background estimation (Sec.~\ref{sec:supervisedback}) for supervised searches.

\subsubsection{Object Tagging}
\label{sec:supervisedsignal}

Jet tagging has driven much of the innovation in signal classification due to their complexity and ubiquity at high energy particle colliders.  A jet can be composed of tens to hundreds of particles and each particle has a four-momentum and attributes such as electric charge.  As a result, jets exist in a high- and variable-dimensional space, and deep learning has been used to study them as images, sequences, trees, sets, and graphs. %
Top quark jet tagging has become a benchmark task for studying new methods, as described in the community report,  Ref.~\cite{Kasieczka:2019dbj}.
Machine learning has also been used extensively to tag single particles such as electrons, muons, and pions in collider, fixed-target, and neutrino experiments (see, e.g.~\cite{MicroBooNE:2020hho}).  When measured by multiple, segmented detectors, these objects can be represented by many features and machine learning can be a powerful tool for a variety of classification and regression tasks. Examples of tagging for composite and single objects in an experimental context will be given below in the Frontier Highlights.  

\subsubsection{Event Classification}
\label{sec:supervisedsignalevent}

Numerous event-level classifiers have been used for enhancing signal sensitivity across HEP.  Boosted decision trees are particularly common.  The first study comparing deep learning with more traditional shallow learning methods was Ref.~\cite{Baldi:2014kfa}.  A conclusion from this study that has since been repeated many times is that deep learning methods can process low-level inputs (e.g.~particle four-vectors) and achieve comparable or superior performance to shallow methods that take as input physics-engineered high-level features.  The move from $\mathcal{O}(10)$ features to many hundreds or thousands of input features first occurred in Refs.~\cite{Aurisano:2016jvx,Bhimji:2017qvb}, which represented entire neutrino/collider events as images and processed them using CNNs.  Similar to the earlier studies, it was found that deep learning on low-level inputs was able to exceed the sensitivity of standard approaches and did not improve when hand-crafted observables constructed from low-level inputs were provided to the neural networks.

\subsubsection{Background Estimation}
\label{sec:supervisedback}

Machine learning has been proposed to enhance each of the background estimation strategies highlighted in Fig.~\ref{fig:schematiclandscape}.  Strategies based completely on simulation can be optimized end-to-end in an \textit{inference-aware} approach~\cite{Wunsch:2020iuh,Elwood:2020pik,Xia:2018kgd,deCastro:2018mgh,Charnock_2018,Alsing:2019dvb,lukas_heinrich_2020_3697981,1807719} such that the learning knows about the final test statistic and hypothesis test.  Hybrid methods that use auxiliary measurements to constrain the simulation can be made \textit{uncertainty-aware}~\cite{Kasieczka:2020vlh,Bollweg:2019skg,Araz:2021wqm,Bellagente:2021yyh,Brehmer:2019xox,Brehmer:2018hga,Brehmer:2018kdj,Brehmer:2018eca,Nachman:2019dol,Ghosh:2021roe} by incorporating aspects of the statistical and systematic uncertainty during training.  Simulation corrections (called \textit{domain adaptation} in machine learning) can be derived from auxiliary measurements in many dimensions using machine learning methods as well~\cite{Rogozhnikov:2016bdp,Andreassen:2019nnm,Cranmer:2015bka,2009.03796,Nachman:2021opi,clavijo2020adversarial,MINERvA:2018smv}.  Matrix methods rely on the independence of features, which can either be assumed physically and then combined with machine learning~\cite{Lin:2019htn} or enforced with machine learning~\cite{Kasieczka:2020pil,Mikuni:2021nwn}.  Bump hunts can be enhanced with machine learning, but they must not sculpt localized features.  This can be achieved with \textit{decorrelation} methods~\cite{Blance:2019ibf,Englert:2018cfo,Louppe:2016ylz,Dolen:2016kst,Moult:2017okx,Stevens:2013dya,Shimmin:2017mfk,Bradshaw:2019ipy,ATL-PHYS-PUB-2018-014,DiscoFever,Wunsch:2019qbo,Rogozhnikov:2014zea,10.1088/2632-2153/ab9023,Kasieczka:2020pil,Kitouni:2020xgb,Estrade:2019gzk,Aguilar-Saavedra:2017rzt,aguilarsaavedra2020mass}, which automatically ensure that classifiers have a controlled dependence on the resonant feature(s).  A well-studied case is when there is no dependence (independence) on the resonant feature.  These tools have also been proposed to reduce uncertainties~\cite{Englert:2018cfo,Wunsch:2019qbo}, although this should be approached with caution~\cite{Ghosh:2021hrh}.  There is also a connection to the machine learning field of \textit{fairness}, which endeavors to make classifiers equal (invariant) or equitable across populations~\cite{chouldechova2018frontiers,mehrabi2019survey}.  Machine learning has also been studied for other aspects of bump hunts, including highly flexible background fits~\cite{Frate:2017mai,DiSipio:2019imz,Chisholm:2021pdn}.

\subsection{Less-Than-Supervised Searches}
\label{sec:lessthan}

Supervised searches are often constructed as simple hypothesis tests, where the presence of signal is one hypothesis and the other is the SM-only hypothesis.  When the signal hypothesis is rejected, then the hypothesized model is excluded, typically at 90\% or 95\% confidence.  In the absence of nuisance parameters, the optimal test statistic for this hypothesis is the likelihood ratio $p_\text{SM+BSM}(x)/p_\text{SM}(x)$~\cite{neyman1933ix}.\footnote{Optimal in this context means that for a fixed probability of rejecting the given hypothesis when it is true (level), the probability for rejecting the given hypothesis when the alternative is true (power) is maximized with the likelihood ratio test statistic.}  In the presence of nuisance parameters, there is no uniformly best test statistic, but the likelihood ratio is still very powerful and widely used. Note that any test statistic that is a monotonic function of the likelihood ratio will have the same statistical properties.

Less-than-supervised searches do not have a particular signal hypothesis.  In this case, it is not possible to construct a test statistic that is optimal for all potential signals.  However, it is still possible to develop methods for discovering new physics signatures that are statistically motivated.  Unsupervised searches typically target events with low $p_\text{SM}$.  Weakly- and semisupervised methods use some label information and are therefore able to construct alternative likelihood ratio statistics such as $p_\text{data}(x)/p_\text{SM}(x)$.  This test statistic is optimal for a data-versus-background hypothesis test, and when ML methods are trained with samples drawn from the exact $p_\text{SM}$, the test statistic is sometimes called the \textit{idealized anomaly detector}.  

Most less-than-supervised searches in HEP differ from the typical anomaly detection setting in industry because $p_\text{SM}$ is usually not zero.  No single event can be labeled as signal with certainty and so HEP often targets \textit{group} or \textit{collective} anomalies as opposed to \textit{point} or \textit{off-manifold} anomalies that are common elsewhere.  Many examples of less-than-supervised methods can be additionally found in the recent LHC Olympics and Dark Machines community challenge reports~\cite{Kasieczka:2021xcg,Aarrestad:2021oeb}.

\subsubsection{Unsupervised}

The strategy of unsupervised anomaly detection methods is to find data points that are far from the bulk of the background.  
A common approach is the autoencoder network architecture~\cite{hinton2006reducing}.
Here, a pair of networks form a lossy compression algorithm. 
One network (the encoder) maps data into a latent space representation, while a second network (the decoder) maps from this latent space back to data. 
These networks are trained to minimize a loss function such as the absolute difference between input data and decoder output.
By limiting the capacity of this transformation, the autoencoder can be prevented from learning the identity function and instead identify relevant features of the data.
If trained on data dominated by a background process, an autoencoder accordingly will learn to minimise the loss for it while returning a higher loss for  previously unseen signal data.

Since the initial proposals of using autoencoders for anomaly detection~\cite{Heimel:2018mkt,Farina:2018fyg, Hajer:2018kqm}, a number of improvements and modifications have been suggested.
An important observation is that autoencoders can be biased by the relative complexity of anomalous and background data, potentially leading to outliers with a lower loss than the background~\cite{Finke:2021sdf,Dillon:2021nxw,Fraser:2021lxm}.
As the latent space in VAEs~\cite{kingma2014autoencoding,Kingma2019} 
is optimised to follow a known distribution for backgrounds, it can also be used as anomaly score~\cite{Cerri:2018anq,Govorkova:2021utb}.

Beyond the autoencoder family, a number of other unsupervised approaches based
on support vectors~\cite{1800445}, latent space dirichlet analysis (LDA)~\cite{Dillon:2019cqt,Caron:2021wmq}, 
clustering~\cite{Mikuni:2020qds}, 
and GANs~\cite{knapp2020adversarially}
have been investigated.

\subsubsection{Weakly and Semi-Supervised}

In contrast to unsupervised searches, weakly- and semisupervised searches use some label information to inform the training.  This can result in an improved sensitivity for BSM particles, at the cost of additional assumptions~\cite{Amram:2020ykb,Collins:2021nxn}.  The aspect that distinguishes weakly supervised learning and semi-supervised learning is the fidelity of the labels.  Weakly supervised learning uses noisy labels while semisupervised learning use noiseless labels, but only for a subset of the training examples.

In particle physics, the usual application of weak supervision is to isolate two sets of data (call them $A$ and $B$) that are each composed of the same two classes $1$ and $0$ (1 for signal and 0 for background).  The probability density of the mixtures $A$ and $B$ are then given by $p_A=\epsilon_A p_1+(1-\epsilon_A)p_0$ and $p_B=\epsilon_B p_1+(1-\epsilon_B)p_0$, for some unknown signal fractions $\epsilon_A$ and $\epsilon_B$. If $\epsilon_A>\epsilon_B$, then mixture $A$ is given the noisy label of \textit{signal-like} and mixture $B$ is given the noisy label of \textit{background-like}.  If a classifier is trained using these noisy labels, it will learn a function monotonically related to $p_A/p_B$ which is itself monotonically related to $p_1/p_0$ and thus optimal for the target task.  

This weakly supervised approach works well only when the $1$ and $0$ classes in $A$ and $B$ are statistically identical. Otherwise, the classifier will be distracted by differences between $A$ and $B$ that are unrelated to the signal.  This means that whatever features are used to construct $A$ and $B$ must not significantly distort the features used for classification.  One widely studied setting where this applies is resonance searches.  Such searches are defined by a feature $m$ (often an invariant mass) that is resonant for a potential signal and without localized features for the background.  The Classification Without Labels approach of Ref.~\cite{Collins:2018epr,Collins:2019jip} proposed using a region near the potential signal to construct $A$ and a sideband region is used to define $B$.  Another option is to build $B$ using pure~\cite{DAgnolo:2018cun,DAgnolo:2019vbw,dAgnolo:2021aun} or data-augmented~\cite{Andreassen:2020nkr,1815227} simulation, which is composed of only the $0$ class by construction.  Simulations can also be used to add signal-like labels for $A$~\cite{Park:2020pak,Khosa:2020qrz}.  Hybrid approaches have also been proposed that use parameterized density estimation from the sideband to estimate the background density in the signal region~\cite{Nachman:2020lpy,Stein:2020rou,Hallin:2021wme} or autoencoders to learn the noisy labels in the first place~\cite{Amram:2020ykb}.  Many of these methods are also naturally robust to correlations between $m$ and the classification features.

\section{Frontier Highlights}
\label{sec:frontiers}

This section presents the status of machine learning in the energy, neutrino, and rare event frontiers. Events at energy frontier experiments (Sec.~\ref{sec:hadron}) occur at the same point in space (center of the detector) at a fixed frequency and directly probe the highest energies with terrestrial experiments. The current energy frontier experiments are at the Large Hadron Collider (LHC). Events in neutrino experiments (Sec.~\ref{sec:neutrinos}) are usually time-synchronous with the reactor or accelerator neutrino source, but can occur anywhere within the detector. The space and time of events in rare event experiments (Sec.~\ref{sec:rare}) are not controlled by the experimenters. These searches for dark matter direct detection, astrophysical neutrinos, as well as other weakly interacting phenomena typically require large detectors built deep underground. 

This section will highlight ML applications across all three steps of the data pipeline, including data acquisition, data reconstruction, and final data analysis.

\subsection{Energy Frontier}
\label{sec:hadron}

The goal of direct new physics searches at the LHC is to test the Standard Model at the energy frontier, by producing new BSM particles, and then endeavoring to detect their distinctive signatures (e.g.\ through their decays to SM particles) over the SM background.
Examples of new physics scenarios searched for at the LHC include supersymmetry (motivated by the hierarchy problem), extra dimensions, black holes, dark matter, new generations of quarks and leptons, and new fundamental force carriers ($Z'$'s).

Machine learning has found widespread use across LHC experiments.  At the lowest levels, machine learning is used for detector calibrations, data acquisition, pattern recognition, denoising, particle identification, and detector simulation.  At the level of individual analyses, machine learning is used for background estimation as well as for constructing final analysis discriminants.  This section will highlight a few examples from across these areas.

\vspace{2mm}

\noindent\textbf{Data acquisition.} One of the key challenges at the LHC is the extreme data rate, with collisions happening at 40~MHz and each event requiring $\mathcal{O}(\text{MB})$ of memory. 
It is not possible to write every event to disk and so significant online processing is required.  
The ATLAS and CMS experiments use a two-stage processing to discard more than 99.99\% of events and reduce the data rate to 1~kHz. 
To achieve the needed low latency, the first stage (termed Level-1-Trigger, L1T) is implemented using Field Programmable Gate Array (FPGA) hardware while a second stage (termed High-Level-Trigger, HLT)
runs on computer hardware such as CPUs or GPUs.
Given the less stringent timing constraints and more powerful hardware available, ML was first applied at the HLT stage.
For example, the CSVv2 algorithm based on a simple neural network for $b$-quark identification
was adopted for use in the HLT by the CMS experiment~\cite{CMS:2017wtu}.
For upcoming data taking periods, plans exist to include 
 machine learning --- including deep networks --- also at L1T~\cite{Duarte:2018ite,Nottbeck:2019rqu,Zabi:2020gjd,Summers:2020xiy,Aarrestad:2021zos,Hong:2021snb,Deiana:2021niw}, to improve the selection efficiency.

A particularly interesting strategy of removing the first hardware stage of the trigger and processing all events by a software trigger stage running at 30~MHz is considered by the LHCb experiment~\cite{LHCbCollaboration:2717938}. Utilising the parallelisation capabilities of GPU-hardware 
allows executing track reconstruction 
as well as physics-based selections at the full input rate of 30~MHz, corresponding to 40~Tbit/s of raw data and thereby reducing the input to subsequent selection stages by a factor of 20-40~\cite{Aaij:2019zbu}.
The presence of GPU-hardware also enables the relatively simple adaptation of highly accurate ML algorithms in the trigger to further improve selection efficiency.

Deep generative models are now being deployed for creating large synthetic datasets that will be used for all aspects of downstream inference tasks~\cite{ATL-SOFT-PUB-2018-001,Chekalina:2018hxi,ATL-SOFT-PUB-2020-006,ATLAS:2021pzo}.

\vspace{2mm}

\noindent\textbf{Object Tagging.} Identifying known SM particles to enrich samples that include a specific physics process of interest is common practice at the LHC, and ML-based taggers are studied --- and in many cases deployed --- for essentially all such object tagging tasks. 
One example of a single-object tagger is tagging jets originating from a $b$-quark ($b$-tagging). 
The DL1 algorithm used by the ATLAS collaboration~\cite{ATLAS:2019bwq} is a fully-connected Deep Neural Network (DNN) achieving a light-flavor false-positive rate of 1/390 while retaining 70\% true positive rate (efficiency) for true $b$-quark jets, greatly outperforming simpler approaches.
These methods are widely used, with a large fraction of analyses requiring at least some kind of flavor information.
At the same time, architectures tailored to using low-level properties and symmetries of data promise further gains in performance~\cite{Bols:2020bkb,ATL-PHYS-PUB-2020-014}.

A similar situation exists for more complex signals. 
Considering detector activities in a larger geometrical region using so-called large-radius jets~\cite{Larkoski:2017jix,1803.06991}
allows for the identification of hadronically decaying, Lorentz-boosted heavy resonances such as $W$, $Z$, or Higgs bosons and top quarks.
For example, Ref.~\cite{Sirunyan:2020lcu} gives an overview of tagging methods considered by the CMS collaboration.
Again, such standard taggers are used across a large number of applications~\cite{CMS:2019ykj,CMS-PAS-B2G-20-004,CMS-PAS-B2G-20-007,CMS-PAS-B2G-21-001} to enrich the relative fraction of the desired particle or to design signal-regions for searches.

Both for narrow and large jets, these methods can be calibrated and scale factors can be provided by comparing recorded data and Monte Carlo simulations~\cite{ATLAS:2019bwq,Sirunyan:2020lcu, CMS:2017wtu, CMS:2021mux,ATL-PHYS-PUB-2021-035}.
Finally, taggers are also developed for hypothetical, unobserved, signal particles. In this case, training has to rely on simulation and calibration in data is only possible for backgrounds~\cite{CMS:2019dqq} putting an additional emphasis on robust background estimation techniques.

\vspace{2mm}

\noindent\textbf{Background Estimation.} Decorrelation strategies are widely used by ATLAS, CMS, and LHCb to enable resonance searches with machine learning; see Ref.~\cite{Sirunyan:2020lcu,ATL-PHYS-PUB-2018-014} for performance studies, and recent examples of physics results in Ref.~\cite{CMS:2021far,LHCb:2020wxx,collaboration2020dijet}. A growing number of searches using control region methods employ machine learning to perform high-dimensional reweighting to better match the background data in signal-sensitive regions~\cite{ATLAS:2018tti,ATLAS:2021ypo}.  The mitigation of data/simulation differences can also be part of the classifier training directly~\cite{CMS:2019dqq}.

\vspace{2mm}

\noindent\textbf{Final Analysis Discriminants.} Shallow, fully supervised machine learning (e.g.~using BDTs) is ubiquitous in searches at the LHC.  For example, ATLAS and CMS have published over 600 papers searching for new particles and the TMVA multivariate analysis package~\cite{Hocker:2007ht} is cited in over 10\% of them.  This does not include the analyses that use per-object ML-based classifiers (described above), which are likely a large fraction of all searches.  A growing number of searches are exploiting deep learning for the final analysis setup, using a variety of combinations of signal sensitivity and background specificity approaches highlighted in Fig.~\ref{fig:schematiclandscape}.  

The most common applications of deep learning to event-level analysis use supervised methods.  For example, the recent ATLAS search using multiple charged leptons in Ref.~\cite{ATLAS:2020tlo} is fully supervised and given the precisely known multilepton final state, simulations are used directly for the background estimation.  A search by CMS in the $bb\tau\tau$ final state uses an event-level, supervised classifier that is mostly calibrated using a control region method~\cite{CMS:2021yci}.  An example of a search using matrix methods for the background estimation is the recent ATLAS search for exotic Higgs boson decay to a $Z$ boson and a light pseudoscalar~\cite{ATLAS:2020pcy}.  In order to maintain sensitivity across pseudoscalar masses, a classification network is combined with a mass regression network before estimating the background using the ABCD method.  There are many resonance searches that use supervised learning, but they are all currently using machine learning through object classifiers.

While there have been many proposed less-than-supervised searches, the number of data results is still limited due to the time required to perform a complete data analysis.  There is a tradition of semi-supervised analyses that use signal simulations and control region data to build analysis discriminants. A recent result of this type using modern machine learning is the diphoton search from ATLAS~\cite{ATLAS:2021jbf} that uses XGBoost~\cite{Chen:2016:XST:2939672.2939785}.  The first weakly supervised analysis was performed by ATLAS in the dijet final state, using the Classification Without Labels protocol~\cite{collaboration2020dijet}. While this initial result used only two features, it was still able to extend the sensitivity of inclusive searches by automatically identifying anomalous regions of phase space.

While this section has focused on the LHC, modern machine learning tools are also being applied at Belle~II (e.g.~for tracking~\cite{BelleIITrackingGroup:2020hpx}), and older machine learning algorithms have a long history at previous colliders and will be essential for the physics program of future colliders.

\subsection{Neutrino Experiments}
\label{sec:neutrinos}

The discovery of neutrino masses (via their flavor oscillation) is one of the only direct observations of BSM physics.  Current searches in neutrino physics seek to understand the origin of neutrino masses and search for deviations from the three-neutrino paradigm, e.g.~as predicted by light sterile neutrinos \cite{Abazajian:2012ys,Dentler:2018sju}, new fundamental force mediators \cite{Bertuzzo:2018itn,Ballett:2018ynz}, extra dimensions \cite{MINOS:2016vvv}, Lorentz and CPT symmetry violation \cite{Kostelecky:2003cr}, or non-standard interactions of neutrinos with matter \cite{Miranda:2015dra,deGouvea:2015ndi}.  A key requirement for these analyses is the classification of the interacting neutrino flavor through identifying interaction final states, which is used to probe deviations to the standard oscillation picture. Identifying the final state objects is also important for a broad set of analyses exploring neutrinos as a portal to dark matter physics.  The phenomenology of these models often predicts non-standard final state objects like long-lived particle decays to $e^+e^-$ pairs \cite{Bertuzzo:2018itn,Ballett:2018ynz}. As a consequence, methods similar to object tagging for colliders are of growing interest and use, particularly for detectors with high spatial and calorimetric resolution.

With the need for increased precision, use of ML has been growing in neutrino experiments, through all stages of the data pipeline. %
Within the last five years, deep learning algorithms have become increasingly popular, and have enabled large improvements in performance and physics reach compared to traditional methods. Deep learning applications now span the full extent of data processing for neutrino experiments, including data acquisition \cite{Jwa:2019zlh,ArgoNeuT:2021xtd,Uboldi:2021jyj,ARIANNA:2021inb}, final data analysis (see, e.g.~\cite{NOvA:2021smv,MicroBooNE:2021jwr}), and in particular data reconstruction (see, e.g.~\cite{Baldi:2018qhe,MicroBooNE:2021ojx,KM3NeT:2020zod}), due to the increasing use of large, high-resolution tracking calorimeters as neutrino detectors.

Compared to collider experiments, %
neutrino detectors usually comprise a large and homogeneous target where neutrino interactions can occur uniformly, leading to interaction signals that are translationally invariant. The features of interest are topological characteristics of illuminated pixels in a fixed segmented (2D or 3D) geometry, with spatially and temporally connected pixels correlating to a specific particle ``track''--the shape and length of the track, and the pixel intensity being indicative of the particle type and kinematic properties. As such, CNNs and other deep learning algorithms associated with computer vision have found strong relevance and thus become a particularly common and important tool in neutrino experiments~\cite{Radovic:2018dip}. A pioneering demonstration of CNNs applied to neutrino event classification based on their topology without the need for detailed reconstruction was performed in \cite{Aurisano:2016jvx}, demonstrating improved physics performance over traditional algorithms for the case of the NO$\nu$A \cite{NOvA:2007rmc} experiment. A comprehensive overview of machine learning in neutrino experiments is provided in Ref.~\cite{Psihas:2020pby}; the subsequent paragraphs highlight some of the more recent advancements.

\vspace{2mm}

\noindent\textbf{Image Recognition.} A detector technology ideally suited for computer vision applications in neutrino physics is that of liquid argon time projection chambers (LArTPCs)---employed by the future DUNE \cite{DUNE:2020lwj}, current MicroBooNE \cite{MicroBooNE:2016pwy}, and upcoming SBN \cite{MicroBooNE:2015bmn} experiments. These detectors function as stereoscopic image streaming devices, offering the possibility of direct application of image recognition at as early as the data acquisition stage. For the time being, however, applications of deep learning algorithms for these experiments have predominantly been focused on data reconstruction and final analysis tasks.

The DUNE collaboration is exploring the use of CNNs for neutrino interaction classification that allows for simultaneous identification and classification of neutrino interaction final states and neutrino interaction type \cite{DUNE:2020gpm}, while kinematics reconstruction with the use of 2D and 3D regression CNNs has been proposed in \cite{Liu:2020pzv}. MicroBooNE was first to demonstrate the successful use and advantages of CNNs for classifying signal vs.~background images, where signal images contain particles produced by a neutrino interaction \cite{MicroBooNE:2016dpb}. Moving more toward implementation of deep learning algorithms for reconstruction tasks rather than end-user analysis tasks, MicroBooNE has further demonstrated multi-particle type identification \cite{MicroBooNE:2016dpb,MicroBooNE:2020hho}, as well as pixel-level object prediction \cite{MicroBooNE:2018kka} through a successful first application of a UNet style CNN architecture and semantic segmentation techniques to LArTPC neutrino data. In an effort to minimize computational needs for CNN applications to (particularly sparse) LArTPC data analysis, MicroBooNE has also demonstrated the use of semantic segmentation with a sparse CNN for event reconstruction \cite{MicroBooNE:2020yze}. This was motivated by Ref.\cite{Domine:2019zhm}, proposing a spatially sparse, UResNet-style architecture for particle-wise segmentation labels for LArTPC-like datasets \cite{Domine:2019zhm}. SBN's near detector (SBND) has also applied a UResNet network for cosmic background removal \cite{SBND:2020eho}, demonstrating scalability of larger CNNs for pixel-level signal-background rejection task with larger images~\cite{SBND:2020eho}. The technique has also been employed by MicroBooNE as part of the experiment’s flagship BSM physics search~\cite{MicroBooNE:2020yze}.

Reference~\cite{Drielsma:2021jdv} recently demonstrated a multi-task, end-to-end optimization of a full data reconstruction chain for imaging detectors using sparse CNN and graph neural networks. This chain consists of multiple neural network modules where each module performs a traditional data reconstruction task (e.g.~clustering of pixels, particle type and momentum estimation, reconstruction of a particle flow and hierarchy among parents and children)~\cite{Domine:2019zhm,Domine:2020tlx,Koh:2020snv,DeepLearnPhysics:2020hut}, and it is being integrated into DUNE’s near-detector analysis and more experiments of this class~\cite{Adams:2020vlj}.

Beyond LArTPC experiments, additional pioneering applications of deep learning have exploited the unique detector geometry, working principle, or raw data structure of a given experiment. 
For example, the NO$\nu$A detector's 2D imaging working principle has also prompted the development and use of a CNN for neutrino event identification and reconstruction \cite{Psihas:2017yuc}, combining two orthogonal visual projections of given neutrino interactions in the detector in a Siamese network structure and allowing the network to learn from independent 2D depictions of 3D energy depositions in neutrino interactions. The Kamland-Zen detector, which is a roughly spherical detector, has made use of spherical CNNs for their physics analyses \cite{KamLAND-Zen:2016pfg}. MINERvA has adapted CNNs with the use of domain adversarial neural networks as a way of mitigating unknown biases in the training inputs \cite{MINERvA:2018smv}. Finally, Daya-Bay, in one of the first demonstrations of applying \textit{unsupervised} DNNs for pattern recognition, has demonstrated the use of dimensionality reduction as a method for interpreting features extracted by a CNN employed in separating neutrino interactions from noise in their detector \cite{Racah:2016gnm}. 

\vspace{2mm}

A unique case of deep learning reconstruction applications is that of the IceCube experiment, whose non-uniform detector configuration makes it less fitting for CNNs. As such, IceCube has turned to the application of graph neural networks, which are capable of dealing with irregular data geometries and sizes more effectively, finding significant improvement in physics performance over traditional algorithms \cite{IceCube:2018gms,IceCube:2021dvc}. 

\vspace{2mm}

\noindent\textbf{Data Acquisition.} With the use of increasingly larger neutrino interaction target volumes and finer readout segmentation, data challenges for neutrino experiments begin to approach those of current collider experiments. For example, DUNE's multiple far detectors \cite{DUNE:2020mra} %
will each generate raw data rates of several terabytes per second, and plan to be operated for at least a decade, requiring also 100\% live-time in order to be sensitive to neutrinos from nearby supernova bursts or other stochastic BSM signals. With ML becoming increasingly common in neutrino experiments, the community is further steering its attention toward hardware acceleration of ML-based inference. GPU-accelerated ML inference as a service for computing in neutrino experiments is discussed in \cite{Wang:2020fjr}, while new developments are also targeting GPU- or FPGA-based acceleration for use of machine learning algorithms such as 1D or 2D CNNs in real-time or online processing of raw LArTPC data at the data acquisition and trigger level \cite{Jwa:2019zlh,ArgoNeuT:2021xtd,Uboldi:2021jyj}. %

\subsection{Rare Event Searches}
\label{sec:rare}

 Dark matter (DM) direct detection and neutrinoless double beta decay ($0\nu\beta\beta$) experiments have observation of BSM processes as their primary goal, most commonly in the form of scattering of weakly interacting massive particles or $0\nu\beta\beta$ via light Majorana neutrino exchange, respectively. Though machine learning in such searches is not as prevalent as in collider physics or neutrino experiments, perhaps due in part to its particular sensitivity to mismodeling at the few-event level, its use has grown in recent years. At present, applications primarily consist of methods to improve either event reconstruction or signal/background discrimination.

\vspace{2mm}

\noindent\textbf{Shallow Discriminators.} Relatively simple methods such as BDTs remain popular, likely due to their robustness and ease of use. BDTs have been employed for effective removal of rare backgrounds in xenon DM detectors such as LUX \cite{LUX-S2, LUX-EFT-Run4, Rossiter-thesis} and XENON1T \cite{XENON1T-nu}, using high-level engineered features as inputs and applying a threshold on the output prior to future stages of analysis. Similar approaches have been used to remove noise pulses in solid state DM detectors such as SuperCDMS \cite{SuperCDMS-2018} and COSINE-100 \cite{COSINE-100-threshold}. In the EXO-200 $0\nu\beta\beta$ experiment, this approach was taken a step further by combining the BDT output with other variables such as measured energy in the final two-dimensional fit \cite{EXO-excited, EXO-upgraded, EXO-complete}. 

\vspace{2mm}

\noindent\textbf{Time Series Analysis.} Machine learning has also been harnessed for more complete analysis of time-series data (waveforms). More accurate extraction of pulse parameters, such as height and start time, has been demonstrated in solid state detectors using PCA \cite{Yu-PCA}, CNNs, and RNNs \cite{wagner2020-thesis}, while efficient (low-dimension) summaries of pulse shape have been achieved using BDTs \cite{LUX-S2} and convolutional autoencoders \cite{Ge-pulse-shape}. Background discrimination via pulse shape has shown promise using either fully-connected neural networks or CNNs for bubble chamber DM experiments such as PICO \cite{PICO:2018fbf, PICO-semi-supervised}, germanium neutrinoless double beta decay detectors \cite{Ge-pulse-shape}, and cryogenic calcium tungstate DM detectors such as CRESST \cite{muhlmann2019-thesis}.

\vspace{2mm}

\noindent\textbf{Image Recognition.} Drawing on successful image recognition techniques in industry and neutrino experiments \cite{MicroBooNE:2016dpb}, rare event searches have implemented 2D CNNs to take full advantage of spatial (or spatiotemporal) information using light and charge detector hit patterns. This approach is particularly effective at topological signal/background discrimination in Xe time projection chambers (TPCs) for $0\nu\beta\beta$ \cite{Ai-gas-TPC-2018}, including NEXT \cite{NEXT-2017, NEXT-2021}, nEXO \cite{nEXO-charge, nEXO-design}, and PandaX-III \cite{PandaX-III-MC}, as well as the KamLAND-Zen $0\nu\beta\beta$ scintillator experiment \cite{Kamland-2019} and even nuclear emulsion DM searches \cite{emulsion-dm}. 2D detector hit patterns have also been used directly as inputs to CNNs, fully-connected networks, or Bayesian optimization methods \cite{bolfi-dm} for more precise position reconstruction in noble element TPCs, such as EXO-200 \cite{EXO-deep}, XENON1T \cite{XENON1T-recon}, and the Ar-based DarkSide DM experiments \cite{DarkSide-recon, DarkSide-CNN}.

\vspace{2mm}

\noindent\textbf{Less-than-supervised Searches.} A common theme across rare event search experiments is the desire for reduced reliance on simulations, through fully data-driven training, semi-supervised, or unsupervised learning approaches. Xe TPC experiments such as EXO-200 and LUX have trained ML models to better reconstruct events with either a missing charge or light signal, using data events with both present to provide labels for supervised training, after re-weighting to reduce reliance on features exclusive to the original training domain such as its energy spectrum \cite{EXO-deep, LUX-S2}. Imperfect labels from selection of data events with unique signatures such as gamma escape peaks can be used to better identify single and multiple scatter events in germanium detectors \cite{Ge-pulse-shape}. Supplementing simulated events with unlabeled data events, using confident network predictions as truth labels in subsequent training iterations, has been shown to boost performance in PICO \cite{PICO-semi-supervised, PICO:2018fbf}; in NEXT, features extracted through convolutional layers were found to agree more closely in simulations and data when augmenting training events by repeating them after physics-invariant transformations such as rotation \cite{NEXT-2021}. Fully unsupervised methods, such as tSNE, PCA, and autoencoders, have been successfully employed to separate unwanted backgrounds through clustering in CRESST \cite{wagner2020-thesis}. 

In many cases, the choice of representation of the data appears to have a greater effect on the results than the algorithm or architecture. PICO has observed improved particle discrimination from a simple fully-connected network applied to the first few components of Fourier space data than a CNN in the time domain \cite{PICO:2018fbf}. EXO-200 sees minimal improvement in discrimination when using a CNN classifier over a BDT with engineered features \cite{EXO-complete}, while DarkSide-20k achieves improved performance using a fully-connected network over a CNN \cite{DarkSide-recon}. In contrast, initial training of a convolutional autoencoder followed by training a fully-connected network on its encoded latent space has shown good results in germanium \cite{Ge-pulse-shape}, suggesting that separating the tasks of representation and classification may be more robust.

In the near future, an increased emphasis on extracting physical insights from unsupervised methods, careful quantification of uncertainties, and clever construction of training sets to reduce domain discrepancies will be crucial to taking full advantage of the benefits of deep learning in rare event searches.

\section{Outlook and Challenges}
\label{sec:outlook}

The development and deployment of machine learning methods in the search for new fundamental physics is becoming urgent given the dearth of evidence from traditional methods.  New particles may be discoverable with existing and near-future experiments, but we may need to explore our data in its natural high-dimensionality and to reduce our model dependence in order to uncover the new phenomena.  New methods are being developed at a rapid rate (inside and outside of experimental collaborations) and we are starting to see many innovative applications to both data processing and data analysis.

Fundamental physics applications of machine learning have unique challenges that may not be solved by industry. We are typically looking for ultra rare and subtle deviations from the Standard Model.  Furthermore, it is often the case that no one datum is uniquely anomalous - only in the context of many examples can we build statistical evidence for a discovery.  At the same time, it will also be essential to integrate state-of-the-art deep learning tools like TensorFlow~\cite{tensorflow2015-whitepaper} and PyTorch~\cite{NEURIPS2019_9015} into analysis workflows in order to make the best use of new techniques.  There are also serious computing challenges associated with training and inference.  For example, the BSM exclusion limits in Ref.~\cite{collaboration2020dijet} required training 20k neural networks.  This is because the event selection depends on the data and the data changes whenever a different amount of signal is injected.  Especially for less-than-supervised searches, a radically new approach to data preservation and analysis reinterpretation will be required.

While this review has focused on \textit{direct} searches for new particles, there are also many \textit{indirect} searches in the form of Standard Model measurements.  Analyses probing small differences between the Standard Model and the data are also increasingly employing machine learning strategies for data corrections in the presence of BSM~\cite{Bellagente:2019uyp,Komiske:2021vym}, effective field theory analysis~\cite{Brehmer:2019xox,Brehmer:2018hga,Brehmer:2018kdj,Brehmer:2018eca,Chatterjee:2021nms,Chen:2020mev,Erbin:2018csv}, reinterpretation~\cite{Caron:2016hib,Bertone:2016mdy,Kronheim:2020vct}, and more~\cite{Feickert:2021ajf}.

Astronomy and cosmology also offer unique and powerful windows into many types of physics beyond the Standard Model (e.g.\ dark matter, dark energy, baryogenesis, axions). These fields also generate enormous and complex datasets, with ongoing and upcoming observatories such as Gaia~\cite{2018A&A...616A...2L}, Large-aperture Synoptic Survey Telescope~\cite{2019ApJ...873..111I}, Laser Interferometer Gravitational-Wave Observatory~\cite{LIGOScientific:2007fwp}, and the Square Kilometer Array~\cite{2020PASA...37....2W}. 
Much effort is already being invested in applying ML methods to accelerate data analysis and discovery in these fields; see, e.g.,\ Ref.~\cite{george_stein_2020_4024768} for a comprehensive list of ML applications to cosmology. The connections between these areas and particle physics will be extremely interesting and important to explore going forward.

While many of the challenges associated with machine learning are technical, a new community mindset will also be required to take full advantage of the new tools.  How can we make discoveries using representations we cannot easily visualize or understand?  Forging physics with statistical learning provides a new path forward toward robust, sensitive, and reliable methods for uncovering the fundamental structure of nature.

\begin{acknowledgments}
S. Kravitz and B. Nachman are supported by the U.S.~Department of Energy, Office of Science under contract DE-AC02-05CH11231.
G. Kasieczka acknowledges the support of the Deutsche Forschungsgemeinschaft (DFG, German Research Foundation) under Germany’s Excellence Strategy – EXC 2121 “Quantum Universe” – 390833306. The work of D. Shih was supported by DOE grant DOE-SC0010008. G. Karagiorgi is supported by the U.S.~National Science Foundation under Grant No.~PHY-1753228.
\end{acknowledgments}

\bibliography{other,HEPML,neutrino}

\begin{thebibliography}{272}%
\makeatletter
\providecommand \@ifxundefined [1]{%
 \@ifx{#1\undefined}
}%
\providecommand \@ifnum [1]{%
 \ifnum #1\expandafter \@firstoftwo
 \else \expandafter \@secondoftwo
 \fi
}%
\providecommand \@ifx [1]{%
 \ifx #1\expandafter \@firstoftwo
 \else \expandafter \@secondoftwo
 \fi
}%
\providecommand \natexlab [1]{#1}%
\providecommand \enquote  [1]{``#1''}%
\providecommand \bibnamefont  [1]{#1}%
\providecommand \bibfnamefont [1]{#1}%
\providecommand \citenamefont [1]{#1}%
\providecommand \href@noop [0]{\@secondoftwo}%
\providecommand \href [0]{\begingroup \@sanitize@url \@href}%
\providecommand \@href[1]{\@@startlink{#1}\@@href}%
\providecommand \@@href[1]{\endgroup#1\@@endlink}%
\providecommand \@sanitize@url [0]{\catcode `\\12\catcode `\$12\catcode
  `\&12\catcode `\#12\catcode `\^12\catcode `\_12\catcode `\%12\relax}%
\providecommand \@@startlink[1]{}%
\providecommand \@@endlink[0]{}%
\providecommand \url  [0]{\begingroup\@sanitize@url \@url }%
\providecommand \@url [1]{\endgroup\@href {#1}{\urlprefix }}%
\providecommand \urlprefix  [0]{URL }%
\providecommand \Eprint [0]{\href }%
\providecommand \doibase [0]{https://doi.org/}%
\providecommand \selectlanguage [0]{\@gobble}%
\providecommand \bibinfo  [0]{\@secondoftwo}%
\providecommand \bibfield  [0]{\@secondoftwo}%
\providecommand \translation [1]{[#1]}%
\providecommand \BibitemOpen [0]{}%
\providecommand \bibitemStop [0]{}%
\providecommand \bibitemNoStop [0]{.\EOS\space}%
\providecommand \EOS [0]{\spacefactor3000\relax}%
\providecommand \BibitemShut  [1]{\csname bibitem#1\endcsname}%
\let\auto@bib@innerbib\@empty
\bibitem [{\citenamefont {Aad}\ \emph {et~al.}(2012)\citenamefont {Aad} \emph
  {et~al.}}]{Aad:2012tfa}%
  \BibitemOpen
  \bibfield  {author} {\bibinfo {author} {\bibfnamefont {G.}~\bibnamefont
  {Aad}} \emph {et~al.} (\bibinfo {collaboration} {ATLAS}),\ }\bibfield
  {title} {\bibinfo {title} {{Observation of a new particle in the search for
  the Standard Model Higgs boson with the ATLAS detector at the LHC}},\ }\href
  {https://doi.org/10.1016/j.physletb.2012.08.020} {\bibfield  {journal}
  {\bibinfo  {journal} {Phys. Lett. B}\ }\textbf {\bibinfo {volume} {716}},\
  \bibinfo {pages} {1} (\bibinfo {year} {2012})},\ \Eprint
  {https://arxiv.org/abs/1207.7214} {arXiv:1207.7214 [hep-ex]} \BibitemShut
  {NoStop}%
\bibitem [{\citenamefont {Chatrchyan}\ \emph {et~al.}(2012)\citenamefont
  {Chatrchyan} \emph {et~al.}}]{Chatrchyan:2012ufa}%
  \BibitemOpen
  \bibfield  {author} {\bibinfo {author} {\bibfnamefont {S.}~\bibnamefont
  {Chatrchyan}} \emph {et~al.} (\bibinfo {collaboration} {CMS}),\ }\bibfield
  {title} {\bibinfo {title} {{Observation of a New Boson at a Mass of 125 GeV
  with the CMS Experiment at the LHC}},\ }\href
  {https://doi.org/10.1016/j.physletb.2012.08.021} {\bibfield  {journal}
  {\bibinfo  {journal} {Phys. Lett. B}\ }\textbf {\bibinfo {volume} {716}},\
  \bibinfo {pages} {30} (\bibinfo {year} {2012})},\ \Eprint
  {https://arxiv.org/abs/1207.7235} {arXiv:1207.7235 [hep-ex]} \BibitemShut
  {NoStop}%
\bibitem [{\citenamefont {Zyla}\ \emph {et~al.}(2020)\citenamefont {Zyla} \emph
  {et~al.}}]{Zyla:2020zbs}%
  \BibitemOpen
  \bibfield  {author} {\bibinfo {author} {\bibfnamefont {P.}~\bibnamefont
  {Zyla}} \emph {et~al.} (\bibinfo {collaboration} {Particle Data Group}),\
  }\bibfield  {title} {\bibinfo {title} {{Review of Particle Physics}},\ }\href
  {https://doi.org/10.1093/ptep/ptaa104} {\bibfield  {journal} {\bibinfo
  {journal} {PTEP}\ }\textbf {\bibinfo {volume} {2020}},\ \bibinfo {pages}
  {083C01} (\bibinfo {year} {2020})}\BibitemShut {NoStop}%
\bibitem [{\citenamefont {Fukuda}\ \emph {et~al.}(1998)\citenamefont {Fukuda}
  \emph {et~al.}}]{Fukuda:1998mi}%
  \BibitemOpen
  \bibfield  {author} {\bibinfo {author} {\bibfnamefont {Y.}~\bibnamefont
  {Fukuda}} \emph {et~al.} (\bibinfo {collaboration} {Super-Kamiokande}),\
  }\bibfield  {title} {\bibinfo {title} {{Evidence for oscillation of
  atmospheric neutrinos}},\ }\href
  {https://doi.org/10.1103/PhysRevLett.81.1562} {\bibfield  {journal} {\bibinfo
   {journal} {Phys. Rev. Lett.}\ }\textbf {\bibinfo {volume} {81}},\ \bibinfo
  {pages} {1562} (\bibinfo {year} {1998})},\ \Eprint
  {https://arxiv.org/abs/hep-ex/9807003} {arXiv:hep-ex/9807003} \BibitemShut
  {NoStop}%
\bibitem [{\citenamefont {Ahmad}\ \emph {et~al.}(2002)\citenamefont {Ahmad}
  \emph {et~al.}}]{Ahmad:2002jz}%
  \BibitemOpen
  \bibfield  {author} {\bibinfo {author} {\bibfnamefont {Q.}~\bibnamefont
  {Ahmad}} \emph {et~al.} (\bibinfo {collaboration} {SNO}),\ }\bibfield
  {title} {\bibinfo {title} {{Direct evidence for neutrino flavor
  transformation from neutral current interactions in the Sudbury Neutrino
  Observatory}},\ }\href {https://doi.org/10.1103/PhysRevLett.89.011301}
  {\bibfield  {journal} {\bibinfo  {journal} {Phys. Rev. Lett.}\ }\textbf
  {\bibinfo {volume} {89}},\ \bibinfo {pages} {011301} (\bibinfo {year}
  {2002})},\ \Eprint {https://arxiv.org/abs/nucl-ex/0204008}
  {arXiv:nucl-ex/0204008} \BibitemShut {NoStop}%
\bibitem [{\citenamefont {Canetti}\ \emph {et~al.}(2012)\citenamefont
  {Canetti}, \citenamefont {Drewes},\ and\ \citenamefont
  {Shaposhnikov}}]{Canetti:2012zc}%
  \BibitemOpen
  \bibfield  {author} {\bibinfo {author} {\bibfnamefont {L.}~\bibnamefont
  {Canetti}}, \bibinfo {author} {\bibfnamefont {M.}~\bibnamefont {Drewes}},\
  and\ \bibinfo {author} {\bibfnamefont {M.}~\bibnamefont {Shaposhnikov}},\
  }\bibfield  {title} {\bibinfo {title} {{Matter and Antimatter in the
  Universe}},\ }\href {https://doi.org/10.1088/1367-2630/14/9/095012}
  {\bibfield  {journal} {\bibinfo  {journal} {New J. Phys.}\ }\textbf {\bibinfo
  {volume} {14}},\ \bibinfo {pages} {095012} (\bibinfo {year} {2012})},\
  \Eprint {https://arxiv.org/abs/1204.4186} {arXiv:1204.4186 [hep-ph]}
  \BibitemShut {NoStop}%
\bibitem [{\citenamefont {Abel}\ \emph {et~al.}(2020)\citenamefont {Abel} \emph
  {et~al.}}]{Abel:2020gbr}%
  \BibitemOpen
  \bibfield  {author} {\bibinfo {author} {\bibfnamefont {C.}~\bibnamefont
  {Abel}} \emph {et~al.} (\bibinfo {collaboration} {nEDM}),\ }\bibfield
  {title} {\bibinfo {title} {{Measurement of the permanent electric dipole
  moment of the neutron}},\ }\href
  {https://doi.org/10.1103/PhysRevLett.124.081803} {\bibfield  {journal}
  {\bibinfo  {journal} {Phys. Rev. Lett.}\ }\textbf {\bibinfo {volume} {124}},\
  \bibinfo {pages} {081803} (\bibinfo {year} {2020})},\ \Eprint
  {https://arxiv.org/abs/2001.11966} {arXiv:2001.11966 [hep-ex]} \BibitemShut
  {NoStop}%
\bibitem [{\citenamefont {Hocker}\ \emph {et~al.}(2007)\citenamefont {Hocker}
  \emph {et~al.}}]{Hocker:2007ht}%
  \BibitemOpen
  \bibfield  {author} {\bibinfo {author} {\bibfnamefont {A.}~\bibnamefont
  {Hocker}} \emph {et~al.},\ }\bibfield  {title} {\bibinfo {title} {{TMVA -
  Toolkit for Multivariate Data Analysis}},\ }\href@noop {} {\  (\bibinfo
  {year} {2007})},\ \Eprint {https://arxiv.org/abs/physics/0703039}
  {arXiv:physics/0703039} \BibitemShut {NoStop}%
\bibitem [{\citenamefont {Radovic}\ \emph {et~al.}(2018)\citenamefont
  {Radovic}, \citenamefont {Williams}, \citenamefont {Rousseau}, \citenamefont
  {Kagan}, \citenamefont {Bonacorsi}, \citenamefont {Himmel}, \citenamefont
  {Aurisano}, \citenamefont {Terao},\ and\ \citenamefont
  {Wongjirad}}]{Radovic:2018dip}%
  \BibitemOpen
  \bibfield  {author} {\bibinfo {author} {\bibfnamefont {A.}~\bibnamefont
  {Radovic}}, \bibinfo {author} {\bibfnamefont {M.}~\bibnamefont {Williams}},
  \bibinfo {author} {\bibfnamefont {D.}~\bibnamefont {Rousseau}}, \bibinfo
  {author} {\bibfnamefont {M.}~\bibnamefont {Kagan}}, \bibinfo {author}
  {\bibfnamefont {D.}~\bibnamefont {Bonacorsi}}, \bibinfo {author}
  {\bibfnamefont {A.}~\bibnamefont {Himmel}}, \bibinfo {author} {\bibfnamefont
  {A.}~\bibnamefont {Aurisano}}, \bibinfo {author} {\bibfnamefont
  {K.}~\bibnamefont {Terao}},\ and\ \bibinfo {author} {\bibfnamefont
  {T.}~\bibnamefont {Wongjirad}},\ }\bibfield  {title} {\bibinfo {title}
  {{Machine learning at the energy and intensity frontiers of particle
  physics}},\ }\href {https://doi.org/10.1038/s41586-018-0361-2} {\bibfield
  {journal} {\bibinfo  {journal} {Nature}\ }\textbf {\bibinfo {volume} {560}},\
  \bibinfo {pages} {41} (\bibinfo {year} {2018})}\BibitemShut {NoStop}%
\bibitem [{\citenamefont {Feickert}\ and\ \citenamefont
  {Nachman}(2021)}]{Feickert:2021ajf}%
  \BibitemOpen
  \bibfield  {author} {\bibinfo {author} {\bibfnamefont {M.}~\bibnamefont
  {Feickert}}\ and\ \bibinfo {author} {\bibfnamefont {B.}~\bibnamefont
  {Nachman}},\ }\bibfield  {title} {\bibinfo {title} {{A Living Review of
  Machine Learning for Particle Physics}},\ }\href@noop {} {\  (\bibinfo {year}
  {2021})},\ \Eprint {https://arxiv.org/abs/2102.02770} {arXiv:2102.02770
  [hep-ph]} \BibitemShut {NoStop}%
\bibitem [{\citenamefont {Fukushima}\ and\ \citenamefont
  {Miyake}(1982)}]{fukushima1982neocognitron}%
  \BibitemOpen
  \bibfield  {author} {\bibinfo {author} {\bibfnamefont {K.}~\bibnamefont
  {Fukushima}}\ and\ \bibinfo {author} {\bibfnamefont {S.}~\bibnamefont
  {Miyake}},\ }\bibfield  {title} {\bibinfo {title} {Neocognitron: A
  self-organizing neural network model for a mechanism of visual pattern
  recognition},\ }in\ \href@noop {} {\emph {\bibinfo {booktitle} {Competition
  and cooperation in neural nets}}}\ (\bibinfo  {publisher} {Springer},\
  \bibinfo {year} {1982})\ pp.\ \bibinfo {pages} {267--285}\BibitemShut
  {NoStop}%
\bibitem [{\citenamefont {LeCun}\ \emph {et~al.}(1989)\citenamefont {LeCun},
  \citenamefont {Boser}, \citenamefont {Denker}, \citenamefont {Henderson},
  \citenamefont {Howard}, \citenamefont {Hubbard},\ and\ \citenamefont
  {Jackel}}]{lecun1989handwritten}%
  \BibitemOpen
  \bibfield  {author} {\bibinfo {author} {\bibfnamefont {Y.}~\bibnamefont
  {LeCun}}, \bibinfo {author} {\bibfnamefont {B.}~\bibnamefont {Boser}},
  \bibinfo {author} {\bibfnamefont {J.}~\bibnamefont {Denker}}, \bibinfo
  {author} {\bibfnamefont {D.}~\bibnamefont {Henderson}}, \bibinfo {author}
  {\bibfnamefont {R.}~\bibnamefont {Howard}}, \bibinfo {author} {\bibfnamefont
  {W.}~\bibnamefont {Hubbard}},\ and\ \bibinfo {author} {\bibfnamefont
  {L.}~\bibnamefont {Jackel}},\ }\bibfield  {title} {\bibinfo {title}
  {Handwritten digit recognition with a back-propagation network},\ }\href@noop
  {} {\bibfield  {journal} {\bibinfo  {journal} {Advances in neural information
  processing systems}\ }\textbf {\bibinfo {volume} {2}} (\bibinfo {year}
  {1989})}\BibitemShut {NoStop}%
\bibitem [{\citenamefont {de~Oliveira}\ \emph {et~al.}(2016)\citenamefont
  {de~Oliveira}, \citenamefont {Kagan}, \citenamefont {Mackey}, \citenamefont
  {Nachman},\ and\ \citenamefont {Schwartzman}}]{deOliveira:2015xxd}%
  \BibitemOpen
  \bibfield  {author} {\bibinfo {author} {\bibfnamefont {L.}~\bibnamefont
  {de~Oliveira}}, \bibinfo {author} {\bibfnamefont {M.}~\bibnamefont {Kagan}},
  \bibinfo {author} {\bibfnamefont {L.}~\bibnamefont {Mackey}}, \bibinfo
  {author} {\bibfnamefont {B.}~\bibnamefont {Nachman}},\ and\ \bibinfo {author}
  {\bibfnamefont {A.}~\bibnamefont {Schwartzman}},\ }\bibfield  {title}
  {\bibinfo {title} {{Jet-images — deep learning edition}},\ }\href
  {https://doi.org/10.1007/JHEP07(2016)069} {\bibfield  {journal} {\bibinfo
  {journal} {JHEP}\ }\textbf {\bibinfo {volume} {07}},\ \bibinfo {pages}
  {069}},\ \Eprint {https://arxiv.org/abs/1511.05190} {arXiv:1511.05190
  [hep-ph]} \BibitemShut {NoStop}%
\bibitem [{\citenamefont {Baldi}\ \emph {et~al.}(2016)\citenamefont {Baldi},
  \citenamefont {Bauer}, \citenamefont {Eng}, \citenamefont {Sadowski},\ and\
  \citenamefont {Whiteson}}]{Baldi:2016fql}%
  \BibitemOpen
  \bibfield  {author} {\bibinfo {author} {\bibfnamefont {P.}~\bibnamefont
  {Baldi}}, \bibinfo {author} {\bibfnamefont {K.}~\bibnamefont {Bauer}},
  \bibinfo {author} {\bibfnamefont {C.}~\bibnamefont {Eng}}, \bibinfo {author}
  {\bibfnamefont {P.}~\bibnamefont {Sadowski}},\ and\ \bibinfo {author}
  {\bibfnamefont {D.}~\bibnamefont {Whiteson}},\ }\bibfield  {title} {\bibinfo
  {title} {{Jet Substructure Classification in High-Energy Physics with Deep
  Neural Networks}},\ }\href {https://doi.org/10.1103/PhysRevD.93.094034}
  {\bibfield  {journal} {\bibinfo  {journal} {Phys. Rev. D}\ }\textbf {\bibinfo
  {volume} {93}},\ \bibinfo {pages} {094034} (\bibinfo {year} {2016})},\
  \Eprint {https://arxiv.org/abs/1603.09349} {arXiv:1603.09349 [hep-ex]}
  \BibitemShut {NoStop}%
\bibitem [{\citenamefont {Aurisano}\ \emph {et~al.}(2016)\citenamefont
  {Aurisano}, \citenamefont {Radovic}, \citenamefont {Rocco}, \citenamefont
  {Himmel}, \citenamefont {Messier}, \citenamefont {Niner}, \citenamefont
  {Pawloski}, \citenamefont {Psihas}, \citenamefont {Sousa},\ and\
  \citenamefont {Vahle}}]{Aurisano:2016jvx}%
  \BibitemOpen
  \bibfield  {author} {\bibinfo {author} {\bibfnamefont {A.}~\bibnamefont
  {Aurisano}}, \bibinfo {author} {\bibfnamefont {A.}~\bibnamefont {Radovic}},
  \bibinfo {author} {\bibfnamefont {D.}~\bibnamefont {Rocco}}, \bibinfo
  {author} {\bibfnamefont {A.}~\bibnamefont {Himmel}}, \bibinfo {author}
  {\bibfnamefont {M.~D.}\ \bibnamefont {Messier}}, \bibinfo {author}
  {\bibfnamefont {E.}~\bibnamefont {Niner}}, \bibinfo {author} {\bibfnamefont
  {G.}~\bibnamefont {Pawloski}}, \bibinfo {author} {\bibfnamefont
  {F.}~\bibnamefont {Psihas}}, \bibinfo {author} {\bibfnamefont
  {A.}~\bibnamefont {Sousa}},\ and\ \bibinfo {author} {\bibfnamefont
  {P.}~\bibnamefont {Vahle}},\ }\bibfield  {title} {\bibinfo {title} {{A
  Convolutional Neural Network Neutrino Event Classifier}},\ }\href
  {https://doi.org/10.1088/1748-0221/11/09/P09001} {\bibfield  {journal}
  {\bibinfo  {journal} {JINST}\ }\textbf {\bibinfo {volume} {11}}\bibfield
  {number} {\bibinfo  {number} { (09)},\ \bibinfo {pages} {P09001}},\ }\Eprint
  {https://arxiv.org/abs/1604.01444} {arXiv:1604.01444 [hep-ex]} \BibitemShut
  {NoStop}%
\bibitem [{\citenamefont {Rumelhart}\ \emph {et~al.}(1986)\citenamefont
  {Rumelhart}, \citenamefont {Hinton},\ and\ \citenamefont
  {Williams}}]{rumelhart1986learning}%
  \BibitemOpen
  \bibfield  {author} {\bibinfo {author} {\bibfnamefont {D.~E.}\ \bibnamefont
  {Rumelhart}}, \bibinfo {author} {\bibfnamefont {G.~E.}\ \bibnamefont
  {Hinton}},\ and\ \bibinfo {author} {\bibfnamefont {R.~J.}\ \bibnamefont
  {Williams}},\ }\bibfield  {title} {\bibinfo {title} {Learning representations
  by back-propagating errors},\ }\href@noop {} {\bibfield  {journal} {\bibinfo
  {journal} {nature}\ }\textbf {\bibinfo {volume} {323}},\ \bibinfo {pages}
  {533} (\bibinfo {year} {1986})}\BibitemShut {NoStop}%
\bibitem [{\citenamefont {Hochreiter}\ and\ \citenamefont
  {Schmidhuber}(1997)}]{LSTM}%
  \BibitemOpen
  \bibfield  {author} {\bibinfo {author} {\bibfnamefont {S.}~\bibnamefont
  {Hochreiter}}\ and\ \bibinfo {author} {\bibfnamefont {J.}~\bibnamefont
  {Schmidhuber}},\ }\bibfield  {title} {\bibinfo {title} {Long short-term
  memory},\ }\href {https://doi.org/10.1162/neco.1997.9.8.1735} {\bibfield
  {journal} {\bibinfo  {journal} {Neural computation}\ }\textbf {\bibinfo
  {volume} {9}},\ \bibinfo {pages} {1735} (\bibinfo {year} {1997})}\BibitemShut
  {NoStop}%
\bibitem [{\citenamefont {Guest}\ \emph {et~al.}(2016)\citenamefont {Guest},
  \citenamefont {Collado}, \citenamefont {Baldi}, \citenamefont {Hsu},
  \citenamefont {Urban},\ and\ \citenamefont {Whiteson}}]{Guest:2016iqz}%
  \BibitemOpen
  \bibfield  {author} {\bibinfo {author} {\bibfnamefont {D.}~\bibnamefont
  {Guest}}, \bibinfo {author} {\bibfnamefont {J.}~\bibnamefont {Collado}},
  \bibinfo {author} {\bibfnamefont {P.}~\bibnamefont {Baldi}}, \bibinfo
  {author} {\bibfnamefont {S.-C.}\ \bibnamefont {Hsu}}, \bibinfo {author}
  {\bibfnamefont {G.}~\bibnamefont {Urban}},\ and\ \bibinfo {author}
  {\bibfnamefont {D.}~\bibnamefont {Whiteson}},\ }\bibfield  {title} {\bibinfo
  {title} {{Jet Flavor Classification in High-Energy Physics with Deep Neural
  Networks}},\ }\href {https://doi.org/10.1103/PhysRevD.94.112002} {\bibfield
  {journal} {\bibinfo  {journal} {Phys. Rev.}\ }\textbf {\bibinfo {volume}
  {D94}},\ \bibinfo {pages} {112002} (\bibinfo {year} {2016})},\ \Eprint
  {https://arxiv.org/abs/1607.08633} {arXiv:1607.08633 [hep-ex]} \BibitemShut
  {NoStop}%
\bibitem [{\citenamefont {Louppe}\ \emph {et~al.}(2019)\citenamefont {Louppe},
  \citenamefont {Cho}, \citenamefont {Becot},\ and\ \citenamefont
  {Cranmer}}]{Louppe:2017ipp}%
  \BibitemOpen
  \bibfield  {author} {\bibinfo {author} {\bibfnamefont {G.}~\bibnamefont
  {Louppe}}, \bibinfo {author} {\bibfnamefont {K.}~\bibnamefont {Cho}},
  \bibinfo {author} {\bibfnamefont {C.}~\bibnamefont {Becot}},\ and\ \bibinfo
  {author} {\bibfnamefont {K.}~\bibnamefont {Cranmer}},\ }\bibfield  {title}
  {\bibinfo {title} {{QCD-Aware Recursive Neural Networks for Jet Physics}},\
  }\href {https://doi.org/10.1007/JHEP01(2019)057} {\bibfield  {journal}
  {\bibinfo  {journal} {JHEP}\ }\textbf {\bibinfo {volume} {01}},\ \bibinfo
  {pages} {057}},\ \Eprint {https://arxiv.org/abs/1702.00748} {arXiv:1702.00748
  [hep-ph]} \BibitemShut {NoStop}%
\bibitem [{\citenamefont {Dolan}\ and\ \citenamefont
  {Ore}(2021)}]{Dolan:2020qkr}%
  \BibitemOpen
  \bibfield  {author} {\bibinfo {author} {\bibfnamefont {M.~J.}\ \bibnamefont
  {Dolan}}\ and\ \bibinfo {author} {\bibfnamefont {A.}~\bibnamefont {Ore}},\
  }\bibfield  {title} {\bibinfo {title} {{Equivariant Energy Flow Networks for
  Jet Tagging}},\ }\href {https://doi.org/10.1103/PhysRevD.103.074022}
  {\bibfield  {journal} {\bibinfo  {journal} {Phys. Rev. D}\ }\textbf {\bibinfo
  {volume} {103}},\ \bibinfo {pages} {074022} (\bibinfo {year} {2021})},\
  \Eprint {https://arxiv.org/abs/2012.00964} {arXiv:2012.00964 [hep-ph]}
  \BibitemShut {NoStop}%
\bibitem [{\citenamefont {Serviansky}\ \emph {et~al.}(2020)\citenamefont
  {Serviansky}, \citenamefont {Segol}, \citenamefont {Shlomi}, \citenamefont
  {Cranmer}, \citenamefont {Gross}, \citenamefont {Maron},\ and\ \citenamefont
  {Lipman}}]{serviansky2020set2graph}%
  \BibitemOpen
  \bibfield  {author} {\bibinfo {author} {\bibfnamefont {H.}~\bibnamefont
  {Serviansky}}, \bibinfo {author} {\bibfnamefont {N.}~\bibnamefont {Segol}},
  \bibinfo {author} {\bibfnamefont {J.}~\bibnamefont {Shlomi}}, \bibinfo
  {author} {\bibfnamefont {K.}~\bibnamefont {Cranmer}}, \bibinfo {author}
  {\bibfnamefont {E.}~\bibnamefont {Gross}}, \bibinfo {author} {\bibfnamefont
  {H.}~\bibnamefont {Maron}},\ and\ \bibinfo {author} {\bibfnamefont
  {Y.}~\bibnamefont {Lipman}},\ }\href@noop {} {\bibinfo {title} {Set2graph:
  Learning graphs from sets}} (\bibinfo {year} {2020}),\ \Eprint
  {https://arxiv.org/abs/2002.08772} {arXiv:2002.08772 [cs.LG]} \BibitemShut
  {NoStop}%
\bibitem [{\citenamefont {Bogatskiy}\ \emph {et~al.}(2020)\citenamefont
  {Bogatskiy}, \citenamefont {Anderson}, \citenamefont {Offermann},
  \citenamefont {Roussi}, \citenamefont {Miller},\ and\ \citenamefont
  {Kondor}}]{Bogatskiy:2020tje}%
  \BibitemOpen
  \bibfield  {author} {\bibinfo {author} {\bibfnamefont {A.}~\bibnamefont
  {Bogatskiy}}, \bibinfo {author} {\bibfnamefont {B.}~\bibnamefont {Anderson}},
  \bibinfo {author} {\bibfnamefont {J.~T.}\ \bibnamefont {Offermann}}, \bibinfo
  {author} {\bibfnamefont {M.}~\bibnamefont {Roussi}}, \bibinfo {author}
  {\bibfnamefont {D.~W.}\ \bibnamefont {Miller}},\ and\ \bibinfo {author}
  {\bibfnamefont {R.}~\bibnamefont {Kondor}},\ }\bibfield  {title} {\bibinfo
  {title} {{Lorentz Group Equivariant Neural Network for Particle Physics}},\
  }\href@noop {} {\  (\bibinfo {year} {2020})},\ \Eprint
  {https://arxiv.org/abs/2006.04780} {arXiv:2006.04780 [hep-ph]} \BibitemShut
  {NoStop}%
\bibitem [{\citenamefont {Shimmin}(2021)}]{Shimmin:2021pkm}%
  \BibitemOpen
  \bibfield  {author} {\bibinfo {author} {\bibfnamefont {C.}~\bibnamefont
  {Shimmin}},\ }\bibfield  {title} {\bibinfo {title} {{Particle Convolution for
  High Energy Physics}}\ }(\bibinfo {year} {2021})\ \Eprint
  {https://arxiv.org/abs/2107.02908} {arXiv:2107.02908 [hep-ph]} \BibitemShut
  {NoStop}%
\bibitem [{\citenamefont {Zaheer}\ \emph {et~al.}(2017)\citenamefont {Zaheer},
  \citenamefont {Kottur}, \citenamefont {Ravanbakhsh}, \citenamefont
  {P{\'{o}}czos}, \citenamefont {Salakhutdinov},\ and\ \citenamefont
  {Smola}}]{DBLP:conf/nips/ZaheerKRPSS17}%
  \BibitemOpen
  \bibfield  {author} {\bibinfo {author} {\bibfnamefont {M.}~\bibnamefont
  {Zaheer}}, \bibinfo {author} {\bibfnamefont {S.}~\bibnamefont {Kottur}},
  \bibinfo {author} {\bibfnamefont {S.}~\bibnamefont {Ravanbakhsh}}, \bibinfo
  {author} {\bibfnamefont {B.}~\bibnamefont {P{\'{o}}czos}}, \bibinfo {author}
  {\bibfnamefont {R.}~\bibnamefont {Salakhutdinov}},\ and\ \bibinfo {author}
  {\bibfnamefont {A.~J.}\ \bibnamefont {Smola}},\ }\bibfield  {title} {\bibinfo
  {title} {Deep sets},\ }in\ \href {http://papers.nips.cc/paper/6931-deep-sets}
  {\emph {\bibinfo {booktitle} {Advances in Neural Information Processing
  Systems 30: Annual Conference on Neural Information Processing Systems 2017,
  4-9 December 2017, Long Beach, CA, {USA}}}}\ (\bibinfo {year} {2017})\ pp.\
  \bibinfo {pages} {3391--3401}\BibitemShut {NoStop}%
\bibitem [{\citenamefont {Komiske}\ \emph {et~al.}(2019)\citenamefont
  {Komiske}, \citenamefont {Metodiev},\ and\ \citenamefont
  {Thaler}}]{Komiske:2018cqr}%
  \BibitemOpen
  \bibfield  {author} {\bibinfo {author} {\bibfnamefont {P.~T.}\ \bibnamefont
  {Komiske}}, \bibinfo {author} {\bibfnamefont {E.~M.}\ \bibnamefont
  {Metodiev}},\ and\ \bibinfo {author} {\bibfnamefont {J.}~\bibnamefont
  {Thaler}},\ }\bibfield  {title} {\bibinfo {title} {{Energy Flow Networks:
  Deep Sets for Particle Jets}},\ }\href
  {https://doi.org/10.1007/JHEP01(2019)121} {\bibfield  {journal} {\bibinfo
  {journal} {JHEP}\ }\textbf {\bibinfo {volume} {01}},\ \bibinfo {pages}
  {121}},\ \Eprint {https://arxiv.org/abs/1810.05165} {arXiv:1810.05165
  [hep-ph]} \BibitemShut {NoStop}%
\bibitem [{\citenamefont {Henrion}\ \emph {et~al.}(2017)\citenamefont
  {Henrion}, \citenamefont {Cranmer}, \citenamefont {Bruna}, \citenamefont
  {Cho}, \citenamefont {Brehmer}, \citenamefont {Louppe},\ and\ \citenamefont
  {Rochette}}]{Henrion:DLPS2017}%
  \BibitemOpen
  \bibfield  {author} {\bibinfo {author} {\bibfnamefont {I.}~\bibnamefont
  {Henrion}}, \bibinfo {author} {\bibfnamefont {K.}~\bibnamefont {Cranmer}},
  \bibinfo {author} {\bibfnamefont {J.}~\bibnamefont {Bruna}}, \bibinfo
  {author} {\bibfnamefont {K.}~\bibnamefont {Cho}}, \bibinfo {author}
  {\bibfnamefont {J.}~\bibnamefont {Brehmer}}, \bibinfo {author} {\bibfnamefont
  {G.}~\bibnamefont {Louppe}},\ and\ \bibinfo {author} {\bibfnamefont
  {G.}~\bibnamefont {Rochette}},\ }\bibfield  {title} {\bibinfo {title}
  {{Neural Message Passing for Jet Physics}},\ }\href
  {{https://dl4physicalsciences.github.io/files/nips_dlps_2017_29.pdf}}
  {\bibfield  {journal} {\bibinfo  {journal} {{Proceedings of the Deep Learning
  for Physical Sciences Workshop at NIPS (2017)}}\ } (\bibinfo {year}
  {2017})}\BibitemShut {NoStop}%
\bibitem [{\citenamefont {Choma}\ \emph
  {et~al.}(2018{\natexlab{a}})\citenamefont {Choma}, \citenamefont {Monti},
  \citenamefont {Gerhardt}, \citenamefont {Palczewski}, \citenamefont
  {Ronaghi}, \citenamefont {Prabhat}, \citenamefont {Bhimji}, \citenamefont
  {Bronstein}, \citenamefont {Klein},\ and\ \citenamefont
  {Bruna}}]{choma2018graph}%
  \BibitemOpen
  \bibfield  {author} {\bibinfo {author} {\bibfnamefont {N.}~\bibnamefont
  {Choma}}, \bibinfo {author} {\bibfnamefont {F.}~\bibnamefont {Monti}},
  \bibinfo {author} {\bibfnamefont {L.}~\bibnamefont {Gerhardt}}, \bibinfo
  {author} {\bibfnamefont {T.}~\bibnamefont {Palczewski}}, \bibinfo {author}
  {\bibfnamefont {Z.}~\bibnamefont {Ronaghi}}, \bibinfo {author} {\bibnamefont
  {Prabhat}}, \bibinfo {author} {\bibfnamefont {W.}~\bibnamefont {Bhimji}},
  \bibinfo {author} {\bibfnamefont {M.~M.}\ \bibnamefont {Bronstein}}, \bibinfo
  {author} {\bibfnamefont {S.~R.}\ \bibnamefont {Klein}},\ and\ \bibinfo
  {author} {\bibfnamefont {J.}~\bibnamefont {Bruna}},\ }\href@noop {} {\bibinfo
  {title} {Graph neural networks for icecube signal classification}} (\bibinfo
  {year} {2018}{\natexlab{a}}),\ \Eprint {https://arxiv.org/abs/1809.06166}
  {arXiv:1809.06166 [cs.LG]} \BibitemShut {NoStop}%
\bibitem [{\citenamefont {Qu}\ and\ \citenamefont
  {Gouskos}(2020)}]{Qu:2019gqs}%
  \BibitemOpen
  \bibfield  {author} {\bibinfo {author} {\bibfnamefont {H.}~\bibnamefont
  {Qu}}\ and\ \bibinfo {author} {\bibfnamefont {L.}~\bibnamefont {Gouskos}},\
  }\bibfield  {title} {\bibinfo {title} {{ParticleNet: Jet Tagging via Particle
  Clouds}},\ }\href {https://doi.org/10.1103/PhysRevD.101.056019} {\bibfield
  {journal} {\bibinfo  {journal} {Phys. Rev. D}\ }\textbf {\bibinfo {volume}
  {101}},\ \bibinfo {pages} {056019} (\bibinfo {year} {2020})},\ \Eprint
  {https://arxiv.org/abs/1902.08570} {arXiv:1902.08570 [hep-ph]} \BibitemShut
  {NoStop}%
\bibitem [{\citenamefont {Shlomi}\ \emph {et~al.}(2021)\citenamefont {Shlomi},
  \citenamefont {Battaglia},\ and\ \citenamefont {Vlimant}}]{1808887}%
  \BibitemOpen
  \bibfield  {author} {\bibinfo {author} {\bibfnamefont {J.}~\bibnamefont
  {Shlomi}}, \bibinfo {author} {\bibfnamefont {P.}~\bibnamefont {Battaglia}},\
  and\ \bibinfo {author} {\bibfnamefont {J.-R.}\ \bibnamefont {Vlimant}},\
  }\bibfield  {title} {\bibinfo {title} {{Graph neural networks in particle
  physics}},\ }\href {https://doi.org/10.1088/2632-2153/abbf9a} {\bibfield
  {journal} {\bibinfo  {journal} {Machine Learning: Science and Technology}\
  }\textbf {\bibinfo {volume} {2}},\ \bibinfo {pages} {021001} (\bibinfo {year}
  {2021})},\ \Eprint {https://arxiv.org/abs/2007.13681} {2007.13681}
  \BibitemShut {NoStop}%
\bibitem [{\citenamefont {Cheong}\ \emph {et~al.}(2020)\citenamefont {Cheong},
  \citenamefont {Cukierman}, \citenamefont {Nachman}, \citenamefont {Safdari},\
  and\ \citenamefont {Schwartzman}}]{Cheong:2019upg}%
  \BibitemOpen
  \bibfield  {author} {\bibinfo {author} {\bibfnamefont {S.}~\bibnamefont
  {Cheong}}, \bibinfo {author} {\bibfnamefont {A.}~\bibnamefont {Cukierman}},
  \bibinfo {author} {\bibfnamefont {B.}~\bibnamefont {Nachman}}, \bibinfo
  {author} {\bibfnamefont {M.}~\bibnamefont {Safdari}},\ and\ \bibinfo {author}
  {\bibfnamefont {A.}~\bibnamefont {Schwartzman}},\ }\bibfield  {title}
  {\bibinfo {title} {{Parametrizing the Detector Response with Neural
  Networks}},\ }\href {https://doi.org/10.1088/1748-0221/15/01/P01030}
  {\bibfield  {journal} {\bibinfo  {journal} {JINST}\ }\textbf {\bibinfo
  {volume} {15}}\bibfield  {number} {\bibinfo  {number} { (01)},\ \bibinfo
  {pages} {P01030}},\ }\Eprint {https://arxiv.org/abs/1910.03773}
  {arXiv:1910.03773 [physics.data-an]} \BibitemShut {NoStop}%
\bibitem [{\citenamefont {Goodfellow}\ \emph {et~al.}(2014)\citenamefont
  {Goodfellow} \emph {et~al.}}]{Goodfellow:2014:GAN:2969033.2969125}%
  \BibitemOpen
  \bibfield  {author} {\bibinfo {author} {\bibfnamefont {I.~J.}\ \bibnamefont
  {Goodfellow}} \emph {et~al.},\ }\bibfield  {title} {\bibinfo {title}
  {{Generative Adversarial Nets}},\ }\href
  {http://dl.acm.org/citation.cfm?id=2969033.2969125} {\bibfield  {journal}
  {\bibinfo  {journal} {Conference on Neural Information Processing Systems}\
  }\textbf {\bibinfo {volume} {2}},\ \bibinfo {pages} {2672} (\bibinfo {year}
  {2014})}\BibitemShut {NoStop}%
\bibitem [{\citenamefont {Creswell}\ \emph {et~al.}(2018)\citenamefont
  {Creswell}, \citenamefont {White}, \citenamefont {Dumoulin}, \citenamefont
  {Arulkumaran}, \citenamefont {Sengupta},\ and\ \citenamefont
  {Bharath}}]{Creswell2018}%
  \BibitemOpen
  \bibfield  {author} {\bibinfo {author} {\bibfnamefont {A.}~\bibnamefont
  {Creswell}}, \bibinfo {author} {\bibfnamefont {T.}~\bibnamefont {White}},
  \bibinfo {author} {\bibfnamefont {V.}~\bibnamefont {Dumoulin}}, \bibinfo
  {author} {\bibfnamefont {K.}~\bibnamefont {Arulkumaran}}, \bibinfo {author}
  {\bibfnamefont {B.}~\bibnamefont {Sengupta}},\ and\ \bibinfo {author}
  {\bibfnamefont {A.~A.}\ \bibnamefont {Bharath}},\ }\bibfield  {title}
  {\bibinfo {title} {Generative adversarial networks: An overview},\ }\href
  {https://doi.org/10.1109/msp.2017.2765202} {\bibfield  {journal} {\bibinfo
  {journal} {IEEE Signal Processing Magazine}\ }\textbf {\bibinfo {volume}
  {35}},\ \bibinfo {pages} {53} (\bibinfo {year} {2018})}\BibitemShut {NoStop}%
\bibitem [{\citenamefont {Kingma}\ and\ \citenamefont
  {Welling}(2014)}]{kingma2014autoencoding}%
  \BibitemOpen
  \bibfield  {author} {\bibinfo {author} {\bibfnamefont {D.~P.}\ \bibnamefont
  {Kingma}}\ and\ \bibinfo {author} {\bibfnamefont {M.}~\bibnamefont
  {Welling}},\ }\bibfield  {title} {\bibinfo {title} {Auto-encoding variational
  bayes},\ }\href@noop {} {\  (\bibinfo {year} {2014})},\ \Eprint
  {https://arxiv.org/abs/1312.6114} {arXiv:1312.6114 [stat.ML]} \BibitemShut
  {NoStop}%
\bibitem [{\citenamefont {Kingma}\ and\ \citenamefont
  {Welling}(2019)}]{Kingma2019}%
  \BibitemOpen
  \bibfield  {author} {\bibinfo {author} {\bibfnamefont {D.~P.}\ \bibnamefont
  {Kingma}}\ and\ \bibinfo {author} {\bibfnamefont {M.}~\bibnamefont
  {Welling}},\ }\bibfield  {title} {\bibinfo {title} {{An Introduction to
  Variational Autoencoders}},\ }\href {https://doi.org/10.1561/2200000056}
  {\bibfield  {journal} {\bibinfo  {journal} {Foundations and Trends in Machine
  Learning}\ }\textbf {\bibinfo {volume} {12}},\ \bibinfo {pages} {307}
  (\bibinfo {year} {2019})}\BibitemShut {NoStop}%
\bibitem [{\citenamefont {Rezende}\ and\ \citenamefont
  {Mohamed}(2015)}]{10.5555/3045118.3045281}%
  \BibitemOpen
  \bibfield  {author} {\bibinfo {author} {\bibfnamefont {D.~J.}\ \bibnamefont
  {Rezende}}\ and\ \bibinfo {author} {\bibfnamefont {S.}~\bibnamefont
  {Mohamed}},\ }\bibfield  {title} {\bibinfo {title} {Variational inference
  with normalizing flows},\ }\href@noop {} {\bibfield  {journal} {\bibinfo
  {journal} {International Conference on Machine Learning}\ }\textbf {\bibinfo
  {volume} {37}},\ \bibinfo {pages} {1530} (\bibinfo {year}
  {2015})}\BibitemShut {NoStop}%
\bibitem [{\citenamefont {Kobyzev}\ \emph {et~al.}(2020)\citenamefont
  {Kobyzev}, \citenamefont {Prince},\ and\ \citenamefont
  {Brubaker}}]{Kobyzev2020}%
  \BibitemOpen
  \bibfield  {author} {\bibinfo {author} {\bibfnamefont {I.}~\bibnamefont
  {Kobyzev}}, \bibinfo {author} {\bibfnamefont {S.}~\bibnamefont {Prince}},\
  and\ \bibinfo {author} {\bibfnamefont {M.}~\bibnamefont {Brubaker}},\
  }\bibfield  {title} {\bibinfo {title} {{Normalizing Flows: An Introduction
  and Review of Current Methods}},\ }\href
  {https://doi.org/10.1109/tpami.2020.2992934} {\bibfield  {journal} {\bibinfo
  {journal} {IEEE Transactions on Pattern Analysis and Machine Intelligence}\
  ,\ \bibinfo {pages} {1}} (\bibinfo {year} {2020})}\BibitemShut {NoStop}%
\bibitem [{\citenamefont {de~Oliveira}\ \emph {et~al.}(2017)\citenamefont
  {de~Oliveira}, \citenamefont {Paganini},\ and\ \citenamefont
  {Nachman}}]{deOliveira:2017pjk}%
  \BibitemOpen
  \bibfield  {author} {\bibinfo {author} {\bibfnamefont {L.}~\bibnamefont
  {de~Oliveira}}, \bibinfo {author} {\bibfnamefont {M.}~\bibnamefont
  {Paganini}},\ and\ \bibinfo {author} {\bibfnamefont {B.}~\bibnamefont
  {Nachman}},\ }\bibfield  {title} {\bibinfo {title} {{Learning Particle
  Physics by Example: Location-Aware Generative Adversarial Networks for
  Physics Synthesis}},\ }\href {https://doi.org/10.1007/s41781-017-0004-6}
  {\bibfield  {journal} {\bibinfo  {journal} {Comput. Softw. Big Sci.}\
  }\textbf {\bibinfo {volume} {1}},\ \bibinfo {pages} {4} (\bibinfo {year}
  {2017})},\ \Eprint {https://arxiv.org/abs/1701.05927} {arXiv:1701.05927
  [stat.ML]} \BibitemShut {NoStop}%
\bibitem [{\citenamefont {Mustafa}\ \emph {et~al.}(2019)\citenamefont
  {Mustafa}, \citenamefont {Bard}, \citenamefont {Bhimji}, \citenamefont
  {Lukić}, \citenamefont {Al-Rfou},\ and\ \citenamefont
  {Kratochvil}}]{Mustafa_2019}%
  \BibitemOpen
  \bibfield  {author} {\bibinfo {author} {\bibfnamefont {M.}~\bibnamefont
  {Mustafa}}, \bibinfo {author} {\bibfnamefont {D.}~\bibnamefont {Bard}},
  \bibinfo {author} {\bibfnamefont {W.}~\bibnamefont {Bhimji}}, \bibinfo
  {author} {\bibfnamefont {Z.}~\bibnamefont {Lukić}}, \bibinfo {author}
  {\bibfnamefont {R.}~\bibnamefont {Al-Rfou}},\ and\ \bibinfo {author}
  {\bibfnamefont {J.~M.}\ \bibnamefont {Kratochvil}},\ }\bibfield  {title}
  {\bibinfo {title} {Cosmogan: creating high-fidelity weak lensing convergence
  maps using generative adversarial networks},\ }\bibfield  {journal} {\bibinfo
   {journal} {Computational Astrophysics and Cosmology}\ }\textbf {\bibinfo
  {volume} {6}},\ \href {https://doi.org/10.1186/s40668-019-0029-9}
  {10.1186/s40668-019-0029-9} (\bibinfo {year} {2019})\BibitemShut {NoStop}%
\bibitem [{ATL(2018{\natexlab{a}})}]{ATL-SOFT-PUB-2018-001}%
  \BibitemOpen
  \bibfield  {title} {\bibinfo {title} {{Deep generative models for fast shower
  simulation in ATLAS}},\ }\href {http://cds.cern.ch/record/2630433} {\bibfield
   {journal} {\bibinfo  {journal} {ATL-SOFT-PUB-2018-001}\ } (\bibinfo {year}
  {2018}{\natexlab{a}})}\BibitemShut {NoStop}%
\bibitem [{\citenamefont {Hajer}\ \emph {et~al.}(2020)\citenamefont {Hajer},
  \citenamefont {Li}, \citenamefont {Liu},\ and\ \citenamefont
  {Wang}}]{Hajer:2018kqm}%
  \BibitemOpen
  \bibfield  {author} {\bibinfo {author} {\bibfnamefont {J.}~\bibnamefont
  {Hajer}}, \bibinfo {author} {\bibfnamefont {Y.-Y.}\ \bibnamefont {Li}},
  \bibinfo {author} {\bibfnamefont {T.}~\bibnamefont {Liu}},\ and\ \bibinfo
  {author} {\bibfnamefont {H.}~\bibnamefont {Wang}},\ }\bibfield  {title}
  {\bibinfo {title} {{Novelty Detection Meets Collider Physics}},\ }\href
  {https://doi.org/10.1103/PhysRevD.101.076015} {\bibfield  {journal} {\bibinfo
   {journal} {Phys. Rev. D}\ }\textbf {\bibinfo {volume} {101}},\ \bibinfo
  {pages} {076015} (\bibinfo {year} {2020})},\ \Eprint
  {https://arxiv.org/abs/1807.10261} {arXiv:1807.10261 [hep-ph]} \BibitemShut
  {NoStop}%
\bibitem [{\citenamefont {Farina}\ \emph {et~al.}(2020)\citenamefont {Farina},
  \citenamefont {Nakai},\ and\ \citenamefont {Shih}}]{Farina:2018fyg}%
  \BibitemOpen
  \bibfield  {author} {\bibinfo {author} {\bibfnamefont {M.}~\bibnamefont
  {Farina}}, \bibinfo {author} {\bibfnamefont {Y.}~\bibnamefont {Nakai}},\ and\
  \bibinfo {author} {\bibfnamefont {D.}~\bibnamefont {Shih}},\ }\bibfield
  {title} {\bibinfo {title} {{Searching for New Physics with Deep
  Autoencoders}},\ }\href {https://doi.org/10.1103/PhysRevD.101.075021}
  {\bibfield  {journal} {\bibinfo  {journal} {Phys. Rev. D}\ }\textbf {\bibinfo
  {volume} {101}},\ \bibinfo {pages} {075021} (\bibinfo {year} {2020})},\
  \Eprint {https://arxiv.org/abs/1808.08992} {arXiv:1808.08992 [hep-ph]}
  \BibitemShut {NoStop}%
\bibitem [{\citenamefont {Heimel}\ \emph {et~al.}(2019)\citenamefont {Heimel},
  \citenamefont {Kasieczka}, \citenamefont {Plehn},\ and\ \citenamefont
  {Thompson}}]{Heimel:2018mkt}%
  \BibitemOpen
  \bibfield  {author} {\bibinfo {author} {\bibfnamefont {T.}~\bibnamefont
  {Heimel}}, \bibinfo {author} {\bibfnamefont {G.}~\bibnamefont {Kasieczka}},
  \bibinfo {author} {\bibfnamefont {T.}~\bibnamefont {Plehn}},\ and\ \bibinfo
  {author} {\bibfnamefont {J.~M.}\ \bibnamefont {Thompson}},\ }\bibfield
  {title} {\bibinfo {title} {{QCD or What?}},\ }\href
  {https://doi.org/10.21468/SciPostPhys.6.3.030} {\bibfield  {journal}
  {\bibinfo  {journal} {SciPost Phys.}\ }\textbf {\bibinfo {volume} {6}},\
  \bibinfo {pages} {030} (\bibinfo {year} {2019})},\ \Eprint
  {https://arxiv.org/abs/1808.08979} {arXiv:1808.08979 [hep-ph]} \BibitemShut
  {NoStop}%
\bibitem [{\citenamefont {Albergo}\ \emph {et~al.}(2019)\citenamefont
  {Albergo}, \citenamefont {Kanwar},\ and\ \citenamefont
  {Shanahan}}]{Albergo:2019eim}%
  \BibitemOpen
  \bibfield  {author} {\bibinfo {author} {\bibfnamefont {M.~S.}\ \bibnamefont
  {Albergo}}, \bibinfo {author} {\bibfnamefont {G.}~\bibnamefont {Kanwar}},\
  and\ \bibinfo {author} {\bibfnamefont {P.~E.}\ \bibnamefont {Shanahan}},\
  }\bibfield  {title} {\bibinfo {title} {{Flow-based generative models for
  Markov chain Monte Carlo in lattice field theory}},\ }\href
  {https://doi.org/10.1103/PhysRevD.100.034515} {\bibfield  {journal} {\bibinfo
   {journal} {Phys. Rev. D}\ }\textbf {\bibinfo {volume} {100}},\ \bibinfo
  {pages} {034515} (\bibinfo {year} {2019})},\ \Eprint
  {https://arxiv.org/abs/1904.12072} {arXiv:1904.12072 [hep-lat]} \BibitemShut
  {NoStop}%
\bibitem [{\citenamefont {Gao}\ \emph {et~al.}(2020{\natexlab{a}})\citenamefont
  {Gao}, \citenamefont {Höche}, \citenamefont {Isaacson}, \citenamefont
  {Krause},\ and\ \citenamefont {Schulz}}]{Gao:2020zvv}%
  \BibitemOpen
  \bibfield  {author} {\bibinfo {author} {\bibfnamefont {C.}~\bibnamefont
  {Gao}}, \bibinfo {author} {\bibfnamefont {S.}~\bibnamefont {Höche}},
  \bibinfo {author} {\bibfnamefont {J.}~\bibnamefont {Isaacson}}, \bibinfo
  {author} {\bibfnamefont {C.}~\bibnamefont {Krause}},\ and\ \bibinfo {author}
  {\bibfnamefont {H.}~\bibnamefont {Schulz}},\ }\bibfield  {title} {\bibinfo
  {title} {{Event Generation with Normalizing Flows}},\ }\href
  {https://doi.org/10.1103/PhysRevD.101.076002} {\bibfield  {journal} {\bibinfo
   {journal} {Phys. Rev. D}\ }\textbf {\bibinfo {volume} {101}},\ \bibinfo
  {pages} {076002} (\bibinfo {year} {2020}{\natexlab{a}})},\ \Eprint
  {https://arxiv.org/abs/2001.10028} {arXiv:2001.10028 [hep-ph]} \BibitemShut
  {NoStop}%
\bibitem [{\citenamefont {Gao}\ \emph {et~al.}(2020{\natexlab{b}})\citenamefont
  {Gao}, \citenamefont {Isaacson},\ and\ \citenamefont {Krause}}]{Gao:2020vdv}%
  \BibitemOpen
  \bibfield  {author} {\bibinfo {author} {\bibfnamefont {C.}~\bibnamefont
  {Gao}}, \bibinfo {author} {\bibfnamefont {J.}~\bibnamefont {Isaacson}},\ and\
  \bibinfo {author} {\bibfnamefont {C.}~\bibnamefont {Krause}},\ }\bibfield
  {title} {\bibinfo {title} {{i-flow: High-dimensional Integration and Sampling
  with Normalizing Flows}},\ }\href {https://doi.org/10.1088/2632-2153/abab62}
  {\bibfield  {journal} {\bibinfo  {journal} {Mach. Learn. Sci. Tech.}\
  }\textbf {\bibinfo {volume} {1}},\ \bibinfo {pages} {045023} (\bibinfo {year}
  {2020}{\natexlab{b}})},\ \Eprint {https://arxiv.org/abs/2001.05486}
  {arXiv:2001.05486 [physics.comp-ph]} \BibitemShut {NoStop}%
\bibitem [{\citenamefont {Bothmann}\ \emph {et~al.}(2020)\citenamefont
  {Bothmann}, \citenamefont {Jan\ss{}en}, \citenamefont {Knobbe}, \citenamefont
  {Schmale},\ and\ \citenamefont {Schumann}}]{Bothmann:2020ywa}%
  \BibitemOpen
  \bibfield  {author} {\bibinfo {author} {\bibfnamefont {E.}~\bibnamefont
  {Bothmann}}, \bibinfo {author} {\bibfnamefont {T.}~\bibnamefont
  {Jan\ss{}en}}, \bibinfo {author} {\bibfnamefont {M.}~\bibnamefont {Knobbe}},
  \bibinfo {author} {\bibfnamefont {T.}~\bibnamefont {Schmale}},\ and\ \bibinfo
  {author} {\bibfnamefont {S.}~\bibnamefont {Schumann}},\ }\bibfield  {title}
  {\bibinfo {title} {{Exploring phase space with Neural Importance Sampling}},\
  }\href {https://doi.org/10.21468/SciPostPhys.8.4.069} {\bibfield  {journal}
  {\bibinfo  {journal} {SciPost Phys.}\ }\textbf {\bibinfo {volume} {8}},\
  \bibinfo {pages} {069} (\bibinfo {year} {2020})},\ \Eprint
  {https://arxiv.org/abs/2001.05478} {arXiv:2001.05478 [hep-ph]} \BibitemShut
  {NoStop}%
\bibitem [{\citenamefont {Nachman}\ and\ \citenamefont
  {Shih}(2020)}]{Nachman:2020lpy}%
  \BibitemOpen
  \bibfield  {author} {\bibinfo {author} {\bibfnamefont {B.}~\bibnamefont
  {Nachman}}\ and\ \bibinfo {author} {\bibfnamefont {D.}~\bibnamefont {Shih}},\
  }\bibfield  {title} {\bibinfo {title} {{Anomaly Detection with Density
  Estimation}},\ }\href {https://doi.org/10.1103/PhysRevD.101.075042}
  {\bibfield  {journal} {\bibinfo  {journal} {Phys. Rev. D}\ }\textbf {\bibinfo
  {volume} {101}},\ \bibinfo {pages} {075042} (\bibinfo {year} {2020})},\
  \Eprint {https://arxiv.org/abs/2001.04990} {arXiv:2001.04990 [hep-ph]}
  \BibitemShut {NoStop}%
\bibitem [{\citenamefont {Zhou}(2017)}]{10.1093/nsr/nwx106}%
  \BibitemOpen
  \bibfield  {author} {\bibinfo {author} {\bibfnamefont {Z.-H.}\ \bibnamefont
  {Zhou}},\ }\bibfield  {title} {\bibinfo {title} {{A brief introduction to
  weakly supervised learning}},\ }\href {https://doi.org/10.1093/nsr/nwx106}
  {\bibfield  {journal} {\bibinfo  {journal} {National Science Review}\
  }\textbf {\bibinfo {volume} {5}},\ \bibinfo {pages} {44} (\bibinfo {year}
  {2017})},\ \Eprint
  {https://arxiv.org/abs/https://academic.oup.com/nsr/article-pdf/5/1/44/31567770/nwx106.pdf}
  {https://academic.oup.com/nsr/article-pdf/5/1/44/31567770/nwx106.pdf}
  \BibitemShut {NoStop}%
\bibitem [{\citenamefont {Dery}\ \emph {et~al.}(2017)\citenamefont {Dery},
  \citenamefont {Nachman}, \citenamefont {Rubbo},\ and\ \citenamefont
  {Schwartzman}}]{Dery:2017fap}%
  \BibitemOpen
  \bibfield  {author} {\bibinfo {author} {\bibfnamefont {L.~M.}\ \bibnamefont
  {Dery}}, \bibinfo {author} {\bibfnamefont {B.}~\bibnamefont {Nachman}},
  \bibinfo {author} {\bibfnamefont {F.}~\bibnamefont {Rubbo}},\ and\ \bibinfo
  {author} {\bibfnamefont {A.}~\bibnamefont {Schwartzman}},\ }\bibfield
  {title} {\bibinfo {title} {{Weakly Supervised Classification in High Energy
  Physics}},\ }\href {https://doi.org/10.1007/JHEP05(2017)145} {\bibfield
  {journal} {\bibinfo  {journal} {JHEP}\ }\textbf {\bibinfo {volume} {05}},\
  \bibinfo {pages} {145}},\ \Eprint {https://arxiv.org/abs/1702.00414}
  {arXiv:1702.00414 [hep-ph]} \BibitemShut {NoStop}%
\bibitem [{\citenamefont {Metodiev}\ \emph {et~al.}(2017)\citenamefont
  {Metodiev}, \citenamefont {Nachman},\ and\ \citenamefont
  {Thaler}}]{Metodiev:2017vrx}%
  \BibitemOpen
  \bibfield  {author} {\bibinfo {author} {\bibfnamefont {E.~M.}\ \bibnamefont
  {Metodiev}}, \bibinfo {author} {\bibfnamefont {B.}~\bibnamefont {Nachman}},\
  and\ \bibinfo {author} {\bibfnamefont {J.}~\bibnamefont {Thaler}},\
  }\bibfield  {title} {\bibinfo {title} {{Classification without labels:
  Learning from mixed samples in high energy physics}},\ }\href
  {https://doi.org/10.1007/JHEP10(2017)174} {\bibfield  {journal} {\bibinfo
  {journal} {JHEP}\ }\textbf {\bibinfo {volume} {10}},\ \bibinfo {pages}
  {174}},\ \Eprint {https://arxiv.org/abs/1708.02949} {arXiv:1708.02949
  [hep-ph]} \BibitemShut {NoStop}%
\bibitem [{\citenamefont {Cohen}\ \emph {et~al.}(2018)\citenamefont {Cohen},
  \citenamefont {Freytsis},\ and\ \citenamefont {Ostdiek}}]{Cohen:2017exh}%
  \BibitemOpen
  \bibfield  {author} {\bibinfo {author} {\bibfnamefont {T.}~\bibnamefont
  {Cohen}}, \bibinfo {author} {\bibfnamefont {M.}~\bibnamefont {Freytsis}},\
  and\ \bibinfo {author} {\bibfnamefont {B.}~\bibnamefont {Ostdiek}},\
  }\bibfield  {title} {\bibinfo {title} {{(Machine) Learning to Do More with
  Less}},\ }\href {https://doi.org/10.1007/JHEP02(2018)034} {\bibfield
  {journal} {\bibinfo  {journal} {JHEP}\ }\textbf {\bibinfo {volume} {02}},\
  \bibinfo {pages} {034}},\ \Eprint {https://arxiv.org/abs/1706.09451}
  {arXiv:1706.09451 [hep-ph]} \BibitemShut {NoStop}%
\bibitem [{\citenamefont {Komiske}\ \emph {et~al.}(2018)\citenamefont
  {Komiske}, \citenamefont {Metodiev}, \citenamefont {Nachman},\ and\
  \citenamefont {Schwartz}}]{Komiske:2018oaa}%
  \BibitemOpen
  \bibfield  {author} {\bibinfo {author} {\bibfnamefont {P.~T.}\ \bibnamefont
  {Komiske}}, \bibinfo {author} {\bibfnamefont {E.~M.}\ \bibnamefont
  {Metodiev}}, \bibinfo {author} {\bibfnamefont {B.}~\bibnamefont {Nachman}},\
  and\ \bibinfo {author} {\bibfnamefont {M.~D.}\ \bibnamefont {Schwartz}},\
  }\bibfield  {title} {\bibinfo {title} {{Learning to classify from impure
  samples with high-dimensional data}},\ }\href
  {https://doi.org/10.1103/PhysRevD.98.011502} {\bibfield  {journal} {\bibinfo
  {journal} {Phys. Rev. D}\ }\textbf {\bibinfo {volume} {98}},\ \bibinfo
  {pages} {011502} (\bibinfo {year} {2018})},\ \Eprint
  {https://arxiv.org/abs/1801.10158} {arXiv:1801.10158 [hep-ph]} \BibitemShut
  {NoStop}%
\bibitem [{\citenamefont {Knuteson}()}]{sleuth}%
  \BibitemOpen
  \bibfield  {author} {\bibinfo {author} {\bibfnamefont {B.}~\bibnamefont
  {Knuteson}},\ }\href
  {https://www-d0.fnal.gov/results/publications_talks/thesis/knuteson/thesis.ps}
  {\bibinfo {title} {{A Quasi-Model-Independent Search for New High $p_T$
  Physics at D0}}},\ \bibinfo {note} {ph.D. thesis, University of California at
  Berkeley (2000)}\BibitemShut {NoStop}%
\bibitem [{\citenamefont {Abbott}\ \emph {et~al.}(2000)\citenamefont {Abbott}
  \emph {et~al.}}]{Abbott:2000fb}%
  \BibitemOpen
  \bibfield  {author} {\bibinfo {author} {\bibfnamefont {B.}~\bibnamefont
  {Abbott}} \emph {et~al.} (\bibinfo {collaboration} {D0}),\ }\bibfield
  {title} {\bibinfo {title} {{Search for new physics in $e\mu X$ data at D\O\
  using Sherlock: A quasi model independent search strategy for new physics}},\
  }\href {https://doi.org/10.1103/PhysRevD.62.092004} {\bibfield  {journal}
  {\bibinfo  {journal} {Phys. Rev.}\ }\textbf {\bibinfo {volume} {D62}},\
  \bibinfo {pages} {092004} (\bibinfo {year} {2000})},\ \Eprint
  {https://arxiv.org/abs/hep-ex/0006011} {arXiv:hep-ex/0006011 [hep-ex]}
  \BibitemShut {NoStop}%
\bibitem [{\citenamefont {Abazov}\ \emph {et~al.}(2001)\citenamefont {Abazov}
  \emph {et~al.}}]{Abbott:2000gx}%
  \BibitemOpen
  \bibfield  {author} {\bibinfo {author} {\bibfnamefont {V.~M.}\ \bibnamefont
  {Abazov}} \emph {et~al.} (\bibinfo {collaboration} {D0}),\ }\bibfield
  {title} {\bibinfo {title} {{A Quasi model independent search for new physics
  at large transverse momentum}},\ }\href
  {https://doi.org/10.1103/PhysRevD.64.012004} {\bibfield  {journal} {\bibinfo
  {journal} {Phys. Rev.}\ }\textbf {\bibinfo {volume} {D64}},\ \bibinfo {pages}
  {012004} (\bibinfo {year} {2001})},\ \Eprint
  {https://arxiv.org/abs/hep-ex/0011067} {arXiv:hep-ex/0011067 [hep-ex]}
  \BibitemShut {NoStop}%
\bibitem [{\citenamefont {Abbott}\ \emph {et~al.}(2001)\citenamefont {Abbott}
  \emph {et~al.}}]{Abbott:2001ke}%
  \BibitemOpen
  \bibfield  {author} {\bibinfo {author} {\bibfnamefont {B.}~\bibnamefont
  {Abbott}} \emph {et~al.} (\bibinfo {collaboration} {D0}),\ }\bibfield
  {title} {\bibinfo {title} {{A quasi-model-independent search for new high
  $p_T$ physics at D\O}},\ }\href {https://doi.org/10.1103/PhysRevLett.86.3712}
  {\bibfield  {journal} {\bibinfo  {journal} {Phys. Rev. Lett.}\ }\textbf
  {\bibinfo {volume} {86}},\ \bibinfo {pages} {3712} (\bibinfo {year}
  {2001})},\ \Eprint {https://arxiv.org/abs/hep-ex/0011071}
  {arXiv:hep-ex/0011071 [hep-ex]} \BibitemShut {NoStop}%
\bibitem [{\citenamefont {Aaron}\ \emph {et~al.}(2009)\citenamefont {Aaron}
  \emph {et~al.}}]{Aaron:2008aa}%
  \BibitemOpen
  \bibfield  {author} {\bibinfo {author} {\bibfnamefont {F.~D.}\ \bibnamefont
  {Aaron}} \emph {et~al.} (\bibinfo {collaboration} {H1}),\ }\bibfield  {title}
  {\bibinfo {title} {{A General Search for New Phenomena at HERA}},\ }\href
  {https://doi.org/10.1016/j.physletb.2009.03.034} {\bibfield  {journal}
  {\bibinfo  {journal} {Phys. Lett.}\ }\textbf {\bibinfo {volume} {B674}},\
  \bibinfo {pages} {257} (\bibinfo {year} {2009})},\ \Eprint
  {https://arxiv.org/abs/0901.0507} {arXiv:0901.0507 [hep-ex]} \BibitemShut
  {NoStop}%
\bibitem [{\citenamefont {Aktas}\ \emph {et~al.}(2004)\citenamefont {Aktas}
  \emph {et~al.}}]{Aktas:2004pz}%
  \BibitemOpen
  \bibfield  {author} {\bibinfo {author} {\bibfnamefont {A.}~\bibnamefont
  {Aktas}} \emph {et~al.} (\bibinfo {collaboration} {H1}),\ }\bibfield  {title}
  {\bibinfo {title} {{A General search for new phenomena in ep scattering at
  HERA}},\ }\href {https://doi.org/10.1016/j.physletb.2004.09.057} {\bibfield
  {journal} {\bibinfo  {journal} {Phys. Lett.}\ }\textbf {\bibinfo {volume}
  {B602}},\ \bibinfo {pages} {14} (\bibinfo {year} {2004})},\ \Eprint
  {https://arxiv.org/abs/hep-ex/0408044} {arXiv:hep-ex/0408044 [hep-ex]}
  \BibitemShut {NoStop}%
\bibitem [{\citenamefont {Cranmer}(2005)}]{Cranmer:2005zn}%
  \BibitemOpen
  \bibfield  {author} {\bibinfo {author} {\bibfnamefont {K.~S.}\ \bibnamefont
  {Cranmer}},\ }\emph {\bibinfo {title} {{Searching for new physics:
  Contributions to LEP and the LHC}}},\ \href
  {http://weblib.cern.ch/abstract?CERN-THESIS-2005-011} {Ph.D. thesis},\
  \bibinfo  {school} {Wisconsin U., Madison} (\bibinfo {year}
  {2005})\BibitemShut {NoStop}%
\bibitem [{\citenamefont {Aaltonen}\ \emph {et~al.}(2008)\citenamefont
  {Aaltonen} \emph {et~al.}}]{Aaltonen:2007dg}%
  \BibitemOpen
  \bibfield  {author} {\bibinfo {author} {\bibfnamefont {T.}~\bibnamefont
  {Aaltonen}} \emph {et~al.} (\bibinfo {collaboration} {CDF}),\ }\bibfield
  {title} {\bibinfo {title} {{Model-Independent and Quasi-Model-Independent
  Search for New Physics at CDF}},\ }\href
  {https://doi.org/10.1103/PhysRevD.78.012002} {\bibfield  {journal} {\bibinfo
  {journal} {Phys. Rev.}\ }\textbf {\bibinfo {volume} {D78}},\ \bibinfo {pages}
  {012002} (\bibinfo {year} {2008})},\ \Eprint
  {https://arxiv.org/abs/0712.1311} {arXiv:0712.1311 [hep-ex]} \BibitemShut
  {NoStop}%
\bibitem [{\citenamefont {Aaltonen}\ \emph {et~al.}(2007)\citenamefont
  {Aaltonen} \emph {et~al.}}]{Aaltonen:2007ab}%
  \BibitemOpen
  \bibfield  {author} {\bibinfo {author} {\bibfnamefont {T.}~\bibnamefont
  {Aaltonen}} \emph {et~al.} (\bibinfo {collaboration} {CDF}),\ }\bibfield
  {title} {\bibinfo {title} {{Model-Independent Global Search for New High-p(T)
  Physics at CDF}}\ }\href {https://doi.org/10.2172/922303} {10.2172/922303}
  (\bibinfo {year} {2007}),\ \Eprint {https://arxiv.org/abs/0712.2534}
  {arXiv:0712.2534 [hep-ex]} \BibitemShut {NoStop}%
\bibitem [{\citenamefont {Aaltonen}\ \emph {et~al.}(2009)\citenamefont
  {Aaltonen} \emph {et~al.}}]{Aaltonen:2008vt}%
  \BibitemOpen
  \bibfield  {author} {\bibinfo {author} {\bibfnamefont {T.}~\bibnamefont
  {Aaltonen}} \emph {et~al.} (\bibinfo {collaboration} {CDF}),\ }\bibfield
  {title} {\bibinfo {title} {{Global Search for New Physics with 2.0 fb$^{-1}$
  at CDF}},\ }\href {https://doi.org/10.1103/PhysRevD.79.011101} {\bibfield
  {journal} {\bibinfo  {journal} {Phys. Rev.}\ }\textbf {\bibinfo {volume}
  {D79}},\ \bibinfo {pages} {011101} (\bibinfo {year} {2009})},\ \Eprint
  {https://arxiv.org/abs/0809.3781} {arXiv:0809.3781 [hep-ex]} \BibitemShut
  {NoStop}%
\bibitem [{\citenamefont {{CMS Collaboration}}(2017)}]{CMS-PAS-EXO-14-016}%
  \BibitemOpen
  \bibfield  {author} {\bibinfo {author} {\bibnamefont {{CMS Collaboration}}},\
  }\bibfield  {title} {\bibinfo {title} {{MUSiC, a Model Unspecific Search for
  New Physics, in $pp$ Collisions at $\sqrt{s}=8$ TeV}},\ }\href@noop {} {\
  (\bibinfo {year} {2017})}\BibitemShut {NoStop}%
\bibitem [{CMS(2011)}]{CMS-PAS-EXO-10-021}%
  \BibitemOpen
  \href {http://cds.cern.ch/record/1360173} {\emph {\bibinfo {title} {{Model
  Unspecific Search for New Physics in pp Collisions at $\sqrt{s} = 7$
  TeV}}}},\ \bibinfo {type} {Tech. Rep.}\ \bibinfo {number}
  {CMS-PAS-EXO-10-021}\ (\bibinfo  {institution} {CERN},\ \bibinfo {address}
  {Geneva},\ \bibinfo {year} {2011})\BibitemShut {NoStop}%
\bibitem [{\citenamefont {{CMS
  Collaboration}}(2020{\natexlab{a}})}]{CMS:2020ohc}%
  \BibitemOpen
  \bibfield  {author} {\bibinfo {author} {\bibnamefont {{CMS Collaboration}}},\
  }\href@noop {} {\bibinfo {title} {{MUSiC, a model unspecific search for new
  physics, in $pp$ collisions at $\sqrt{s}=13$ TeV}}},\ \bibinfo {howpublished}
  {\url{https://cds.cern.ch/record/2718811}} (\bibinfo {year}
  {2020}{\natexlab{a}})\BibitemShut {NoStop}%
\bibitem [{\citenamefont {Sirunyan}\ \emph
  {et~al.}(2021{\natexlab{a}})\citenamefont {Sirunyan} \emph
  {et~al.}}]{Sirunyan:2020jwk}%
  \BibitemOpen
  \bibfield  {author} {\bibinfo {author} {\bibfnamefont {A.~M.}\ \bibnamefont
  {Sirunyan}} \emph {et~al.} (\bibinfo {collaboration} {CMS}),\ }\bibfield
  {title} {\bibinfo {title} {{MUSiC: a model-unspecific search for new physics
  in proton\textendash{}proton collisions at $\sqrt{s} = 13\,\text {TeV} $}},\
  }\href {https://doi.org/10.1140/epjc/s10052-021-09236-z} {\bibfield
  {journal} {\bibinfo  {journal} {Eur. Phys. J. C}\ }\textbf {\bibinfo {volume}
  {81}},\ \bibinfo {pages} {629} (\bibinfo {year} {2021}{\natexlab{a}})},\
  \Eprint {https://arxiv.org/abs/2010.02984} {arXiv:2010.02984 [hep-ex]}
  \BibitemShut {NoStop}%
\bibitem [{\citenamefont {Aaboud}\ \emph {et~al.}(2019)\citenamefont {Aaboud}
  \emph {et~al.}}]{Aaboud:2018ufy}%
  \BibitemOpen
  \bibfield  {author} {\bibinfo {author} {\bibfnamefont {M.}~\bibnamefont
  {Aaboud}} \emph {et~al.} (\bibinfo {collaboration} {ATLAS}),\ }\bibfield
  {title} {\bibinfo {title} {{A strategy for a general search for new phenomena
  using data-derived signal regions and its application within the ATLAS
  experiment}},\ }\href {https://doi.org/10.1140/epjc/s10052-019-6540-y}
  {\bibfield  {journal} {\bibinfo  {journal} {Eur. Phys. J.}\ }\textbf
  {\bibinfo {volume} {C79}},\ \bibinfo {pages} {120} (\bibinfo {year}
  {2019})},\ \Eprint {https://arxiv.org/abs/1807.07447} {arXiv:1807.07447
  [hep-ex]} \BibitemShut {NoStop}%
\bibitem [{ATL(2014)}]{ATLAS-CONF-2014-006}%
  \BibitemOpen
  \bibfield  {title} {\bibinfo {title} {{A general search for new phenomena
  with the ATLAS detector in pp collisions at $\sqrt{s}=8$ TeV}},\ }\href
  {https://cds.cern.ch/record/1666536} {\bibfield  {journal} {\bibinfo
  {journal} {ATLAS-CONF-2014-006}\ } (\bibinfo {year} {2014})}\BibitemShut
  {NoStop}%
\bibitem [{ATL(2012)}]{ATLAS-CONF-2012-107}%
  \BibitemOpen
  \bibfield  {title} {\bibinfo {title} {{A general search for new phenomena
  with the ATLAS detector in pp collisions at sort(s)=7 TeV.}},\ }\href
  {https://cds.cern.ch/record/1472686} {\bibfield  {journal} {\bibinfo
  {journal} {ATLAS-CONF-2012-107}\ } (\bibinfo {year} {2012})}\BibitemShut
  {NoStop}%
\bibitem [{\citenamefont {Butter}\ \emph {et~al.}(2019)\citenamefont {Butter}
  \emph {et~al.}}]{Kasieczka:2019dbj}%
  \BibitemOpen
  \bibfield  {author} {\bibinfo {author} {\bibfnamefont {A.}~\bibnamefont
  {Butter}} \emph {et~al.},\ }\bibfield  {title} {\bibinfo {title} {{The
  Machine Learning Landscape of Top Taggers}},\ }\href
  {https://doi.org/10.21468/SciPostPhys.7.1.014} {\bibfield  {journal}
  {\bibinfo  {journal} {SciPost Phys.}\ }\textbf {\bibinfo {volume} {7}},\
  \bibinfo {pages} {014} (\bibinfo {year} {2019})},\ \Eprint
  {https://arxiv.org/abs/1902.09914} {arXiv:1902.09914 [hep-ph]} \BibitemShut
  {NoStop}%
\bibitem [{\citenamefont {Abratenko}\ \emph
  {et~al.}(2021{\natexlab{a}})\citenamefont {Abratenko} \emph
  {et~al.}}]{MicroBooNE:2020hho}%
  \BibitemOpen
  \bibfield  {author} {\bibinfo {author} {\bibfnamefont {P.}~\bibnamefont
  {Abratenko}} \emph {et~al.} (\bibinfo {collaboration} {MicroBooNE}),\
  }\bibfield  {title} {\bibinfo {title} {{Convolutional neural network for
  multiple particle identification in the MicroBooNE liquid argon time
  projection chamber}},\ }\href {https://doi.org/10.1103/PhysRevD.103.092003}
  {\bibfield  {journal} {\bibinfo  {journal} {Phys. Rev. D}\ }\textbf {\bibinfo
  {volume} {103}},\ \bibinfo {pages} {092003} (\bibinfo {year}
  {2021}{\natexlab{a}})},\ \Eprint {https://arxiv.org/abs/2010.08653}
  {arXiv:2010.08653 [hep-ex]} \BibitemShut {NoStop}%
\bibitem [{\citenamefont {Baldi}\ \emph {et~al.}(2014)\citenamefont {Baldi},
  \citenamefont {Sadowski},\ and\ \citenamefont {Whiteson}}]{Baldi:2014kfa}%
  \BibitemOpen
  \bibfield  {author} {\bibinfo {author} {\bibfnamefont {P.}~\bibnamefont
  {Baldi}}, \bibinfo {author} {\bibfnamefont {P.}~\bibnamefont {Sadowski}},\
  and\ \bibinfo {author} {\bibfnamefont {D.}~\bibnamefont {Whiteson}},\
  }\bibfield  {title} {\bibinfo {title} {{Searching for Exotic Particles in
  High-Energy Physics with Deep Learning}},\ }\href
  {https://doi.org/10.1038/ncomms5308} {\bibfield  {journal} {\bibinfo
  {journal} {Nature Commun.}\ }\textbf {\bibinfo {volume} {5}},\ \bibinfo
  {pages} {4308} (\bibinfo {year} {2014})},\ \Eprint
  {https://arxiv.org/abs/1402.4735} {arXiv:1402.4735 [hep-ph]} \BibitemShut
  {NoStop}%
\bibitem [{\citenamefont {Bhimji}\ \emph {et~al.}(2018)\citenamefont {Bhimji},
  \citenamefont {Farrell}, \citenamefont {Kurth}, \citenamefont {Paganini},
  \citenamefont {Prabhat},\ and\ \citenamefont {Racah}}]{Bhimji:2017qvb}%
  \BibitemOpen
  \bibfield  {author} {\bibinfo {author} {\bibfnamefont {W.}~\bibnamefont
  {Bhimji}}, \bibinfo {author} {\bibfnamefont {S.~A.}\ \bibnamefont {Farrell}},
  \bibinfo {author} {\bibfnamefont {T.}~\bibnamefont {Kurth}}, \bibinfo
  {author} {\bibfnamefont {M.}~\bibnamefont {Paganini}}, \bibinfo {author}
  {\bibnamefont {Prabhat}},\ and\ \bibinfo {author} {\bibfnamefont
  {E.}~\bibnamefont {Racah}},\ }\bibfield  {title} {\bibinfo {title} {{Deep
  Neural Networks for Physics Analysis on low-level whole-detector data at the
  LHC}},\ }\href {https://doi.org/10.1088/1742-6596/1085/4/042034} {\bibfield
  {journal} {\bibinfo  {journal} {J. Phys. Conf. Ser.}\ }\textbf {\bibinfo
  {volume} {1085}},\ \bibinfo {pages} {042034} (\bibinfo {year} {2018})},\
  \Eprint {https://arxiv.org/abs/1711.03573} {arXiv:1711.03573 [hep-ex]}
  \BibitemShut {NoStop}%
\bibitem [{\citenamefont {Wunsch}\ \emph {et~al.}(2021)\citenamefont {Wunsch},
  \citenamefont {J\"orger}, \citenamefont {Wolf},\ and\ \citenamefont
  {Quast}}]{Wunsch:2020iuh}%
  \BibitemOpen
  \bibfield  {author} {\bibinfo {author} {\bibfnamefont {S.}~\bibnamefont
  {Wunsch}}, \bibinfo {author} {\bibfnamefont {S.}~\bibnamefont {J\"orger}},
  \bibinfo {author} {\bibfnamefont {R.}~\bibnamefont {Wolf}},\ and\ \bibinfo
  {author} {\bibfnamefont {G.}~\bibnamefont {Quast}},\ }\bibfield  {title}
  {\bibinfo {title} {{Optimal statistical inference in the presence of
  systematic uncertainties using neural network optimization based on binned
  Poisson likelihoods with nuisance parameters}},\ }\href
  {https://doi.org/10.1007/s41781-020-00049-5} {\bibfield  {journal} {\bibinfo
  {journal} {Comput. Softw. Big Sci.}\ }\textbf {\bibinfo {volume} {5}},\
  \bibinfo {pages} {4} (\bibinfo {year} {2021})},\ \Eprint
  {https://arxiv.org/abs/2003.07186} {arXiv:2003.07186 [physics.data-an]}
  \BibitemShut {NoStop}%
\bibitem [{\citenamefont {Elwood}\ \emph {et~al.}(2020)\citenamefont {Elwood},
  \citenamefont {Kr\"ucker},\ and\ \citenamefont
  {Shchedrolosiev}}]{Elwood:2020pik}%
  \BibitemOpen
  \bibfield  {author} {\bibinfo {author} {\bibfnamefont {A.}~\bibnamefont
  {Elwood}}, \bibinfo {author} {\bibfnamefont {D.}~\bibnamefont {Kr\"ucker}},\
  and\ \bibinfo {author} {\bibfnamefont {M.}~\bibnamefont {Shchedrolosiev}},\
  }\bibfield  {title} {\bibinfo {title} {{Direct optimization of the discovery
  significance in machine learning for new physics searches in particle
  colliders}},\ }\href {https://doi.org/10.1088/1742-6596/1525/1/012110}
  {\bibfield  {journal} {\bibinfo  {journal} {J. Phys. Conf. Ser.}\ }\textbf
  {\bibinfo {volume} {1525}},\ \bibinfo {pages} {012110} (\bibinfo {year}
  {2020})}\BibitemShut {NoStop}%
\bibitem [{\citenamefont {Xia}(2019)}]{Xia:2018kgd}%
  \BibitemOpen
  \bibfield  {author} {\bibinfo {author} {\bibfnamefont {L.-G.}\ \bibnamefont
  {Xia}},\ }\bibfield  {title} {\bibinfo {title} {{QBDT, a new boosting
  decision tree method with systematical uncertainties into training for High
  Energy Physics}},\ }\href {https://doi.org/10.1016/j.nima.2019.03.088}
  {\bibfield  {journal} {\bibinfo  {journal} {Nucl. Instrum. Meth.}\ }\textbf
  {\bibinfo {volume} {A930}},\ \bibinfo {pages} {15} (\bibinfo {year}
  {2019})},\ \Eprint {https://arxiv.org/abs/1810.08387} {arXiv:1810.08387
  [physics.data-an]} \BibitemShut {NoStop}%
\bibitem [{\citenamefont {De~Castro}\ and\ \citenamefont
  {Dorigo}(2019)}]{deCastro:2018mgh}%
  \BibitemOpen
  \bibfield  {author} {\bibinfo {author} {\bibfnamefont {P.}~\bibnamefont
  {De~Castro}}\ and\ \bibinfo {author} {\bibfnamefont {T.}~\bibnamefont
  {Dorigo}},\ }\bibfield  {title} {\bibinfo {title} {{INFERNO: Inference-Aware
  Neural Optimisation}},\ }\href {https://doi.org/10.1016/j.cpc.2019.06.007}
  {\bibfield  {journal} {\bibinfo  {journal} {Comput. Phys. Commun.}\ }\textbf
  {\bibinfo {volume} {244}},\ \bibinfo {pages} {170} (\bibinfo {year}
  {2019})},\ \Eprint {https://arxiv.org/abs/1806.04743} {arXiv:1806.04743
  [stat.ML]} \BibitemShut {NoStop}%
\bibitem [{\citenamefont {Charnock}\ \emph {et~al.}(2018)\citenamefont
  {Charnock}, \citenamefont {Lavaux},\ and\ \citenamefont
  {Wandelt}}]{Charnock_2018}%
  \BibitemOpen
  \bibfield  {author} {\bibinfo {author} {\bibfnamefont {T.}~\bibnamefont
  {Charnock}}, \bibinfo {author} {\bibfnamefont {G.}~\bibnamefont {Lavaux}},\
  and\ \bibinfo {author} {\bibfnamefont {B.~D.}\ \bibnamefont {Wandelt}},\
  }\bibfield  {title} {\bibinfo {title} {Automatic physical inference with
  information maximizing neural networks},\ }\bibfield  {journal} {\bibinfo
  {journal} {Physical Review D}\ }\textbf {\bibinfo {volume} {97}},\ \href
  {https://doi.org/10.1103/physrevd.97.083004} {10.1103/physrevd.97.083004}
  (\bibinfo {year} {2018})\BibitemShut {NoStop}%
\bibitem [{\citenamefont {Alsing}\ and\ \citenamefont
  {Wandelt}(2019)}]{Alsing:2019dvb}%
  \BibitemOpen
  \bibfield  {author} {\bibinfo {author} {\bibfnamefont {J.}~\bibnamefont
  {Alsing}}\ and\ \bibinfo {author} {\bibfnamefont {B.}~\bibnamefont
  {Wandelt}},\ }\bibfield  {title} {\bibinfo {title} {{Nuisance hardened data
  compression for fast likelihood-free inference}},\ }\href
  {https://doi.org/10.1093/mnras/stz1900} {\bibfield  {journal} {\bibinfo
  {journal} {Mon. Not. Roy. Astron. Soc.}\ }\textbf {\bibinfo {volume} {488}},\
  \bibinfo {pages} {5093} (\bibinfo {year} {2019})},\ \Eprint
  {https://arxiv.org/abs/1903.01473} {arXiv:1903.01473 [astro-ph.CO]}
  \BibitemShut {NoStop}%
\bibitem [{\citenamefont {Heinrich}\ and\ \citenamefont
  {Simpson}(2020)}]{lukas_heinrich_2020_3697981}%
  \BibitemOpen
  \bibfield  {author} {\bibinfo {author} {\bibfnamefont {L.}~\bibnamefont
  {Heinrich}}\ and\ \bibinfo {author} {\bibfnamefont {N.}~\bibnamefont
  {Simpson}},\ }\href {https://doi.org/10.5281/zenodo.3697981} {\bibinfo
  {title} {pyhf/neos: initial zenodo release}} (\bibinfo {year}
  {2020})\BibitemShut {NoStop}%
\bibitem [{\citenamefont {Dorigo}\ and\ \citenamefont
  {de~Castro}(2020)}]{1807719}%
  \BibitemOpen
  \bibfield  {author} {\bibinfo {author} {\bibfnamefont {T.}~\bibnamefont
  {Dorigo}}\ and\ \bibinfo {author} {\bibfnamefont {P.}~\bibnamefont
  {de~Castro}},\ }\bibfield  {title} {\bibinfo {title} {{Dealing with Nuisance
  Parameters using Machine Learning in High Energy Physics: a Review}},\
  }\href@noop {} {\  (\bibinfo {year} {2020})},\ \Eprint
  {https://arxiv.org/abs/2007.09121} {arXiv:2007.09121 [stat.ML]} \BibitemShut
  {NoStop}%
\bibitem [{\citenamefont {Kasieczka}\ \emph
  {et~al.}(2020{\natexlab{a}})\citenamefont {Kasieczka}, \citenamefont
  {Luchmann}, \citenamefont {Otterpohl},\ and\ \citenamefont
  {Plehn}}]{Kasieczka:2020vlh}%
  \BibitemOpen
  \bibfield  {author} {\bibinfo {author} {\bibfnamefont {G.}~\bibnamefont
  {Kasieczka}}, \bibinfo {author} {\bibfnamefont {M.}~\bibnamefont {Luchmann}},
  \bibinfo {author} {\bibfnamefont {F.}~\bibnamefont {Otterpohl}},\ and\
  \bibinfo {author} {\bibfnamefont {T.}~\bibnamefont {Plehn}},\ }\bibfield
  {title} {\bibinfo {title} {{Per-Object Systematics using Deep-Learned
  Calibration}}\ }\href {https://doi.org/10.21468/SciPostPhys.9.6.089}
  {10.21468/SciPostPhys.9.6.089} (\bibinfo {year} {2020}{\natexlab{a}}),\
  \Eprint {https://arxiv.org/abs/2003.11099} {arXiv:2003.11099 [hep-ph]}
  \BibitemShut {NoStop}%
\bibitem [{\citenamefont {Bollweg}\ \emph {et~al.}(2020)\citenamefont
  {Bollweg}, \citenamefont {Haußmann}, \citenamefont {Kasieczka},
  \citenamefont {Luchmann}, \citenamefont {Plehn},\ and\ \citenamefont
  {Thompson}}]{Bollweg:2019skg}%
  \BibitemOpen
  \bibfield  {author} {\bibinfo {author} {\bibfnamefont {S.}~\bibnamefont
  {Bollweg}}, \bibinfo {author} {\bibfnamefont {M.}~\bibnamefont {Haußmann}},
  \bibinfo {author} {\bibfnamefont {G.}~\bibnamefont {Kasieczka}}, \bibinfo
  {author} {\bibfnamefont {M.}~\bibnamefont {Luchmann}}, \bibinfo {author}
  {\bibfnamefont {T.}~\bibnamefont {Plehn}},\ and\ \bibinfo {author}
  {\bibfnamefont {J.}~\bibnamefont {Thompson}},\ }\bibfield  {title} {\bibinfo
  {title} {{Deep-Learning Jets with Uncertainties and More}},\ }\href
  {https://doi.org/10.21468/SciPostPhys.8.1.006} {\bibfield  {journal}
  {\bibinfo  {journal} {SciPost Phys.}\ }\textbf {\bibinfo {volume} {8}},\
  \bibinfo {pages} {006} (\bibinfo {year} {2020})},\ \Eprint
  {https://arxiv.org/abs/1904.10004} {arXiv:1904.10004 [hep-ph]} \BibitemShut
  {NoStop}%
\bibitem [{\citenamefont {Araz}\ and\ \citenamefont
  {Spannowsky}(2021)}]{Araz:2021wqm}%
  \BibitemOpen
  \bibfield  {author} {\bibinfo {author} {\bibfnamefont {J.~Y.}\ \bibnamefont
  {Araz}}\ and\ \bibinfo {author} {\bibfnamefont {M.}~\bibnamefont
  {Spannowsky}},\ }\bibfield  {title} {\bibinfo {title} {{Combine and Conquer:
  Event Reconstruction with Bayesian Ensemble Neural Networks}},\ }\href
  {https://doi.org/10.1007/JHEP04(2021)296} {\bibfield  {journal} {\bibinfo
  {journal} {JHEP}\ }\textbf {\bibinfo {volume} {04}},\ \bibinfo {pages}
  {296}},\ \Eprint {https://arxiv.org/abs/2102.01078} {arXiv:2102.01078
  [hep-ph]} \BibitemShut {NoStop}%
\bibitem [{\citenamefont {Bellagente}\ \emph {et~al.}(2021)\citenamefont
  {Bellagente}, \citenamefont {Hau\ss{}mann}, \citenamefont {Luchmann},\ and\
  \citenamefont {Plehn}}]{Bellagente:2021yyh}%
  \BibitemOpen
  \bibfield  {author} {\bibinfo {author} {\bibfnamefont {M.}~\bibnamefont
  {Bellagente}}, \bibinfo {author} {\bibfnamefont {M.}~\bibnamefont
  {Hau\ss{}mann}}, \bibinfo {author} {\bibfnamefont {M.}~\bibnamefont
  {Luchmann}},\ and\ \bibinfo {author} {\bibfnamefont {T.}~\bibnamefont
  {Plehn}},\ }\bibfield  {title} {\bibinfo {title} {{Understanding
  Event-Generation Networks via Uncertainties}},\ }\href@noop {} {\  (\bibinfo
  {year} {2021})},\ \Eprint {https://arxiv.org/abs/2104.04543}
  {arXiv:2104.04543 [hep-ph]} \BibitemShut {NoStop}%
\bibitem [{\citenamefont {Brehmer}\ \emph
  {et~al.}(2020{\natexlab{a}})\citenamefont {Brehmer}, \citenamefont {Kling},
  \citenamefont {Espejo},\ and\ \citenamefont {Cranmer}}]{Brehmer:2019xox}%
  \BibitemOpen
  \bibfield  {author} {\bibinfo {author} {\bibfnamefont {J.}~\bibnamefont
  {Brehmer}}, \bibinfo {author} {\bibfnamefont {F.}~\bibnamefont {Kling}},
  \bibinfo {author} {\bibfnamefont {I.}~\bibnamefont {Espejo}},\ and\ \bibinfo
  {author} {\bibfnamefont {K.}~\bibnamefont {Cranmer}},\ }\bibfield  {title}
  {\bibinfo {title} {{MadMiner: Machine learning-based inference for particle
  physics}},\ }\href {https://doi.org/10.1007/s41781-020-0035-2} {\bibfield
  {journal} {\bibinfo  {journal} {Comput. Softw. Big Sci.}\ }\textbf {\bibinfo
  {volume} {4}},\ \bibinfo {pages} {3} (\bibinfo {year}
  {2020}{\natexlab{a}})},\ \Eprint {https://arxiv.org/abs/1907.10621}
  {arXiv:1907.10621 [hep-ph]} \BibitemShut {NoStop}%
\bibitem [{\citenamefont {Brehmer}\ \emph
  {et~al.}(2020{\natexlab{b}})\citenamefont {Brehmer}, \citenamefont {Louppe},
  \citenamefont {Pavez},\ and\ \citenamefont {Cranmer}}]{Brehmer:2018hga}%
  \BibitemOpen
  \bibfield  {author} {\bibinfo {author} {\bibfnamefont {J.}~\bibnamefont
  {Brehmer}}, \bibinfo {author} {\bibfnamefont {G.}~\bibnamefont {Louppe}},
  \bibinfo {author} {\bibfnamefont {J.}~\bibnamefont {Pavez}},\ and\ \bibinfo
  {author} {\bibfnamefont {K.}~\bibnamefont {Cranmer}},\ }\bibfield  {title}
  {\bibinfo {title} {{Mining gold from implicit models to improve
  likelihood-free inference}},\ }\href
  {https://doi.org/10.1073/pnas.1915980117} {\bibfield  {journal} {\bibinfo
  {journal} {Proc. Nat. Acad. Sci.}\ ,\ \bibinfo {pages} {201915980}} (\bibinfo
  {year} {2020}{\natexlab{b}})},\ \Eprint {https://arxiv.org/abs/1805.12244}
  {arXiv:1805.12244 [stat.ML]} \BibitemShut {NoStop}%
\bibitem [{\citenamefont {Brehmer}\ \emph
  {et~al.}(2018{\natexlab{a}})\citenamefont {Brehmer}, \citenamefont {Cranmer},
  \citenamefont {Louppe},\ and\ \citenamefont {Pavez}}]{Brehmer:2018kdj}%
  \BibitemOpen
  \bibfield  {author} {\bibinfo {author} {\bibfnamefont {J.}~\bibnamefont
  {Brehmer}}, \bibinfo {author} {\bibfnamefont {K.}~\bibnamefont {Cranmer}},
  \bibinfo {author} {\bibfnamefont {G.}~\bibnamefont {Louppe}},\ and\ \bibinfo
  {author} {\bibfnamefont {J.}~\bibnamefont {Pavez}},\ }\bibfield  {title}
  {\bibinfo {title} {{Constraining Effective Field Theories with Machine
  Learning}}\ }\href {https://doi.org/10.1103/PhysRevLett.121.111801}
  {10.1103/PhysRevLett.121.111801} (\bibinfo {year} {2018}{\natexlab{a}}),\
  \Eprint {https://arxiv.org/abs/1805.00013} {arXiv:1805.00013 [hep-ph]}
  \BibitemShut {NoStop}%
\bibitem [{\citenamefont {Brehmer}\ \emph
  {et~al.}(2018{\natexlab{b}})\citenamefont {Brehmer}, \citenamefont {Cranmer},
  \citenamefont {Louppe},\ and\ \citenamefont {Pavez}}]{Brehmer:2018eca}%
  \BibitemOpen
  \bibfield  {author} {\bibinfo {author} {\bibfnamefont {J.}~\bibnamefont
  {Brehmer}}, \bibinfo {author} {\bibfnamefont {K.}~\bibnamefont {Cranmer}},
  \bibinfo {author} {\bibfnamefont {G.}~\bibnamefont {Louppe}},\ and\ \bibinfo
  {author} {\bibfnamefont {J.}~\bibnamefont {Pavez}},\ }\bibfield  {title}
  {\bibinfo {title} {{A Guide to Constraining Effective Field Theories with
  Machine Learning}}\ }\href {https://doi.org/10.1103/PhysRevD.98.052004}
  {10.1103/PhysRevD.98.052004} (\bibinfo {year} {2018}{\natexlab{b}}),\ \Eprint
  {https://arxiv.org/abs/1805.00020} {arXiv:1805.00020 [hep-ph]} \BibitemShut
  {NoStop}%
\bibitem [{\citenamefont {Nachman}(2019)}]{Nachman:2019dol}%
  \BibitemOpen
  \bibfield  {author} {\bibinfo {author} {\bibfnamefont {B.}~\bibnamefont
  {Nachman}},\ }\bibfield  {title} {\bibinfo {title} {{A guide for deploying
  Deep Learning in LHC searches: How to achieve optimality and account for
  uncertainty}}\ }\href {https://doi.org/10.21468/SciPostPhys.8.6.090}
  {10.21468/SciPostPhys.8.6.090} (\bibinfo {year} {2019}),\ \Eprint
  {https://arxiv.org/abs/1909.03081} {arXiv:1909.03081 [hep-ph]} \BibitemShut
  {NoStop}%
\bibitem [{\citenamefont {Ghosh}\ \emph {et~al.}(2021)\citenamefont {Ghosh},
  \citenamefont {Nachman},\ and\ \citenamefont {Whiteson}}]{Ghosh:2021roe}%
  \BibitemOpen
  \bibfield  {author} {\bibinfo {author} {\bibfnamefont {A.}~\bibnamefont
  {Ghosh}}, \bibinfo {author} {\bibfnamefont {B.}~\bibnamefont {Nachman}},\
  and\ \bibinfo {author} {\bibfnamefont {D.}~\bibnamefont {Whiteson}},\
  }\bibfield  {title} {\bibinfo {title} {{Uncertainty Aware Learning for High
  Energy Physics}},\ }\href@noop {} {\  (\bibinfo {year} {2021})},\ \Eprint
  {https://arxiv.org/abs/2105.08742} {arXiv:2105.08742 [hep-ex]} \BibitemShut
  {NoStop}%
\bibitem [{\citenamefont {Rogozhnikov}(2016)}]{Rogozhnikov:2016bdp}%
  \BibitemOpen
  \bibfield  {author} {\bibinfo {author} {\bibfnamefont {A.}~\bibnamefont
  {Rogozhnikov}},\ }\bibfield  {title} {\bibinfo {title} {{Reweighting with
  Boosted Decision Trees}},\ }\href
  {https://doi.org/10.1088/1742-6596/762/1/012036} {\bibfield  {journal}
  {\bibinfo  {journal} {{Proceedings, 17th International Workshop on Advanced
  Computing and Analysis Techniques in Physics Research (ACAT 2016):
  Valparaiso, Chile, January 18-22, 2016}}\ }\textbf {\bibinfo {volume}
  {762}},\ \bibinfo {pages} {012036} (\bibinfo {year} {2016})},\ \Eprint
  {https://arxiv.org/abs/1608.05806} {arXiv:1608.05806 [physics.data-an]}
  \BibitemShut {NoStop}%
\bibitem [{\citenamefont {Andreassen}\ and\ \citenamefont
  {Nachman}(2020)}]{Andreassen:2019nnm}%
  \BibitemOpen
  \bibfield  {author} {\bibinfo {author} {\bibfnamefont {A.}~\bibnamefont
  {Andreassen}}\ and\ \bibinfo {author} {\bibfnamefont {B.}~\bibnamefont
  {Nachman}},\ }\bibfield  {title} {\bibinfo {title} {{Neural Networks for Full
  Phase-space Reweighting and Parameter Tuning}},\ }\href
  {https://doi.org/10.1103/PhysRevD.101.091901} {\bibfield  {journal} {\bibinfo
   {journal} {Phys. Rev. D}\ }\textbf {\bibinfo {volume} {101}},\ \bibinfo
  {pages} {091901} (\bibinfo {year} {2020})},\ \Eprint
  {https://arxiv.org/abs/1907.08209} {arXiv:1907.08209 [hep-ph]} \BibitemShut
  {NoStop}%
\bibitem [{\citenamefont {Cranmer}\ \emph {et~al.}(2015)\citenamefont
  {Cranmer}, \citenamefont {Pavez},\ and\ \citenamefont
  {Louppe}}]{Cranmer:2015bka}%
  \BibitemOpen
  \bibfield  {author} {\bibinfo {author} {\bibfnamefont {K.}~\bibnamefont
  {Cranmer}}, \bibinfo {author} {\bibfnamefont {J.}~\bibnamefont {Pavez}},\
  and\ \bibinfo {author} {\bibfnamefont {G.}~\bibnamefont {Louppe}},\
  }\bibfield  {title} {\bibinfo {title} {{Approximating Likelihood Ratios with
  Calibrated Discriminative Classifiers}},\ }\href@noop {} {\  (\bibinfo {year}
  {2015})},\ \Eprint {https://arxiv.org/abs/1506.02169} {arXiv:1506.02169
  [stat.AP]} \BibitemShut {NoStop}%
\bibitem [{\citenamefont {Diefenbacher}\ \emph {et~al.}(2020)\citenamefont
  {Diefenbacher}, \citenamefont {Eren}, \citenamefont {Kasieczka},
  \citenamefont {Korol}, \citenamefont {Nachman},\ and\ \citenamefont
  {Shih}}]{2009.03796}%
  \BibitemOpen
  \bibfield  {author} {\bibinfo {author} {\bibfnamefont {S.}~\bibnamefont
  {Diefenbacher}}, \bibinfo {author} {\bibfnamefont {E.}~\bibnamefont {Eren}},
  \bibinfo {author} {\bibfnamefont {G.}~\bibnamefont {Kasieczka}}, \bibinfo
  {author} {\bibfnamefont {A.}~\bibnamefont {Korol}}, \bibinfo {author}
  {\bibfnamefont {B.}~\bibnamefont {Nachman}},\ and\ \bibinfo {author}
  {\bibfnamefont {D.}~\bibnamefont {Shih}},\ }\bibfield  {title} {\bibinfo
  {title} {{DCTRGAN: Improving the Precision of Generative Models with
  Reweighting}},\ }\href {https://doi.org/10.1088/1748-0221/15/11/p11004}
  {\bibfield  {journal} {\bibinfo  {journal} {Journal of Instrumentation}\
  }\textbf {\bibinfo {volume} {15}},\ \bibinfo {pages} {P11004}},\ \Eprint
  {https://arxiv.org/abs/2009.03796} {arXiv:2009.03796 [hep-ph]} \BibitemShut
  {NoStop}%
\bibitem [{\citenamefont {Nachman}\ and\ \citenamefont
  {Thaler}(2021)}]{Nachman:2021opi}%
  \BibitemOpen
  \bibfield  {author} {\bibinfo {author} {\bibfnamefont {B.}~\bibnamefont
  {Nachman}}\ and\ \bibinfo {author} {\bibfnamefont {J.}~\bibnamefont
  {Thaler}},\ }\bibfield  {title} {\bibinfo {title} {{Neural Conditional
  Reweighting}},\ }\href@noop {} {\  (\bibinfo {year} {2021})},\ \Eprint
  {https://arxiv.org/abs/2107.08979} {arXiv:2107.08979 [physics.data-an]}
  \BibitemShut {NoStop}%
\bibitem [{\citenamefont {Clavijo}\ \emph {et~al.}(2020)\citenamefont
  {Clavijo}, \citenamefont {Glaysher},\ and\ \citenamefont
  {Katzy}}]{clavijo2020adversarial}%
  \BibitemOpen
  \bibfield  {author} {\bibinfo {author} {\bibfnamefont {J.~M.}\ \bibnamefont
  {Clavijo}}, \bibinfo {author} {\bibfnamefont {P.}~\bibnamefont {Glaysher}},\
  and\ \bibinfo {author} {\bibfnamefont {J.~M.}\ \bibnamefont {Katzy}},\
  }\bibfield  {title} {\bibinfo {title} {{Adversarial domain adaptation to
  reduce sample bias of a high energy physics classifier}},\ }\href@noop {} {\
  (\bibinfo {year} {2020})},\ \Eprint {https://arxiv.org/abs/2005.00568}
  {arXiv:2005.00568 [stat.ML]} \BibitemShut {NoStop}%
\bibitem [{\citenamefont {Perdue}\ \emph {et~al.}(2018)\citenamefont {Perdue}
  \emph {et~al.}}]{MINERvA:2018smv}%
  \BibitemOpen
  \bibfield  {author} {\bibinfo {author} {\bibfnamefont {G.~N.}\ \bibnamefont
  {Perdue}} \emph {et~al.} (\bibinfo {collaboration} {MINERvA}),\ }\bibfield
  {title} {\bibinfo {title} {{Reducing model bias in a deep learning classifier
  using domain adversarial neural networks in the MINERvA experiment}},\ }\href
  {https://doi.org/10.1088/1748-0221/13/11/P11020} {\bibfield  {journal}
  {\bibinfo  {journal} {JINST}\ }\textbf {\bibinfo {volume} {13}}\bibfield
  {number} {\bibinfo  {number} { (11)},\ \bibinfo {pages} {P11020}},\ }\Eprint
  {https://arxiv.org/abs/1808.08332} {arXiv:1808.08332 [physics.data-an]}
  \BibitemShut {NoStop}%
\bibitem [{\citenamefont {Lin}\ \emph {et~al.}(2019)\citenamefont {Lin},
  \citenamefont {Bhimji},\ and\ \citenamefont {Nachman}}]{Lin:2019htn}%
  \BibitemOpen
  \bibfield  {author} {\bibinfo {author} {\bibfnamefont {J.}~\bibnamefont
  {Lin}}, \bibinfo {author} {\bibfnamefont {W.}~\bibnamefont {Bhimji}},\ and\
  \bibinfo {author} {\bibfnamefont {B.}~\bibnamefont {Nachman}},\ }\bibfield
  {title} {\bibinfo {title} {{Machine Learning Templates for QCD Factorization
  in the Search for Physics Beyond the Standard Model}},\ }\href
  {https://doi.org/10.1007/JHEP05(2019)181} {\bibfield  {journal} {\bibinfo
  {journal} {JHEP}\ }\textbf {\bibinfo {volume} {05}},\ \bibinfo {pages}
  {181}},\ \Eprint {https://arxiv.org/abs/1903.02556} {arXiv:1903.02556
  [hep-ph]} \BibitemShut {NoStop}%
\bibitem [{\citenamefont {Kasieczka}\ \emph
  {et~al.}(2020{\natexlab{b}})\citenamefont {Kasieczka}, \citenamefont
  {Nachman}, \citenamefont {Schwartz},\ and\ \citenamefont
  {Shih}}]{Kasieczka:2020pil}%
  \BibitemOpen
  \bibfield  {author} {\bibinfo {author} {\bibfnamefont {G.}~\bibnamefont
  {Kasieczka}}, \bibinfo {author} {\bibfnamefont {B.}~\bibnamefont {Nachman}},
  \bibinfo {author} {\bibfnamefont {M.~D.}\ \bibnamefont {Schwartz}},\ and\
  \bibinfo {author} {\bibfnamefont {D.}~\bibnamefont {Shih}},\ }\bibfield
  {title} {\bibinfo {title} {{ABCDisCo: Automating the ABCD Method with Machine
  Learning}}\ }\href {https://doi.org/10.1103/PhysRevD.103.035021}
  {10.1103/PhysRevD.103.035021} (\bibinfo {year} {2020}{\natexlab{b}}),\
  \Eprint {https://arxiv.org/abs/2007.14400} {arXiv:2007.14400 [hep-ph]}
  \BibitemShut {NoStop}%
\bibitem [{\citenamefont {Mikuni}\ \emph {et~al.}(2021)\citenamefont {Mikuni},
  \citenamefont {Nachman},\ and\ \citenamefont {Shih}}]{Mikuni:2021nwn}%
  \BibitemOpen
  \bibfield  {author} {\bibinfo {author} {\bibfnamefont {V.}~\bibnamefont
  {Mikuni}}, \bibinfo {author} {\bibfnamefont {B.}~\bibnamefont {Nachman}},\
  and\ \bibinfo {author} {\bibfnamefont {D.}~\bibnamefont {Shih}},\ }\bibfield
  {title} {\bibinfo {title} {{Online-compatible Unsupervised Non-resonant
  Anomaly Detection}},\ }\href@noop {} {\  (\bibinfo {year} {2021})},\ \Eprint
  {https://arxiv.org/abs/2111.06417} {arXiv:2111.06417 [cs.LG]} \BibitemShut
  {NoStop}%
\bibitem [{\citenamefont {Blance}\ \emph {et~al.}(2019)\citenamefont {Blance},
  \citenamefont {Spannowsky},\ and\ \citenamefont {Waite}}]{Blance:2019ibf}%
  \BibitemOpen
  \bibfield  {author} {\bibinfo {author} {\bibfnamefont {A.}~\bibnamefont
  {Blance}}, \bibinfo {author} {\bibfnamefont {M.}~\bibnamefont {Spannowsky}},\
  and\ \bibinfo {author} {\bibfnamefont {P.}~\bibnamefont {Waite}},\ }\bibfield
   {title} {\bibinfo {title} {{Adversarially-trained autoencoders for robust
  unsupervised new physics searches}},\ }\href
  {https://doi.org/10.1007/JHEP10(2019)047} {\bibfield  {journal} {\bibinfo
  {journal} {JHEP}\ }\textbf {\bibinfo {volume} {10}},\ \bibinfo {pages}
  {047}},\ \Eprint {https://arxiv.org/abs/1905.10384} {arXiv:1905.10384
  [hep-ph]} \BibitemShut {NoStop}%
\bibitem [{\citenamefont {Englert}\ \emph {et~al.}(2019)\citenamefont
  {Englert}, \citenamefont {Galler}, \citenamefont {Harris},\ and\
  \citenamefont {Spannowsky}}]{Englert:2018cfo}%
  \BibitemOpen
  \bibfield  {author} {\bibinfo {author} {\bibfnamefont {C.}~\bibnamefont
  {Englert}}, \bibinfo {author} {\bibfnamefont {P.}~\bibnamefont {Galler}},
  \bibinfo {author} {\bibfnamefont {P.}~\bibnamefont {Harris}},\ and\ \bibinfo
  {author} {\bibfnamefont {M.}~\bibnamefont {Spannowsky}},\ }\bibfield  {title}
  {\bibinfo {title} {{Machine Learning Uncertainties with Adversarial Neural
  Networks}},\ }\href {https://doi.org/10.1140/epjc/s10052-018-6511-8}
  {\bibfield  {journal} {\bibinfo  {journal} {Eur. Phys. J.}\ }\textbf
  {\bibinfo {volume} {C79}},\ \bibinfo {pages} {4} (\bibinfo {year} {2019})},\
  \Eprint {https://arxiv.org/abs/1807.08763} {arXiv:1807.08763 [hep-ph]}
  \BibitemShut {NoStop}%
\bibitem [{\citenamefont {Louppe}\ \emph {et~al.}(2017)\citenamefont {Louppe},
  \citenamefont {Kagan},\ and\ \citenamefont {Cranmer}}]{Louppe:2016ylz}%
  \BibitemOpen
  \bibfield  {author} {\bibinfo {author} {\bibfnamefont {G.}~\bibnamefont
  {Louppe}}, \bibinfo {author} {\bibfnamefont {M.}~\bibnamefont {Kagan}},\ and\
  \bibinfo {author} {\bibfnamefont {K.}~\bibnamefont {Cranmer}},\ }\bibfield
  {title} {\bibinfo {title} {{Learning to Pivot with Adversarial Networks}},\
  }in\ \href
  {https://papers.nips.cc/paper/2017/hash/48ab2f9b45957ab574cf005eb8a76760-Abstract.html}
  {\emph {\bibinfo {booktitle} {{Advances in Neural Information Processing
  Systems}}}},\ Vol.~\bibinfo {volume} {30},\ \bibinfo {editor} {edited by\
  \bibinfo {editor} {\bibfnamefont {I.}~\bibnamefont {Guyon}}, \bibinfo
  {editor} {\bibfnamefont {U.~V.}\ \bibnamefont {Luxburg}}, \bibinfo {editor}
  {\bibfnamefont {S.}~\bibnamefont {Bengio}}, \bibinfo {editor} {\bibfnamefont
  {H.}~\bibnamefont {Wallach}}, \bibinfo {editor} {\bibfnamefont
  {R.}~\bibnamefont {Fergus}}, \bibinfo {editor} {\bibfnamefont
  {S.}~\bibnamefont {Vishwanathan}},\ and\ \bibinfo {editor} {\bibfnamefont
  {R.}~\bibnamefont {Garnett}}}\ (\bibinfo  {publisher} {Curran Associates,
  Inc.},\ \bibinfo {year} {2017})\ \Eprint {https://arxiv.org/abs/1611.01046}
  {arXiv:1611.01046 [stat.ME]} \BibitemShut {NoStop}%
\bibitem [{\citenamefont {Dolen}\ \emph {et~al.}(2016)\citenamefont {Dolen},
  \citenamefont {Harris}, \citenamefont {Marzani}, \citenamefont {Rappoccio},\
  and\ \citenamefont {Tran}}]{Dolen:2016kst}%
  \BibitemOpen
  \bibfield  {author} {\bibinfo {author} {\bibfnamefont {J.}~\bibnamefont
  {Dolen}}, \bibinfo {author} {\bibfnamefont {P.}~\bibnamefont {Harris}},
  \bibinfo {author} {\bibfnamefont {S.}~\bibnamefont {Marzani}}, \bibinfo
  {author} {\bibfnamefont {S.}~\bibnamefont {Rappoccio}},\ and\ \bibinfo
  {author} {\bibfnamefont {N.}~\bibnamefont {Tran}},\ }\bibfield  {title}
  {\bibinfo {title} {{Thinking outside the ROCs: Designing Decorrelated Taggers
  (DDT) for jet substructure}},\ }\href
  {https://doi.org/10.1007/JHEP05(2016)156} {\bibfield  {journal} {\bibinfo
  {journal} {JHEP}\ }\textbf {\bibinfo {volume} {05}},\ \bibinfo {pages}
  {156}},\ \Eprint {https://arxiv.org/abs/1603.00027} {arXiv:1603.00027
  [hep-ph]} \BibitemShut {NoStop}%
\bibitem [{\citenamefont {Moult}\ \emph {et~al.}(2018)\citenamefont {Moult},
  \citenamefont {Nachman},\ and\ \citenamefont {Neill}}]{Moult:2017okx}%
  \BibitemOpen
  \bibfield  {author} {\bibinfo {author} {\bibfnamefont {I.}~\bibnamefont
  {Moult}}, \bibinfo {author} {\bibfnamefont {B.}~\bibnamefont {Nachman}},\
  and\ \bibinfo {author} {\bibfnamefont {D.}~\bibnamefont {Neill}},\ }\bibfield
   {title} {\bibinfo {title} {{Convolved Substructure: Analytically
  Decorrelating Jet Substructure Observables}},\ }\href
  {https://doi.org/10.1007/JHEP05(2018)002} {\bibfield  {journal} {\bibinfo
  {journal} {JHEP}\ }\textbf {\bibinfo {volume} {05}},\ \bibinfo {pages}
  {002}},\ \Eprint {https://arxiv.org/abs/1710.06859} {arXiv:1710.06859
  [hep-ph]} \BibitemShut {NoStop}%
\bibitem [{\citenamefont {Stevens}\ and\ \citenamefont
  {Williams}(2013)}]{Stevens:2013dya}%
  \BibitemOpen
  \bibfield  {author} {\bibinfo {author} {\bibfnamefont {J.}~\bibnamefont
  {Stevens}}\ and\ \bibinfo {author} {\bibfnamefont {M.}~\bibnamefont
  {Williams}},\ }\bibfield  {title} {\bibinfo {title} {{uBoost: A boosting
  method for producing uniform selection efficiencies from multivariate
  classifiers}},\ }\href {https://doi.org/10.1088/1748-0221/8/12/P12013}
  {\bibfield  {journal} {\bibinfo  {journal} {JINST}\ }\textbf {\bibinfo
  {volume} {8}},\ \bibinfo {pages} {P12013}},\ \Eprint
  {https://arxiv.org/abs/1305.7248} {arXiv:1305.7248 [nucl-ex]} \BibitemShut
  {NoStop}%
\bibitem [{\citenamefont {Shimmin}\ \emph {et~al.}(2017)\citenamefont
  {Shimmin}, \citenamefont {Sadowski}, \citenamefont {Baldi}, \citenamefont
  {Weik}, \citenamefont {Whiteson}, \citenamefont {Goul},\ and\ \citenamefont
  {Søgaard}}]{Shimmin:2017mfk}%
  \BibitemOpen
  \bibfield  {author} {\bibinfo {author} {\bibfnamefont {C.}~\bibnamefont
  {Shimmin}}, \bibinfo {author} {\bibfnamefont {P.}~\bibnamefont {Sadowski}},
  \bibinfo {author} {\bibfnamefont {P.}~\bibnamefont {Baldi}}, \bibinfo
  {author} {\bibfnamefont {E.}~\bibnamefont {Weik}}, \bibinfo {author}
  {\bibfnamefont {D.}~\bibnamefont {Whiteson}}, \bibinfo {author}
  {\bibfnamefont {E.}~\bibnamefont {Goul}},\ and\ \bibinfo {author}
  {\bibfnamefont {A.}~\bibnamefont {Søgaard}},\ }\bibfield  {title} {\bibinfo
  {title} {{Decorrelated Jet Substructure Tagging using Adversarial Neural
  Networks}}\ }\href {https://doi.org/10.1103/PhysRevD.96.074034}
  {10.1103/PhysRevD.96.074034} (\bibinfo {year} {2017}),\ \Eprint
  {https://arxiv.org/abs/1703.03507} {arXiv:1703.03507 [hep-ex]} \BibitemShut
  {NoStop}%
\bibitem [{\citenamefont {Bradshaw}\ \emph {et~al.}(2019)\citenamefont
  {Bradshaw}, \citenamefont {Mishra}, \citenamefont {Mitridate},\ and\
  \citenamefont {Ostdiek}}]{Bradshaw:2019ipy}%
  \BibitemOpen
  \bibfield  {author} {\bibinfo {author} {\bibfnamefont {L.}~\bibnamefont
  {Bradshaw}}, \bibinfo {author} {\bibfnamefont {R.~K.}\ \bibnamefont
  {Mishra}}, \bibinfo {author} {\bibfnamefont {A.}~\bibnamefont {Mitridate}},\
  and\ \bibinfo {author} {\bibfnamefont {B.}~\bibnamefont {Ostdiek}},\
  }\bibfield  {title} {\bibinfo {title} {{Mass Agnostic Jet Taggers}}\ }\href
  {https://doi.org/10.21468/SciPostPhys.8.1.011} {10.21468/SciPostPhys.8.1.011}
  (\bibinfo {year} {2019}),\ \Eprint {https://arxiv.org/abs/1908.08959}
  {arXiv:1908.08959 [hep-ph]} \BibitemShut {NoStop}%
\bibitem [{ATL(2018{\natexlab{b}})}]{ATL-PHYS-PUB-2018-014}%
  \BibitemOpen
  \bibfield  {title} {\bibinfo {title} {{Performance of mass-decorrelated jet
  substructure observables for hadronic two-body decay tagging in ATLAS}},\
  }\href {http://cds.cern.ch/record/2630973} {\bibfield  {journal} {\bibinfo
  {journal} {ATL-PHYS-PUB-2018-014}\ } (\bibinfo {year}
  {2018}{\natexlab{b}})}\BibitemShut {NoStop}%
\bibitem [{\citenamefont {Kasieczka}\ and\ \citenamefont
  {Shih}(2020)}]{DiscoFever}%
  \BibitemOpen
  \bibfield  {author} {\bibinfo {author} {\bibfnamefont {G.}~\bibnamefont
  {Kasieczka}}\ and\ \bibinfo {author} {\bibfnamefont {D.}~\bibnamefont
  {Shih}},\ }\bibfield  {title} {\bibinfo {title} {{DisCo Fever: Robust
  Networks Through Distance Correlation}}\ }\href
  {https://doi.org/10.1103/PhysRevLett.125.122001}
  {10.1103/PhysRevLett.125.122001} (\bibinfo {year} {2020}),\ \Eprint
  {https://arxiv.org/abs/2001.05310} {arXiv:2001.05310 [hep-ph]} \BibitemShut
  {NoStop}%
\bibitem [{\citenamefont {Wunsch}\ \emph {et~al.}(2019)\citenamefont {Wunsch},
  \citenamefont {J\'{o}rger}, \citenamefont {Wolf},\ and\ \citenamefont
  {Quast}}]{Wunsch:2019qbo}%
  \BibitemOpen
  \bibfield  {author} {\bibinfo {author} {\bibfnamefont {S.}~\bibnamefont
  {Wunsch}}, \bibinfo {author} {\bibfnamefont {S.}~\bibnamefont {J\'{o}rger}},
  \bibinfo {author} {\bibfnamefont {R.}~\bibnamefont {Wolf}},\ and\ \bibinfo
  {author} {\bibfnamefont {G.}~\bibnamefont {Quast}},\ }\bibfield  {title}
  {\bibinfo {title} {{Reducing the dependence of the neural network function to
  systematic uncertainties in the input space}}\ }\href
  {https://doi.org/10.1007/s41781-020-00037-9} {10.1007/s41781-020-00037-9}
  (\bibinfo {year} {2019}),\ \Eprint {https://arxiv.org/abs/1907.11674}
  {arXiv:1907.11674 [physics.data-an]} \BibitemShut {NoStop}%
\bibitem [{\citenamefont {Rogozhnikov}\ \emph {et~al.}(2015)\citenamefont
  {Rogozhnikov}, \citenamefont {Bukva}, \citenamefont {Gligorov}, \citenamefont
  {Ustyuzhanin},\ and\ \citenamefont {Williams}}]{Rogozhnikov:2014zea}%
  \BibitemOpen
  \bibfield  {author} {\bibinfo {author} {\bibfnamefont {A.}~\bibnamefont
  {Rogozhnikov}}, \bibinfo {author} {\bibfnamefont {A.}~\bibnamefont {Bukva}},
  \bibinfo {author} {\bibfnamefont {V.~V.}\ \bibnamefont {Gligorov}}, \bibinfo
  {author} {\bibfnamefont {A.}~\bibnamefont {Ustyuzhanin}},\ and\ \bibinfo
  {author} {\bibfnamefont {M.}~\bibnamefont {Williams}},\ }\bibfield  {title}
  {\bibinfo {title} {{New approaches for boosting to uniformity}},\ }\href
  {https://doi.org/10.1088/1748-0221/10/03/T03002} {\bibfield  {journal}
  {\bibinfo  {journal} {JINST}\ }\textbf {\bibinfo {volume} {10}}\bibfield
  {number} {\bibinfo  {number} { (03)},\ \bibinfo {pages} {T03002}},\ }\Eprint
  {https://arxiv.org/abs/1410.4140} {arXiv:1410.4140 [hep-ex]} \BibitemShut
  {NoStop}%
\bibitem [{\citenamefont {{CMS
  Collaboration}}(2020{\natexlab{b}})}]{10.1088/2632-2153/ab9023}%
  \BibitemOpen
  \bibfield  {author} {\bibinfo {author} {\bibnamefont {{CMS Collaboration}}},\
  }\bibfield  {title} {\bibinfo {title} {{A deep neural network to search for
  new long-lived particles decaying to jets}},\ }\bibfield  {journal} {\bibinfo
   {journal} {Machine Learning: Science and Technology}\ }\href
  {https://doi.org/10.1088/2632-2153/ab9023} {10.1088/2632-2153/ab9023}
  (\bibinfo {year} {2020}{\natexlab{b}}),\ \Eprint
  {https://arxiv.org/abs/1912.12238} {1912.12238} \BibitemShut {NoStop}%
\bibitem [{\citenamefont {Kitouni}\ \emph {et~al.}(2020)\citenamefont
  {Kitouni}, \citenamefont {Nachman}, \citenamefont {Weisser},\ and\
  \citenamefont {Williams}}]{Kitouni:2020xgb}%
  \BibitemOpen
  \bibfield  {author} {\bibinfo {author} {\bibfnamefont {O.}~\bibnamefont
  {Kitouni}}, \bibinfo {author} {\bibfnamefont {B.}~\bibnamefont {Nachman}},
  \bibinfo {author} {\bibfnamefont {C.}~\bibnamefont {Weisser}},\ and\ \bibinfo
  {author} {\bibfnamefont {M.}~\bibnamefont {Williams}},\ }\bibfield  {title}
  {\bibinfo {title} {{Enhancing searches for resonances with machine learning
  and moment decomposition}},\ }\href@noop {} {\  (\bibinfo {year} {2020})},\
  \Eprint {https://arxiv.org/abs/2010.09745} {arXiv:2010.09745 [hep-ph]}
  \BibitemShut {NoStop}%
\bibitem [{\citenamefont {Estrade}\ \emph {et~al.}(2019)\citenamefont
  {Estrade}, \citenamefont {Germain}, \citenamefont {Guyon},\ and\
  \citenamefont {Rousseau}}]{Estrade:2019gzk}%
  \BibitemOpen
  \bibfield  {author} {\bibinfo {author} {\bibfnamefont {V.}~\bibnamefont
  {Estrade}}, \bibinfo {author} {\bibfnamefont {C.}~\bibnamefont {Germain}},
  \bibinfo {author} {\bibfnamefont {I.}~\bibnamefont {Guyon}},\ and\ \bibinfo
  {author} {\bibfnamefont {D.}~\bibnamefont {Rousseau}},\ }\bibfield  {title}
  {\bibinfo {title} {{Systematic aware learning - A case study in High Energy
  Physics}},\ }\href {https://doi.org/10.1051/epjconf/201921406024} {\bibfield
  {journal} {\bibinfo  {journal} {EPJ Web Conf.}\ }\textbf {\bibinfo {volume}
  {214}},\ \bibinfo {pages} {06024} (\bibinfo {year} {2019})}\BibitemShut
  {NoStop}%
\bibitem [{\citenamefont {Aguilar-Saavedra}\ \emph {et~al.}(2017)\citenamefont
  {Aguilar-Saavedra}, \citenamefont {Collins},\ and\ \citenamefont
  {Mishra}}]{Aguilar-Saavedra:2017rzt}%
  \BibitemOpen
  \bibfield  {author} {\bibinfo {author} {\bibfnamefont {J.~A.}\ \bibnamefont
  {Aguilar-Saavedra}}, \bibinfo {author} {\bibfnamefont {J.~H.}\ \bibnamefont
  {Collins}},\ and\ \bibinfo {author} {\bibfnamefont {R.~K.}\ \bibnamefont
  {Mishra}},\ }\bibfield  {title} {\bibinfo {title} {{A generic anti-QCD jet
  tagger}},\ }\href {https://doi.org/10.1007/JHEP11(2017)163} {\bibfield
  {journal} {\bibinfo  {journal} {JHEP}\ }\textbf {\bibinfo {volume} {11}},\
  \bibinfo {pages} {163}},\ \Eprint {https://arxiv.org/abs/1709.01087}
  {arXiv:1709.01087 [hep-ph]} \BibitemShut {NoStop}%
\bibitem [{\citenamefont {Aguilar-Saavedra}\ \emph {et~al.}(2020)\citenamefont
  {Aguilar-Saavedra}, \citenamefont {Joaquim},\ and\ \citenamefont
  {Seabra}}]{aguilarsaavedra2020mass}%
  \BibitemOpen
  \bibfield  {author} {\bibinfo {author} {\bibfnamefont {J.~A.}\ \bibnamefont
  {Aguilar-Saavedra}}, \bibinfo {author} {\bibfnamefont {F.~R.}\ \bibnamefont
  {Joaquim}},\ and\ \bibinfo {author} {\bibfnamefont {J.~F.}\ \bibnamefont
  {Seabra}},\ }\bibfield  {title} {\bibinfo {title} {{Mass Unspecific
  Supervised Tagging (MUST) for boosted jets}}\ }\href
  {https://doi.org/10.1007/JHEP03(2021)012} {10.1007/JHEP03(2021)012} (\bibinfo
  {year} {2020}),\ \Eprint {https://arxiv.org/abs/2008.12792} {arXiv:2008.12792
  [hep-ph]} \BibitemShut {NoStop}%
\bibitem [{\citenamefont {Ghosh}\ and\ \citenamefont
  {Nachman}(2021)}]{Ghosh:2021hrh}%
  \BibitemOpen
  \bibfield  {author} {\bibinfo {author} {\bibfnamefont {A.}~\bibnamefont
  {Ghosh}}\ and\ \bibinfo {author} {\bibfnamefont {B.}~\bibnamefont
  {Nachman}},\ }\bibfield  {title} {\bibinfo {title} {{A Cautionary Tale of
  Decorrelating Theory Uncertainties}},\ }\href@noop {} {\  (\bibinfo {year}
  {2021})},\ \Eprint {https://arxiv.org/abs/2109.08159} {arXiv:2109.08159
  [hep-ph]} \BibitemShut {NoStop}%
\bibitem [{\citenamefont {Chouldechova}\ and\ \citenamefont
  {Roth}(2018)}]{chouldechova2018frontiers}%
  \BibitemOpen
  \bibfield  {author} {\bibinfo {author} {\bibfnamefont {A.}~\bibnamefont
  {Chouldechova}}\ and\ \bibinfo {author} {\bibfnamefont {A.}~\bibnamefont
  {Roth}},\ }\href@noop {} {\bibinfo {title} {The frontiers of fairness in
  machine learning}} (\bibinfo {year} {2018}),\ \Eprint
  {https://arxiv.org/abs/1810.08810} {arXiv:1810.08810 [cs.LG]} \BibitemShut
  {NoStop}%
\bibitem [{\citenamefont {Mehrabi}\ \emph {et~al.}(2019)\citenamefont
  {Mehrabi}, \citenamefont {Morstatter}, \citenamefont {Saxena}, \citenamefont
  {Lerman},\ and\ \citenamefont {Galstyan}}]{mehrabi2019survey}%
  \BibitemOpen
  \bibfield  {author} {\bibinfo {author} {\bibfnamefont {N.}~\bibnamefont
  {Mehrabi}}, \bibinfo {author} {\bibfnamefont {F.}~\bibnamefont {Morstatter}},
  \bibinfo {author} {\bibfnamefont {N.}~\bibnamefont {Saxena}}, \bibinfo
  {author} {\bibfnamefont {K.}~\bibnamefont {Lerman}},\ and\ \bibinfo {author}
  {\bibfnamefont {A.}~\bibnamefont {Galstyan}},\ }\href@noop {} {\bibinfo
  {title} {A survey on bias and fairness in machine learning}} (\bibinfo {year}
  {2019}),\ \Eprint {https://arxiv.org/abs/1908.09635} {arXiv:1908.09635
  [cs.LG]} \BibitemShut {NoStop}%
\bibitem [{\citenamefont {Frate}\ \emph {et~al.}(2017)\citenamefont {Frate},
  \citenamefont {Cranmer}, \citenamefont {Kalia}, \citenamefont
  {Vandenberg-Rodes},\ and\ \citenamefont {Whiteson}}]{Frate:2017mai}%
  \BibitemOpen
  \bibfield  {author} {\bibinfo {author} {\bibfnamefont {M.}~\bibnamefont
  {Frate}}, \bibinfo {author} {\bibfnamefont {K.}~\bibnamefont {Cranmer}},
  \bibinfo {author} {\bibfnamefont {S.}~\bibnamefont {Kalia}}, \bibinfo
  {author} {\bibfnamefont {A.}~\bibnamefont {Vandenberg-Rodes}},\ and\ \bibinfo
  {author} {\bibfnamefont {D.}~\bibnamefont {Whiteson}},\ }\bibfield  {title}
  {\bibinfo {title} {{Modeling Smooth Backgrounds and Generic Localized Signals
  with Gaussian Processes}},\ }\href@noop {} {\  (\bibinfo {year} {2017})},\
  \Eprint {https://arxiv.org/abs/1709.05681} {arXiv:1709.05681
  [physics.data-an]} \BibitemShut {NoStop}%
\bibitem [{\citenamefont {Di~Sipio}\ \emph {et~al.}(2019)\citenamefont
  {Di~Sipio}, \citenamefont {Faucci~Giannelli}, \citenamefont
  {Ketabchi~Haghighat},\ and\ \citenamefont {Palazzo}}]{DiSipio:2019imz}%
  \BibitemOpen
  \bibfield  {author} {\bibinfo {author} {\bibfnamefont {R.}~\bibnamefont
  {Di~Sipio}}, \bibinfo {author} {\bibfnamefont {M.}~\bibnamefont
  {Faucci~Giannelli}}, \bibinfo {author} {\bibfnamefont {S.}~\bibnamefont
  {Ketabchi~Haghighat}},\ and\ \bibinfo {author} {\bibfnamefont
  {S.}~\bibnamefont {Palazzo}},\ }\bibfield  {title} {\bibinfo {title}
  {{DijetGAN: A Generative-Adversarial Network Approach for the Simulation of
  QCD Dijet Events at the LHC}}\ }\href
  {https://doi.org/10.1007/JHEP08(2019)110} {10.1007/JHEP08(2019)110} (\bibinfo
  {year} {2019}),\ \Eprint {https://arxiv.org/abs/1903.02433} {arXiv:1903.02433
  [hep-ex]} \BibitemShut {NoStop}%
\bibitem [{\citenamefont {Chisholm}\ \emph {et~al.}(2021)\citenamefont
  {Chisholm}, \citenamefont {Neep}, \citenamefont {Nikolopoulos}, \citenamefont
  {Owen}, \citenamefont {Reynolds},\ and\ \citenamefont
  {Silva}}]{Chisholm:2021pdn}%
  \BibitemOpen
  \bibfield  {author} {\bibinfo {author} {\bibfnamefont {A.}~\bibnamefont
  {Chisholm}}, \bibinfo {author} {\bibfnamefont {T.}~\bibnamefont {Neep}},
  \bibinfo {author} {\bibfnamefont {K.}~\bibnamefont {Nikolopoulos}}, \bibinfo
  {author} {\bibfnamefont {R.}~\bibnamefont {Owen}}, \bibinfo {author}
  {\bibfnamefont {E.}~\bibnamefont {Reynolds}},\ and\ \bibinfo {author}
  {\bibfnamefont {J.}~\bibnamefont {Silva}},\ }\bibfield  {title} {\bibinfo
  {title} {{Non-Parametric Data-Driven Background Modelling using Conditional
  Probabilities}},\ }\href@noop {} {\  (\bibinfo {year} {2021})},\ \Eprint
  {https://arxiv.org/abs/2112.00650} {arXiv:2112.00650 [hep-ex]} \BibitemShut
  {NoStop}%
\bibitem [{\citenamefont {Neyman}\ and\ \citenamefont
  {Pearson}(1933)}]{neyman1933ix}%
  \BibitemOpen
  \bibfield  {author} {\bibinfo {author} {\bibfnamefont {J.}~\bibnamefont
  {Neyman}}\ and\ \bibinfo {author} {\bibfnamefont {E.~S.}\ \bibnamefont
  {Pearson}},\ }\bibfield  {title} {\bibinfo {title} {On the problem of the
  most efficient tests of statistical hypotheses},\ }\href@noop {} {\bibfield
  {journal} {\bibinfo  {journal} {Phil. Trans. R. Soc. Lond. A}\ }\textbf
  {\bibinfo {volume} {231}},\ \bibinfo {pages} {289} (\bibinfo {year}
  {1933})}\BibitemShut {NoStop}%
\bibitem [{\citenamefont {Kasieczka}\ \emph {et~al.}(2021)\citenamefont
  {Kasieczka} \emph {et~al.}}]{Kasieczka:2021xcg}%
  \BibitemOpen
  \bibfield  {author} {\bibinfo {author} {\bibfnamefont {G.}~\bibnamefont
  {Kasieczka}} \emph {et~al.},\ }\bibfield  {title} {\bibinfo {title} {{The LHC
  Olympics 2020: A Community Challenge for Anomaly Detection in High Energy
  Physics}},\ }\href@noop {} {\  (\bibinfo {year} {2021})},\ \Eprint
  {https://arxiv.org/abs/2101.08320} {arXiv:2101.08320 [hep-ph]} \BibitemShut
  {NoStop}%
\bibitem [{\citenamefont {Aarrestad}\ \emph
  {et~al.}(2021{\natexlab{a}})\citenamefont {Aarrestad} \emph
  {et~al.}}]{Aarrestad:2021oeb}%
  \BibitemOpen
  \bibfield  {author} {\bibinfo {author} {\bibfnamefont {T.}~\bibnamefont
  {Aarrestad}} \emph {et~al.},\ }\bibfield  {title} {\bibinfo {title} {{The
  Dark Machines Anomaly Score Challenge: Benchmark Data and Model Independent
  Event Classification for the Large Hadron Collider}},\ }\href@noop {} {\
  (\bibinfo {year} {2021}{\natexlab{a}})},\ \Eprint
  {https://arxiv.org/abs/2105.14027} {arXiv:2105.14027 [hep-ph]} \BibitemShut
  {NoStop}%
\bibitem [{\citenamefont {Hinton}\ and\ \citenamefont
  {Salakhutdinov}(2006)}]{hinton2006reducing}%
  \BibitemOpen
  \bibfield  {author} {\bibinfo {author} {\bibfnamefont {G.~E.}\ \bibnamefont
  {Hinton}}\ and\ \bibinfo {author} {\bibfnamefont {R.~R.}\ \bibnamefont
  {Salakhutdinov}},\ }\bibfield  {title} {\bibinfo {title} {Reducing the
  dimensionality of data with neural networks},\ }\href@noop {} {\bibfield
  {journal} {\bibinfo  {journal} {science}\ }\textbf {\bibinfo {volume}
  {313}},\ \bibinfo {pages} {504} (\bibinfo {year} {2006})}\BibitemShut
  {NoStop}%
\bibitem [{\citenamefont {Finke}\ \emph {et~al.}(2021)\citenamefont {Finke},
  \citenamefont {Kr\"amer}, \citenamefont {Morandini}, \citenamefont {M\"uck},\
  and\ \citenamefont {Oleksiyuk}}]{Finke:2021sdf}%
  \BibitemOpen
  \bibfield  {author} {\bibinfo {author} {\bibfnamefont {T.}~\bibnamefont
  {Finke}}, \bibinfo {author} {\bibfnamefont {M.}~\bibnamefont {Kr\"amer}},
  \bibinfo {author} {\bibfnamefont {A.}~\bibnamefont {Morandini}}, \bibinfo
  {author} {\bibfnamefont {A.}~\bibnamefont {M\"uck}},\ and\ \bibinfo {author}
  {\bibfnamefont {I.}~\bibnamefont {Oleksiyuk}},\ }\bibfield  {title} {\bibinfo
  {title} {{Autoencoders for unsupervised anomaly detection in high energy
  physics}},\ }\href@noop {} {\  (\bibinfo {year} {2021})},\ \Eprint
  {https://arxiv.org/abs/2104.09051} {arXiv:2104.09051 [physics.data-an]}
  \BibitemShut {NoStop}%
\bibitem [{\citenamefont {Dillon}\ \emph {et~al.}(2021)\citenamefont {Dillon},
  \citenamefont {Plehn}, \citenamefont {Sauer},\ and\ \citenamefont
  {Sorrenson}}]{Dillon:2021nxw}%
  \BibitemOpen
  \bibfield  {author} {\bibinfo {author} {\bibfnamefont {B.~M.}\ \bibnamefont
  {Dillon}}, \bibinfo {author} {\bibfnamefont {T.}~\bibnamefont {Plehn}},
  \bibinfo {author} {\bibfnamefont {C.}~\bibnamefont {Sauer}},\ and\ \bibinfo
  {author} {\bibfnamefont {P.}~\bibnamefont {Sorrenson}},\ }\bibfield  {title}
  {\bibinfo {title} {{Better Latent Spaces for Better Autoencoders}},\
  }\href@noop {} {\  (\bibinfo {year} {2021})},\ \Eprint
  {https://arxiv.org/abs/2104.08291} {arXiv:2104.08291 [hep-ph]} \BibitemShut
  {NoStop}%
\bibitem [{\citenamefont {Fraser}\ \emph {et~al.}(2021)\citenamefont {Fraser},
  \citenamefont {Homiller}, \citenamefont {Mishra}, \citenamefont {Ostdiek},\
  and\ \citenamefont {Schwartz}}]{Fraser:2021lxm}%
  \BibitemOpen
  \bibfield  {author} {\bibinfo {author} {\bibfnamefont {K.}~\bibnamefont
  {Fraser}}, \bibinfo {author} {\bibfnamefont {S.}~\bibnamefont {Homiller}},
  \bibinfo {author} {\bibfnamefont {R.~K.}\ \bibnamefont {Mishra}}, \bibinfo
  {author} {\bibfnamefont {B.}~\bibnamefont {Ostdiek}},\ and\ \bibinfo {author}
  {\bibfnamefont {M.~D.}\ \bibnamefont {Schwartz}},\ }\bibfield  {title}
  {\bibinfo {title} {{Challenges for Unsupervised Anomaly Detection in Particle
  Physics}},\ }\href@noop {} {\  (\bibinfo {year} {2021})},\ \Eprint
  {https://arxiv.org/abs/2110.06948} {arXiv:2110.06948 [cs.LG]} \BibitemShut
  {NoStop}%
\bibitem [{\citenamefont {Cerri}\ \emph {et~al.}(2019)\citenamefont {Cerri},
  \citenamefont {Nguyen}, \citenamefont {Pierini}, \citenamefont {Spiropulu},\
  and\ \citenamefont {Vlimant}}]{Cerri:2018anq}%
  \BibitemOpen
  \bibfield  {author} {\bibinfo {author} {\bibfnamefont {O.}~\bibnamefont
  {Cerri}}, \bibinfo {author} {\bibfnamefont {T.~Q.}\ \bibnamefont {Nguyen}},
  \bibinfo {author} {\bibfnamefont {M.}~\bibnamefont {Pierini}}, \bibinfo
  {author} {\bibfnamefont {M.}~\bibnamefont {Spiropulu}},\ and\ \bibinfo
  {author} {\bibfnamefont {J.-R.}\ \bibnamefont {Vlimant}},\ }\bibfield
  {title} {\bibinfo {title} {{Variational Autoencoders for New Physics Mining
  at the Large Hadron Collider}},\ }\href
  {https://doi.org/10.1007/JHEP05(2019)036} {\bibfield  {journal} {\bibinfo
  {journal} {JHEP}\ }\textbf {\bibinfo {volume} {05}},\ \bibinfo {pages}
  {036}},\ \Eprint {https://arxiv.org/abs/1811.10276} {arXiv:1811.10276
  [hep-ex]} \BibitemShut {NoStop}%
\bibitem [{\citenamefont {Govorkova}\ \emph {et~al.}(2021)\citenamefont
  {Govorkova} \emph {et~al.}}]{Govorkova:2021utb}%
  \BibitemOpen
  \bibfield  {author} {\bibinfo {author} {\bibfnamefont {E.}~\bibnamefont
  {Govorkova}} \emph {et~al.},\ }\bibfield  {title} {\bibinfo {title}
  {{Autoencoders on FPGAs for real-time, unsupervised new physics detection at
  40 MHz at the Large Hadron Collider}},\ }\href@noop {} {\  (\bibinfo {year}
  {2021})},\ \Eprint {https://arxiv.org/abs/2108.03986} {arXiv:2108.03986
  [physics.ins-det]} \BibitemShut {NoStop}%
\bibitem [{\citenamefont {Romao}\ \emph {et~al.}(2020)\citenamefont {Romao},
  \citenamefont {Castro},\ and\ \citenamefont {Pedro}}]{1800445}%
  \BibitemOpen
  \bibfield  {author} {\bibinfo {author} {\bibfnamefont {M.~C.}\ \bibnamefont
  {Romao}}, \bibinfo {author} {\bibfnamefont {N.}~\bibnamefont {Castro}},\ and\
  \bibinfo {author} {\bibfnamefont {R.}~\bibnamefont {Pedro}},\ }\bibfield
  {title} {\bibinfo {title} {{Finding New Physics without learning about it:
  Anomaly Detection as a tool for Searches at Colliders}}\ }\href
  {https://doi.org/10.1140/epjc/s10052-020-08807-w}
  {10.1140/epjc/s10052-020-08807-w} (\bibinfo {year} {2020}),\ \Eprint
  {https://arxiv.org/abs/2006.05432} {arXiv:2006.05432 [hep-ph]} \BibitemShut
  {NoStop}%
\bibitem [{\citenamefont {Dillon}\ \emph {et~al.}(2019)\citenamefont {Dillon},
  \citenamefont {Faroughy},\ and\ \citenamefont {Kamenik}}]{Dillon:2019cqt}%
  \BibitemOpen
  \bibfield  {author} {\bibinfo {author} {\bibfnamefont {B.~M.}\ \bibnamefont
  {Dillon}}, \bibinfo {author} {\bibfnamefont {D.~A.}\ \bibnamefont
  {Faroughy}},\ and\ \bibinfo {author} {\bibfnamefont {J.~F.}\ \bibnamefont
  {Kamenik}},\ }\bibfield  {title} {\bibinfo {title} {{Uncovering latent jet
  substructure}},\ }\href {https://doi.org/10.1103/PhysRevD.100.056002}
  {\bibfield  {journal} {\bibinfo  {journal} {Phys. Rev.}\ }\textbf {\bibinfo
  {volume} {D100}},\ \bibinfo {pages} {056002} (\bibinfo {year} {2019})},\
  \Eprint {https://arxiv.org/abs/1904.04200} {arXiv:1904.04200 [hep-ph]}
  \BibitemShut {NoStop}%
\bibitem [{\citenamefont {Caron}\ \emph {et~al.}(2021)\citenamefont {Caron},
  \citenamefont {Hendriks},\ and\ \citenamefont {Verheyen}}]{Caron:2021wmq}%
  \BibitemOpen
  \bibfield  {author} {\bibinfo {author} {\bibfnamefont {S.}~\bibnamefont
  {Caron}}, \bibinfo {author} {\bibfnamefont {L.}~\bibnamefont {Hendriks}},\
  and\ \bibinfo {author} {\bibfnamefont {R.}~\bibnamefont {Verheyen}},\
  }\bibfield  {title} {\bibinfo {title} {{Rare and Different: Anomaly Scores
  from a combination of likelihood and out-of-distribution models to detect new
  physics at the LHC}},\ }\href@noop {} {\  (\bibinfo {year} {2021})},\ \Eprint
  {https://arxiv.org/abs/2106.10164} {arXiv:2106.10164 [hep-ph]} \BibitemShut
  {NoStop}%
\bibitem [{\citenamefont {Mikuni}\ and\ \citenamefont
  {Canelli}(2020)}]{Mikuni:2020qds}%
  \BibitemOpen
  \bibfield  {author} {\bibinfo {author} {\bibfnamefont {V.}~\bibnamefont
  {Mikuni}}\ and\ \bibinfo {author} {\bibfnamefont {F.}~\bibnamefont
  {Canelli}},\ }\bibfield  {title} {\bibinfo {title} {{Unsupervised clustering
  for collider physics}},\ }\href@noop {} {\  (\bibinfo {year} {2020})},\
  \Eprint {https://arxiv.org/abs/2010.07106} {arXiv:2010.07106
  [physics.data-an]} \BibitemShut {NoStop}%
\bibitem [{\citenamefont {Knapp}\ \emph {et~al.}(2020)\citenamefont {Knapp},
  \citenamefont {Dissertori}, \citenamefont {Cerri}, \citenamefont {Nguyen},
  \citenamefont {Vlimant},\ and\ \citenamefont
  {Pierini}}]{knapp2020adversarially}%
  \BibitemOpen
  \bibfield  {author} {\bibinfo {author} {\bibfnamefont {O.}~\bibnamefont
  {Knapp}}, \bibinfo {author} {\bibfnamefont {G.}~\bibnamefont {Dissertori}},
  \bibinfo {author} {\bibfnamefont {O.}~\bibnamefont {Cerri}}, \bibinfo
  {author} {\bibfnamefont {T.~Q.}\ \bibnamefont {Nguyen}}, \bibinfo {author}
  {\bibfnamefont {J.-R.}\ \bibnamefont {Vlimant}},\ and\ \bibinfo {author}
  {\bibfnamefont {M.}~\bibnamefont {Pierini}},\ }\bibfield  {title} {\bibinfo
  {title} {{Adversarially Learned Anomaly Detection on CMS Open Data:
  re-discovering the top quark}}\ }\href
  {https://doi.org/10.1140/epjp/s13360-021-01109-4}
  {10.1140/epjp/s13360-021-01109-4} (\bibinfo {year} {2020}),\ \Eprint
  {https://arxiv.org/abs/2005.01598} {arXiv:2005.01598 [hep-ex]} \BibitemShut
  {NoStop}%
\bibitem [{\citenamefont {Amram}\ and\ \citenamefont
  {Suarez}(2020)}]{Amram:2020ykb}%
  \BibitemOpen
  \bibfield  {author} {\bibinfo {author} {\bibfnamefont {O.}~\bibnamefont
  {Amram}}\ and\ \bibinfo {author} {\bibfnamefont {C.~M.}\ \bibnamefont
  {Suarez}},\ }\bibfield  {title} {\bibinfo {title} {{Tag N' Train: A Technique
  to Train Improved Classifiers on Unlabeled Data}}\ }\href
  {https://doi.org/10.1007/JHEP01(2021)153} {10.1007/JHEP01(2021)153} (\bibinfo
  {year} {2020}),\ \Eprint {https://arxiv.org/abs/2002.12376} {arXiv:2002.12376
  [hep-ph]} \BibitemShut {NoStop}%
\bibitem [{\citenamefont {Collins}\ \emph {et~al.}(2021)\citenamefont
  {Collins}, \citenamefont {Mart\'\i{}n-Ramiro}, \citenamefont {Nachman},\ and\
  \citenamefont {Shih}}]{Collins:2021nxn}%
  \BibitemOpen
  \bibfield  {author} {\bibinfo {author} {\bibfnamefont {J.~H.}\ \bibnamefont
  {Collins}}, \bibinfo {author} {\bibfnamefont {P.}~\bibnamefont
  {Mart\'\i{}n-Ramiro}}, \bibinfo {author} {\bibfnamefont {B.}~\bibnamefont
  {Nachman}},\ and\ \bibinfo {author} {\bibfnamefont {D.}~\bibnamefont
  {Shih}},\ }\bibfield  {title} {\bibinfo {title} {{Comparing weak- and
  unsupervised methods for resonant anomaly detection}},\ }\href
  {https://doi.org/10.1140/epjc/s10052-021-09389-x} {\bibfield  {journal}
  {\bibinfo  {journal} {Eur. Phys. J. C}\ }\textbf {\bibinfo {volume} {81}},\
  \bibinfo {pages} {617} (\bibinfo {year} {2021})},\ \Eprint
  {https://arxiv.org/abs/2104.02092} {arXiv:2104.02092 [hep-ph]} \BibitemShut
  {NoStop}%
\bibitem [{\citenamefont {Collins}\ \emph {et~al.}(2018)\citenamefont
  {Collins}, \citenamefont {Howe},\ and\ \citenamefont
  {Nachman}}]{Collins:2018epr}%
  \BibitemOpen
  \bibfield  {author} {\bibinfo {author} {\bibfnamefont {J.~H.}\ \bibnamefont
  {Collins}}, \bibinfo {author} {\bibfnamefont {K.}~\bibnamefont {Howe}},\ and\
  \bibinfo {author} {\bibfnamefont {B.}~\bibnamefont {Nachman}},\ }\bibfield
  {title} {\bibinfo {title} {{Anomaly Detection for Resonant New Physics with
  Machine Learning}},\ }\href {https://doi.org/10.1103/PhysRevLett.121.241803}
  {\bibfield  {journal} {\bibinfo  {journal} {Phys. Rev. Lett.}\ }\textbf
  {\bibinfo {volume} {121}},\ \bibinfo {pages} {241803} (\bibinfo {year}
  {2018})},\ \Eprint {https://arxiv.org/abs/1805.02664} {arXiv:1805.02664
  [hep-ph]} \BibitemShut {NoStop}%
\bibitem [{\citenamefont {Collins}\ \emph {et~al.}(2019)\citenamefont
  {Collins}, \citenamefont {Howe},\ and\ \citenamefont
  {Nachman}}]{Collins:2019jip}%
  \BibitemOpen
  \bibfield  {author} {\bibinfo {author} {\bibfnamefont {J.~H.}\ \bibnamefont
  {Collins}}, \bibinfo {author} {\bibfnamefont {K.}~\bibnamefont {Howe}},\ and\
  \bibinfo {author} {\bibfnamefont {B.}~\bibnamefont {Nachman}},\ }\bibfield
  {title} {\bibinfo {title} {{Extending the search for new resonances with
  machine learning}},\ }\href {https://doi.org/10.1103/PhysRevD.99.014038}
  {\bibfield  {journal} {\bibinfo  {journal} {Phys. Rev.}\ }\textbf {\bibinfo
  {volume} {D99}},\ \bibinfo {pages} {014038} (\bibinfo {year} {2019})},\
  \Eprint {https://arxiv.org/abs/1902.02634} {arXiv:1902.02634 [hep-ph]}
  \BibitemShut {NoStop}%
\bibitem [{\citenamefont {D'Agnolo}\ and\ \citenamefont
  {Wulzer}(2019)}]{DAgnolo:2018cun}%
  \BibitemOpen
  \bibfield  {author} {\bibinfo {author} {\bibfnamefont {R.~T.}\ \bibnamefont
  {D'Agnolo}}\ and\ \bibinfo {author} {\bibfnamefont {A.}~\bibnamefont
  {Wulzer}},\ }\bibfield  {title} {\bibinfo {title} {{Learning New Physics from
  a Machine}},\ }\href {https://doi.org/10.1103/PhysRevD.99.015014} {\bibfield
  {journal} {\bibinfo  {journal} {Phys. Rev.}\ }\textbf {\bibinfo {volume}
  {D99}},\ \bibinfo {pages} {015014} (\bibinfo {year} {2019})},\ \Eprint
  {https://arxiv.org/abs/1806.02350} {arXiv:1806.02350 [hep-ph]} \BibitemShut
  {NoStop}%
\bibitem [{\citenamefont {D'Agnolo}\ \emph {et~al.}(2019)\citenamefont
  {D'Agnolo}, \citenamefont {Grosso}, \citenamefont {Pierini}, \citenamefont
  {Wulzer},\ and\ \citenamefont {Zanetti}}]{DAgnolo:2019vbw}%
  \BibitemOpen
  \bibfield  {author} {\bibinfo {author} {\bibfnamefont {R.~T.}\ \bibnamefont
  {D'Agnolo}}, \bibinfo {author} {\bibfnamefont {G.}~\bibnamefont {Grosso}},
  \bibinfo {author} {\bibfnamefont {M.}~\bibnamefont {Pierini}}, \bibinfo
  {author} {\bibfnamefont {A.}~\bibnamefont {Wulzer}},\ and\ \bibinfo {author}
  {\bibfnamefont {M.}~\bibnamefont {Zanetti}},\ }\bibfield  {title} {\bibinfo
  {title} {{Learning Multivariate New Physics}}\ }\href
  {https://doi.org/10.1140/epjc/s10052-021-08853-y}
  {10.1140/epjc/s10052-021-08853-y} (\bibinfo {year} {2019}),\ \Eprint
  {https://arxiv.org/abs/1912.12155} {arXiv:1912.12155 [hep-ph]} \BibitemShut
  {NoStop}%
\bibitem [{\citenamefont {d'Agnolo}\ \emph {et~al.}(2021)\citenamefont
  {d'Agnolo}, \citenamefont {Grosso}, \citenamefont {Pierini}, \citenamefont
  {Wulzer},\ and\ \citenamefont {Zanetti}}]{dAgnolo:2021aun}%
  \BibitemOpen
  \bibfield  {author} {\bibinfo {author} {\bibfnamefont {R.~T.}\ \bibnamefont
  {d'Agnolo}}, \bibinfo {author} {\bibfnamefont {G.}~\bibnamefont {Grosso}},
  \bibinfo {author} {\bibfnamefont {M.}~\bibnamefont {Pierini}}, \bibinfo
  {author} {\bibfnamefont {A.}~\bibnamefont {Wulzer}},\ and\ \bibinfo {author}
  {\bibfnamefont {M.}~\bibnamefont {Zanetti}},\ }\bibfield  {title} {\bibinfo
  {title} {{Learning New Physics from an Imperfect Machine}},\ }\href@noop {}
  {\  (\bibinfo {year} {2021})},\ \Eprint {https://arxiv.org/abs/2111.13633}
  {arXiv:2111.13633 [hep-ph]} \BibitemShut {NoStop}%
\bibitem [{\citenamefont {Andreassen}\ \emph {et~al.}(2020)\citenamefont
  {Andreassen}, \citenamefont {Nachman},\ and\ \citenamefont
  {Shih}}]{Andreassen:2020nkr}%
  \BibitemOpen
  \bibfield  {author} {\bibinfo {author} {\bibfnamefont {A.}~\bibnamefont
  {Andreassen}}, \bibinfo {author} {\bibfnamefont {B.}~\bibnamefont
  {Nachman}},\ and\ \bibinfo {author} {\bibfnamefont {D.}~\bibnamefont
  {Shih}},\ }\bibfield  {title} {\bibinfo {title} {{Simulation Assisted
  Likelihood-free Anomaly Detection}},\ }\href
  {https://doi.org/10.1103/PhysRevD.101.095004} {\bibfield  {journal} {\bibinfo
   {journal} {Phys. Rev. D}\ }\textbf {\bibinfo {volume} {101}},\ \bibinfo
  {pages} {095004} (\bibinfo {year} {2020})},\ \Eprint
  {https://arxiv.org/abs/2001.05001} {arXiv:2001.05001 [hep-ph]} \BibitemShut
  {NoStop}%
\bibitem [{\citenamefont {Benkendorfer}\ \emph {et~al.}(2021)\citenamefont
  {Benkendorfer}, \citenamefont {Pottier},\ and\ \citenamefont
  {Nachman}}]{1815227}%
  \BibitemOpen
  \bibfield  {author} {\bibinfo {author} {\bibfnamefont {K.}~\bibnamefont
  {Benkendorfer}}, \bibinfo {author} {\bibfnamefont {L.~L.}\ \bibnamefont
  {Pottier}},\ and\ \bibinfo {author} {\bibfnamefont {B.}~\bibnamefont
  {Nachman}},\ }\bibfield  {title} {\bibinfo {title} {{Simulation-assisted
  decorrelation for resonant anomaly detection}},\ }\href
  {https://doi.org/10.1103/PhysRevD.104.035003} {\bibfield  {journal} {\bibinfo
   {journal} {Phys. Rev. D}\ }\textbf {\bibinfo {volume} {104}},\ \bibinfo
  {pages} {035003} (\bibinfo {year} {2021})},\ \Eprint
  {https://arxiv.org/abs/2009.02205} {arXiv:2009.02205 [hep-ph]} \BibitemShut
  {NoStop}%
\bibitem [{\citenamefont {Park}\ \emph {et~al.}(2020)\citenamefont {Park},
  \citenamefont {Rankin}, \citenamefont {Udrescu}, \citenamefont {Yunus},\ and\
  \citenamefont {Harris}}]{Park:2020pak}%
  \BibitemOpen
  \bibfield  {author} {\bibinfo {author} {\bibfnamefont {S.~E.}\ \bibnamefont
  {Park}}, \bibinfo {author} {\bibfnamefont {D.}~\bibnamefont {Rankin}},
  \bibinfo {author} {\bibfnamefont {S.-M.}\ \bibnamefont {Udrescu}}, \bibinfo
  {author} {\bibfnamefont {M.}~\bibnamefont {Yunus}},\ and\ \bibinfo {author}
  {\bibfnamefont {P.}~\bibnamefont {Harris}},\ }\bibfield  {title} {\bibinfo
  {title} {{Quasi Anomalous Knowledge: Searching for new physics with embedded
  knowledge}},\ }\href {https://doi.org/10.1007/JHEP06(2021)030} {\bibfield
  {journal} {\bibinfo  {journal} {JHEP}\ }\textbf {\bibinfo {volume} {21}},\
  \bibinfo {pages} {030}},\ \Eprint {https://arxiv.org/abs/2011.03550}
  {arXiv:2011.03550 [hep-ph]} \BibitemShut {NoStop}%
\bibitem [{\citenamefont {Khosa}\ and\ \citenamefont
  {Sanz}(2020)}]{Khosa:2020qrz}%
  \BibitemOpen
  \bibfield  {author} {\bibinfo {author} {\bibfnamefont {C.~K.}\ \bibnamefont
  {Khosa}}\ and\ \bibinfo {author} {\bibfnamefont {V.}~\bibnamefont {Sanz}},\
  }\bibfield  {title} {\bibinfo {title} {{Anomaly Awareness}},\ }\href@noop {}
  {\  (\bibinfo {year} {2020})},\ \Eprint {https://arxiv.org/abs/2007.14462}
  {arXiv:2007.14462 [cs.LG]} \BibitemShut {NoStop}%
\bibitem [{\citenamefont {Stein}\ \emph {et~al.}(2020)\citenamefont {Stein},
  \citenamefont {Seljak},\ and\ \citenamefont {Dai}}]{Stein:2020rou}%
  \BibitemOpen
  \bibfield  {author} {\bibinfo {author} {\bibfnamefont {G.}~\bibnamefont
  {Stein}}, \bibinfo {author} {\bibfnamefont {U.}~\bibnamefont {Seljak}},\ and\
  \bibinfo {author} {\bibfnamefont {B.}~\bibnamefont {Dai}},\ }\bibfield
  {title} {\bibinfo {title} {{Unsupervised in-distribution anomaly detection of
  new physics through conditional density estimation}},\ }\href@noop {} {\
  (\bibinfo {year} {2020})},\ \Eprint {https://arxiv.org/abs/2012.11638}
  {arXiv:2012.11638 [cs.LG]} \BibitemShut {NoStop}%
\bibitem [{\citenamefont {Hallin}\ \emph {et~al.}(2021)\citenamefont {Hallin},
  \citenamefont {Isaacson}, \citenamefont {Kasieczka}, \citenamefont {Krause},
  \citenamefont {Nachman}, \citenamefont {Quadfasel}, \citenamefont
  {Schlaffer}, \citenamefont {Shih},\ and\ \citenamefont
  {Sommerhalder}}]{Hallin:2021wme}%
  \BibitemOpen
  \bibfield  {author} {\bibinfo {author} {\bibfnamefont {A.}~\bibnamefont
  {Hallin}}, \bibinfo {author} {\bibfnamefont {J.}~\bibnamefont {Isaacson}},
  \bibinfo {author} {\bibfnamefont {G.}~\bibnamefont {Kasieczka}}, \bibinfo
  {author} {\bibfnamefont {C.}~\bibnamefont {Krause}}, \bibinfo {author}
  {\bibfnamefont {B.}~\bibnamefont {Nachman}}, \bibinfo {author} {\bibfnamefont
  {T.}~\bibnamefont {Quadfasel}}, \bibinfo {author} {\bibfnamefont
  {M.}~\bibnamefont {Schlaffer}}, \bibinfo {author} {\bibfnamefont
  {D.}~\bibnamefont {Shih}},\ and\ \bibinfo {author} {\bibfnamefont
  {M.}~\bibnamefont {Sommerhalder}},\ }\bibfield  {title} {\bibinfo {title}
  {{Classifying Anomalies THrough Outer Density Estimation (CATHODE)}},\
  }\href@noop {} {\  (\bibinfo {year} {2021})},\ \Eprint
  {https://arxiv.org/abs/2109.00546} {arXiv:2109.00546 [hep-ph]} \BibitemShut
  {NoStop}%
\bibitem [{\citenamefont {Sirunyan}\ \emph {et~al.}(2018)\citenamefont
  {Sirunyan} \emph {et~al.}}]{CMS:2017wtu}%
  \BibitemOpen
  \bibfield  {author} {\bibinfo {author} {\bibfnamefont {A.~M.}\ \bibnamefont
  {Sirunyan}} \emph {et~al.} (\bibinfo {collaboration} {CMS}),\ }\bibfield
  {title} {\bibinfo {title} {{Identification of heavy-flavour jets with the CMS
  detector in pp collisions at 13 TeV}},\ }\href
  {https://doi.org/10.1088/1748-0221/13/05/P05011} {\bibfield  {journal}
  {\bibinfo  {journal} {JINST}\ }\textbf {\bibinfo {volume} {13}}\bibfield
  {number} {\bibinfo  {number} { (05)},\ \bibinfo {pages} {P05011}},\ }\Eprint
  {https://arxiv.org/abs/1712.07158} {arXiv:1712.07158 [physics.ins-det]}
  \BibitemShut {NoStop}%
\bibitem [{\citenamefont {Duarte}\ \emph {et~al.}(2018)\citenamefont {Duarte}
  \emph {et~al.}}]{Duarte:2018ite}%
  \BibitemOpen
  \bibfield  {author} {\bibinfo {author} {\bibfnamefont {J.}~\bibnamefont
  {Duarte}} \emph {et~al.},\ }\bibfield  {title} {\bibinfo {title} {{Fast
  inference of deep neural networks in FPGAs for particle physics}},\ }\href
  {https://doi.org/10.1088/1748-0221/13/07/P07027} {\bibfield  {journal}
  {\bibinfo  {journal} {JINST}\ }\textbf {\bibinfo {volume} {13}}\bibfield
  {number} {\bibinfo  {number} { (07)},\ \bibinfo {pages} {P07027}},\ }\Eprint
  {https://arxiv.org/abs/1804.06913} {arXiv:1804.06913 [physics.ins-det]}
  \BibitemShut {NoStop}%
\bibitem [{\citenamefont {Nottbeck}\ \emph {et~al.}(2019)\citenamefont
  {Nottbeck}, \citenamefont {Schmitt},\ and\ \citenamefont
  {B\"uscher}}]{Nottbeck:2019rqu}%
  \BibitemOpen
  \bibfield  {author} {\bibinfo {author} {\bibfnamefont {N.}~\bibnamefont
  {Nottbeck}}, \bibinfo {author} {\bibfnamefont {C.}~\bibnamefont {Schmitt}},\
  and\ \bibinfo {author} {\bibfnamefont {V.}~\bibnamefont {B\"uscher}},\
  }\bibfield  {title} {\bibinfo {title} {{Implementation of high-performance,
  sub-microsecond deep neural networks on FPGAs for trigger applications}},\
  }\href {https://doi.org/10.1088/1748-0221/14/09/p09014} {\bibfield  {journal}
  {\bibinfo  {journal} {JINST}\ }\textbf {\bibinfo {volume} {14}}\bibfield
  {number} {\bibinfo  {number} { (09)},\ \bibinfo {pages} {P09014}},\ }\Eprint
  {https://arxiv.org/abs/1903.10201} {arXiv:1903.10201 [physics.ins-det]}
  \BibitemShut {NoStop}%
\bibitem [{\citenamefont {Zabi}\ \emph {et~al.}(2020)\citenamefont {Zabi},
  \citenamefont {Berryhill}, \citenamefont {Perez},\ and\ \citenamefont
  {Tapper}}]{Zabi:2020gjd}%
  \BibitemOpen
  \bibfield  {author} {\bibinfo {author} {\bibfnamefont {A.}~\bibnamefont
  {Zabi}}, \bibinfo {author} {\bibfnamefont {J.~W.}\ \bibnamefont {Berryhill}},
  \bibinfo {author} {\bibfnamefont {E.}~\bibnamefont {Perez}},\ and\ \bibinfo
  {author} {\bibfnamefont {A.~D.}\ \bibnamefont {Tapper}} (\bibinfo
  {collaboration} {CMS}),\ }\bibfield  {title} {\bibinfo {title} {{The Phase-2
  Upgrade of the CMS Level-1 Trigger}},\ }\href@noop {} {\  (\bibinfo {year}
  {2020})}\BibitemShut {NoStop}%
\bibitem [{\citenamefont {Summers}\ \emph {et~al.}(2020)\citenamefont {Summers}
  \emph {et~al.}}]{Summers:2020xiy}%
  \BibitemOpen
  \bibfield  {author} {\bibinfo {author} {\bibfnamefont {S.}~\bibnamefont
  {Summers}} \emph {et~al.},\ }\bibfield  {title} {\bibinfo {title} {{Fast
  inference of Boosted Decision Trees in FPGAs for particle physics}},\ }\href
  {https://doi.org/10.1088/1748-0221/15/05/P05026} {\bibfield  {journal}
  {\bibinfo  {journal} {JINST}\ }\textbf {\bibinfo {volume} {15}}\bibfield
  {number} {\bibinfo  {number} { (05)},\ \bibinfo {pages} {P05026}},\ }\Eprint
  {https://arxiv.org/abs/2002.02534} {arXiv:2002.02534 [physics.comp-ph]}
  \BibitemShut {NoStop}%
\bibitem [{\citenamefont {Aarrestad}\ \emph
  {et~al.}(2021{\natexlab{b}})\citenamefont {Aarrestad} \emph
  {et~al.}}]{Aarrestad:2021zos}%
  \BibitemOpen
  \bibfield  {author} {\bibinfo {author} {\bibfnamefont {T.}~\bibnamefont
  {Aarrestad}} \emph {et~al.},\ }\bibfield  {title} {\bibinfo {title} {{Fast
  convolutional neural networks on FPGAs with hls4ml}},\ }\href
  {https://doi.org/10.1088/2632-2153/ac0ea1} {\bibfield  {journal} {\bibinfo
  {journal} {Mach. Learn. Sci. Tech.}\ }\textbf {\bibinfo {volume} {2}},\
  \bibinfo {pages} {045015} (\bibinfo {year} {2021}{\natexlab{b}})},\ \Eprint
  {https://arxiv.org/abs/2101.05108} {arXiv:2101.05108 [cs.LG]} \BibitemShut
  {NoStop}%
\bibitem [{\citenamefont {Hong}\ \emph {et~al.}(2021)\citenamefont {Hong},
  \citenamefont {Carlson}, \citenamefont {Eubanks}, \citenamefont {Racz},
  \citenamefont {Roche}, \citenamefont {Stelzer},\ and\ \citenamefont
  {Stumpp}}]{Hong:2021snb}%
  \BibitemOpen
  \bibfield  {author} {\bibinfo {author} {\bibfnamefont {T.~M.}\ \bibnamefont
  {Hong}}, \bibinfo {author} {\bibfnamefont {B.}~\bibnamefont {Carlson}},
  \bibinfo {author} {\bibfnamefont {B.}~\bibnamefont {Eubanks}}, \bibinfo
  {author} {\bibfnamefont {S.}~\bibnamefont {Racz}}, \bibinfo {author}
  {\bibfnamefont {S.}~\bibnamefont {Roche}}, \bibinfo {author} {\bibfnamefont
  {J.}~\bibnamefont {Stelzer}},\ and\ \bibinfo {author} {\bibfnamefont
  {D.}~\bibnamefont {Stumpp}},\ }\bibfield  {title} {\bibinfo {title}
  {{Nanosecond machine learning event classification with boosted decision
  trees in FPGA for high energy physics}},\ }\href
  {https://doi.org/10.1088/1748-0221/16/08/P08016} {\bibfield  {journal}
  {\bibinfo  {journal} {JINST}\ }\textbf {\bibinfo {volume} {16}}\bibfield
  {number} {\bibinfo  {number} { (08)},\ \bibinfo {pages} {P08016}},\ }\Eprint
  {https://arxiv.org/abs/2104.03408} {arXiv:2104.03408 [hep-ex]} \BibitemShut
  {NoStop}%
\bibitem [{\citenamefont {Deiana}\ \emph {et~al.}(2021)\citenamefont {Deiana}
  \emph {et~al.}}]{Deiana:2021niw}%
  \BibitemOpen
  \bibfield  {author} {\bibinfo {author} {\bibfnamefont {A.~M.}\ \bibnamefont
  {Deiana}} \emph {et~al.},\ }\bibfield  {title} {\bibinfo {title}
  {{Applications and Techniques for Fast Machine Learning in Science}},\
  }\href@noop {} {\  (\bibinfo {year} {2021})},\ \Eprint
  {https://arxiv.org/abs/2110.13041} {arXiv:2110.13041 [cs.LG]} \BibitemShut
  {NoStop}%
\bibitem [{\citenamefont
  {LHCb~Collaboration}(2020)}]{LHCbCollaboration:2717938}%
  \BibitemOpen
  \bibfield  {author} {\bibinfo {author} {\bibfnamefont {C.~M.}\ \bibnamefont
  {LHCb~Collaboration}},\ }\href {https://cds.cern.ch/record/2717938} {\emph
  {\bibinfo {title} {{LHCb Upgrade GPU High Level Trigger Technical Design
  Report}}}},\ \bibinfo {type} {Tech. Rep.}\ (\bibinfo  {institution} {CERN},\
  \bibinfo {address} {Geneva},\ \bibinfo {year} {2020})\BibitemShut {NoStop}%
\bibitem [{\citenamefont {Aaij}\ \emph {et~al.}(2020)\citenamefont {Aaij} \emph
  {et~al.}}]{Aaij:2019zbu}%
  \BibitemOpen
  \bibfield  {author} {\bibinfo {author} {\bibfnamefont {R.}~\bibnamefont
  {Aaij}} \emph {et~al.},\ }\bibfield  {title} {\bibinfo {title} {{Allen: A
  high level trigger on GPUs for LHCb}},\ }\href
  {https://doi.org/10.1007/s41781-020-00039-7} {\bibfield  {journal} {\bibinfo
  {journal} {Comput. Softw. Big Sci.}\ }\textbf {\bibinfo {volume} {4}},\
  \bibinfo {pages} {7} (\bibinfo {year} {2020})},\ \Eprint
  {https://arxiv.org/abs/1912.09161} {arXiv:1912.09161 [physics.ins-det]}
  \BibitemShut {NoStop}%
\bibitem [{\citenamefont {Chekalina}\ \emph {et~al.}(2018)\citenamefont
  {Chekalina}, \citenamefont {Orlova}, \citenamefont {Ratnikov}, \citenamefont
  {Ulyanov}, \citenamefont {Ustyuzhanin},\ and\ \citenamefont
  {Zakharov}}]{Chekalina:2018hxi}%
  \BibitemOpen
  \bibfield  {author} {\bibinfo {author} {\bibfnamefont {V.}~\bibnamefont
  {Chekalina}}, \bibinfo {author} {\bibfnamefont {E.}~\bibnamefont {Orlova}},
  \bibinfo {author} {\bibfnamefont {F.}~\bibnamefont {Ratnikov}}, \bibinfo
  {author} {\bibfnamefont {D.}~\bibnamefont {Ulyanov}}, \bibinfo {author}
  {\bibfnamefont {A.}~\bibnamefont {Ustyuzhanin}},\ and\ \bibinfo {author}
  {\bibfnamefont {E.}~\bibnamefont {Zakharov}},\ }\bibfield  {title} {\bibinfo
  {title} {{Generative Models for Fast Calorimeter Simulation.LHCb case}},\
  }\bibfield  {journal} {\bibinfo  {journal} {{CHEP 2018}}\ }\href
  {https://doi.org/10.1051/epjconf/201921402034} {10.1051/epjconf/201921402034}
  (\bibinfo {year} {2018}),\ \Eprint {https://arxiv.org/abs/1812.01319}
  {arXiv:1812.01319 [physics.data-an]} \BibitemShut {NoStop}%
\bibitem [{ATL(2020{\natexlab{a}})}]{ATL-SOFT-PUB-2020-006}%
  \BibitemOpen
  \href {http://cds.cern.ch/record/2746032} {\emph {\bibinfo {title} {{Fast
  simulation of the ATLAS calorimeter system with Generative Adversarial
  Networks}}}},\ \bibinfo {type} {Tech. Rep.}\ (\bibinfo  {institution}
  {CERN},\ \bibinfo {address} {Geneva},\ \bibinfo {year} {2020})\BibitemShut
  {NoStop}%
\bibitem [{\citenamefont {Aad}\ \emph {et~al.}(2021{\natexlab{a}})\citenamefont
  {Aad} \emph {et~al.}}]{ATLAS:2021pzo}%
  \BibitemOpen
  \bibfield  {author} {\bibinfo {author} {\bibfnamefont {G.}~\bibnamefont
  {Aad}} \emph {et~al.} (\bibinfo {collaboration} {ATLAS}),\ }\bibfield
  {title} {\bibinfo {title} {{AtlFast3: the next generation of fast simulation
  in ATLAS}},\ }\href@noop {} {\  (\bibinfo {year} {2021}{\natexlab{a}})},\
  \Eprint {https://arxiv.org/abs/2109.02551} {arXiv:2109.02551 [hep-ex]}
  \BibitemShut {NoStop}%
\bibitem [{\citenamefont {Aad}\ \emph {et~al.}(2019)\citenamefont {Aad} \emph
  {et~al.}}]{ATLAS:2019bwq}%
  \BibitemOpen
  \bibfield  {author} {\bibinfo {author} {\bibfnamefont {G.}~\bibnamefont
  {Aad}} \emph {et~al.} (\bibinfo {collaboration} {ATLAS}),\ }\bibfield
  {title} {\bibinfo {title} {{ATLAS b-jet identification performance and
  efficiency measurement with $t{\bar{t}}$ events in pp collisions at
  $\sqrt{s}=13$ TeV}},\ }\href {https://doi.org/10.1140/epjc/s10052-019-7450-8}
  {\bibfield  {journal} {\bibinfo  {journal} {Eur. Phys. J. C}\ }\textbf
  {\bibinfo {volume} {79}},\ \bibinfo {pages} {970} (\bibinfo {year} {2019})},\
  \Eprint {https://arxiv.org/abs/1907.05120} {arXiv:1907.05120 [hep-ex]}
  \BibitemShut {NoStop}%
\bibitem [{\citenamefont {Bols}\ \emph {et~al.}(2020)\citenamefont {Bols},
  \citenamefont {Kieseler}, \citenamefont {Verzetti}, \citenamefont {Stoye},\
  and\ \citenamefont {Stakia}}]{Bols:2020bkb}%
  \BibitemOpen
  \bibfield  {author} {\bibinfo {author} {\bibfnamefont {E.}~\bibnamefont
  {Bols}}, \bibinfo {author} {\bibfnamefont {J.}~\bibnamefont {Kieseler}},
  \bibinfo {author} {\bibfnamefont {M.}~\bibnamefont {Verzetti}}, \bibinfo
  {author} {\bibfnamefont {M.}~\bibnamefont {Stoye}},\ and\ \bibinfo {author}
  {\bibfnamefont {A.}~\bibnamefont {Stakia}},\ }\bibfield  {title} {\bibinfo
  {title} {{Jet Flavour Classification Using DeepJet}}\ }\href
  {https://doi.org/10.1088/1748-0221/15/12/P12012}
  {10.1088/1748-0221/15/12/P12012} (\bibinfo {year} {2020}),\ \Eprint
  {https://arxiv.org/abs/2008.10519} {arXiv:2008.10519 [hep-ex]} \BibitemShut
  {NoStop}%
\bibitem [{ATL(2020{\natexlab{b}})}]{ATL-PHYS-PUB-2020-014}%
  \BibitemOpen
  \href {https://cds.cern.ch/record/2718948} {\emph {\bibinfo {title} {{Deep
  Sets based Neural Networks for Impact Parameter Flavour Tagging in
  ATLAS}}}},\ \bibinfo {type} {Tech. Rep.}\ (\bibinfo  {institution} {CERN},\
  \bibinfo {address} {Geneva},\ \bibinfo {year} {2020})\BibitemShut {NoStop}%
\bibitem [{\citenamefont {Larkoski}\ \emph {et~al.}(2020)\citenamefont
  {Larkoski}, \citenamefont {Moult},\ and\ \citenamefont
  {Nachman}}]{Larkoski:2017jix}%
  \BibitemOpen
  \bibfield  {author} {\bibinfo {author} {\bibfnamefont {A.~J.}\ \bibnamefont
  {Larkoski}}, \bibinfo {author} {\bibfnamefont {I.}~\bibnamefont {Moult}},\
  and\ \bibinfo {author} {\bibfnamefont {B.}~\bibnamefont {Nachman}},\
  }\bibfield  {title} {\bibinfo {title} {{Jet Substructure at the Large Hadron
  Collider: A Review of Recent Advances in Theory and Machine Learning}},\
  }\href {https://doi.org/10.1016/j.physrep.2019.11.001} {\bibfield  {journal}
  {\bibinfo  {journal} {Phys. Rept.}\ }\textbf {\bibinfo {volume} {841}},\
  \bibinfo {pages} {1} (\bibinfo {year} {2020})},\ \Eprint
  {https://arxiv.org/abs/1709.04464} {arXiv:1709.04464 [hep-ph]} \BibitemShut
  {NoStop}%
\bibitem [{\citenamefont {R.~Kogler}(2019)}]{1803.06991}%
  \BibitemOpen
  \bibfield  {author} {\bibinfo {author} {\bibfnamefont {A.~S. e. e.~a.}\
  \bibnamefont {R.~Kogler}, \bibfnamefont {B.~Nachman}},\ }\bibfield  {title}
  {\bibinfo {title} {{Jet Substructure at the Large Hadron Collider:
  Experimental Review}},\ }\href {https://doi.org/10.1103/RevModPhys.91.045003}
  {\bibfield  {journal} {\bibinfo  {journal} {Rev. Mod. Phys. 91}\ }\textbf
  {\bibinfo {volume} {91}},\ \bibinfo {pages} {045003} (\bibinfo {year}
  {2019})},\ \Eprint {https://arxiv.org/abs/1803.06991} {arXiv:1803.06991
  [hep-ph]} \BibitemShut {NoStop}%
\bibitem [{\citenamefont {Sirunyan}\ \emph
  {et~al.}(2020{\natexlab{a}})\citenamefont {Sirunyan} \emph
  {et~al.}}]{Sirunyan:2020lcu}%
  \BibitemOpen
  \bibfield  {author} {\bibinfo {author} {\bibfnamefont {A.~M.}\ \bibnamefont
  {Sirunyan}} \emph {et~al.} (\bibinfo {collaboration} {CMS}),\ }\bibfield
  {title} {\bibinfo {title} {{Identification of heavy, energetic, hadronically
  decaying particles using machine-learning techniques}},\ }\href
  {https://doi.org/10.1088/1748-0221/15/06/P06005} {\bibfield  {journal}
  {\bibinfo  {journal} {JINST}\ }\textbf {\bibinfo {volume} {15}}\bibfield
  {number} {\bibinfo  {number} { (06)},\ \bibinfo {pages} {P06005}},\ }\Eprint
  {https://arxiv.org/abs/2004.08262} {arXiv:2004.08262 [hep-ex]} \BibitemShut
  {NoStop}%
\bibitem [{\citenamefont {Sirunyan}\ \emph
  {et~al.}(2020{\natexlab{b}})\citenamefont {Sirunyan} \emph
  {et~al.}}]{CMS:2019ykj}%
  \BibitemOpen
  \bibfield  {author} {\bibinfo {author} {\bibfnamefont {A.~M.}\ \bibnamefont
  {Sirunyan}} \emph {et~al.} (\bibinfo {collaboration} {CMS}),\ }\bibfield
  {title} {\bibinfo {title} {{Search for dark matter particles produced in
  association with a Higgs boson in proton-proton collisions at $
  \sqrt{\mathrm{s}} $ = 13 TeV}},\ }\href
  {https://doi.org/10.1007/JHEP03(2020)025} {\bibfield  {journal} {\bibinfo
  {journal} {JHEP}\ }\textbf {\bibinfo {volume} {03}},\ \bibinfo {pages}
  {025}},\ \Eprint {https://arxiv.org/abs/1908.01713} {arXiv:1908.01713
  [hep-ex]} \BibitemShut {NoStop}%
\bibitem [{CMS(2021{\natexlab{a}})}]{CMS-PAS-B2G-20-004}%
  \BibitemOpen
  \href {https://cds.cern.ch/record/2777083} {\emph {\bibinfo {title} {{Search
  for resonant Higgs boson pair production in four b quark final state using
  large-area jets in proton-proton collisions at
  $\sqrt{s}=13~\mathrm{TeV}$}}}},\ \bibinfo {type} {Tech. Rep.}\ (\bibinfo
  {institution} {CERN},\ \bibinfo {address} {Geneva},\ \bibinfo {year}
  {2021})\BibitemShut {NoStop}%
\bibitem [{CMS(2021{\natexlab{b}})}]{CMS-PAS-B2G-20-007}%
  \BibitemOpen
  \href {http://cds.cern.ch/record/2777173} {\emph {\bibinfo {title} {{Search
  for heavy resonances decaying to a pair of boosted Higgs bosons in final
  states with leptons and a bottom quark-antiquark pair at $\sqrt{s} = 13$
  TeV}}}},\ \bibinfo {type} {Tech. Rep.}\ (\bibinfo  {institution} {CERN},\
  \bibinfo {address} {Geneva},\ \bibinfo {year} {2021})\BibitemShut {NoStop}%
\bibitem [{CMS(2021{\natexlab{c}})}]{CMS-PAS-B2G-21-001}%
  \BibitemOpen
  \href {https://cds.cern.ch/record/2776802} {\emph {\bibinfo {title} {{Search
  for Higgs boson pair production via vector boson fusion with highly
  Lorentz-boosted Higgs bosons in the four b quark final state at $\sqrt{s} =
  13$ TeV}}}},\ \bibinfo {type} {Tech. Rep.}\ (\bibinfo  {institution} {CERN},\
  \bibinfo {address} {Geneva},\ \bibinfo {year} {2021})\BibitemShut {NoStop}%
\bibitem [{\citenamefont {Sirunyan}\ \emph
  {et~al.}(2021{\natexlab{b}})\citenamefont {Sirunyan} \emph
  {et~al.}}]{CMS:2021mux}%
  \BibitemOpen
  \bibfield  {author} {\bibinfo {author} {\bibfnamefont {A.~M.}\ \bibnamefont
  {Sirunyan}} \emph {et~al.} (\bibinfo {collaboration} {CMS}),\ }\bibfield
  {title} {\bibinfo {title} {{Search for W' bosons decaying to a top and a
  bottom quark at s=13TeV in the hadronic final state}},\ }\href
  {https://doi.org/10.1016/j.physletb.2021.136535} {\bibfield  {journal}
  {\bibinfo  {journal} {Phys. Lett. B}\ }\textbf {\bibinfo {volume} {820}},\
  \bibinfo {pages} {136535} (\bibinfo {year} {2021}{\natexlab{b}})},\ \Eprint
  {https://arxiv.org/abs/2104.04831} {arXiv:2104.04831 [hep-ex]} \BibitemShut
  {NoStop}%
\bibitem [{ATL(2021{\natexlab{a}})}]{ATL-PHYS-PUB-2021-035}%
  \BibitemOpen
  \href {http://cds.cern.ch/record/2777811} {\emph {\bibinfo {title}
  {{Efficiency corrections for a tagger for boosted $H\rightarrow b\bar{b}$
  decays in $pp$ collisions at $\sqrt{s}=13$ TeV with the ATLAS detector}}}},\
  \bibinfo {type} {Tech. Rep.}\ (\bibinfo  {institution} {CERN},\ \bibinfo
  {address} {Geneva},\ \bibinfo {year} {2021})\BibitemShut {NoStop}%
\bibitem [{\citenamefont {Sirunyan}\ \emph
  {et~al.}(2020{\natexlab{c}})\citenamefont {Sirunyan} \emph
  {et~al.}}]{CMS:2019dqq}%
  \BibitemOpen
  \bibfield  {author} {\bibinfo {author} {\bibfnamefont {A.~M.}\ \bibnamefont
  {Sirunyan}} \emph {et~al.} (\bibinfo {collaboration} {CMS}),\ }\bibfield
  {title} {\bibinfo {title} {{A deep neural network to search for new
  long-lived particles decaying to jets}},\ }\href
  {https://doi.org/10.1088/2632-2153/ab9023} {\bibfield  {journal} {\bibinfo
  {journal} {Mach. Learn. Sci. Tech.}\ }\textbf {\bibinfo {volume} {1}},\
  \bibinfo {pages} {035012} (\bibinfo {year} {2020}{\natexlab{c}})},\ \Eprint
  {https://arxiv.org/abs/1912.12238} {arXiv:1912.12238 [hep-ex]} \BibitemShut
  {NoStop}%
\bibitem [{\citenamefont {Tumasyan}\ \emph
  {et~al.}(2021{\natexlab{a}})\citenamefont {Tumasyan} \emph
  {et~al.}}]{CMS:2021far}%
  \BibitemOpen
  \bibfield  {author} {\bibinfo {author} {\bibfnamefont {A.}~\bibnamefont
  {Tumasyan}} \emph {et~al.} (\bibinfo {collaboration} {CMS}),\ }\bibfield
  {title} {\bibinfo {title} {{Search for new particles in events with energetic
  jets and large missing transverse momentum in proton-proton collisions at
  $\sqrt{s} = $ 13 TeV}},\ }\href@noop {} {\  (\bibinfo {year}
  {2021}{\natexlab{a}})},\ \Eprint {https://arxiv.org/abs/2107.13021}
  {arXiv:2107.13021 [hep-ex]} \BibitemShut {NoStop}%
\bibitem [{\citenamefont {Aaij}\ \emph {et~al.}(2021)\citenamefont {Aaij} \emph
  {et~al.}}]{LHCb:2020wxx}%
  \BibitemOpen
  \bibfield  {author} {\bibinfo {author} {\bibfnamefont {R.}~\bibnamefont
  {Aaij}} \emph {et~al.} (\bibinfo {collaboration} {LHCb}),\ }\bibfield
  {title} {\bibinfo {title} {{Search for heavy neutral leptons in
  $W^+\to\mu^{+}\mu^{\pm}\text{jet}$ decays}},\ }\href
  {https://doi.org/10.1140/epjc/s10052-021-08973-5} {\bibfield  {journal}
  {\bibinfo  {journal} {Eur. Phys. J. C}\ }\textbf {\bibinfo {volume} {81}},\
  \bibinfo {pages} {248} (\bibinfo {year} {2021})},\ \Eprint
  {https://arxiv.org/abs/2011.05263} {arXiv:2011.05263 [hep-ex]} \BibitemShut
  {NoStop}%
\bibitem [{\citenamefont {{ATLAS
  Collaboration}}(2020)}]{collaboration2020dijet}%
  \BibitemOpen
  \bibfield  {author} {\bibinfo {author} {\bibnamefont {{ATLAS
  Collaboration}}},\ }\bibfield  {title} {\bibinfo {title} {{Dijet resonance
  search with weak supervision using 13 TeV pp collisions in the ATLAS
  detector}}\ }\href {https://doi.org/10.1103/PhysRevLett.125.131801}
  {10.1103/PhysRevLett.125.131801} (\bibinfo {year} {2020}),\ \Eprint
  {https://arxiv.org/abs/2005.02983} {arXiv:2005.02983 [hep-ex]} \BibitemShut
  {NoStop}%
\bibitem [{\citenamefont {Aaboud}\ \emph {et~al.}(2018)\citenamefont {Aaboud}
  \emph {et~al.}}]{ATLAS:2018tti}%
  \BibitemOpen
  \bibfield  {author} {\bibinfo {author} {\bibfnamefont {M.}~\bibnamefont
  {Aaboud}} \emph {et~al.} (\bibinfo {collaboration} {ATLAS}),\ }\bibfield
  {title} {\bibinfo {title} {{Search for pair production of higgsinos in final
  states with at least three $b$-tagged jets in $\sqrt{s} = 13$ TeV $pp$
  collisions using the ATLAS detector}},\ }\href
  {https://doi.org/10.1103/PhysRevD.98.092002} {\bibfield  {journal} {\bibinfo
  {journal} {Phys. Rev. D}\ }\textbf {\bibinfo {volume} {98}},\ \bibinfo
  {pages} {092002} (\bibinfo {year} {2018})},\ \Eprint
  {https://arxiv.org/abs/1806.04030} {arXiv:1806.04030 [hep-ex]} \BibitemShut
  {NoStop}%
\bibitem [{ATL(2021{\natexlab{b}})}]{ATLAS:2021ypo}%
  \BibitemOpen
  \bibfield  {title} {\bibinfo {title} {{Search for Higgs boson decays into two
  spin-0 particles in the $bb\mu\mu$ final state with the ATLAS detector in
  $pp$ collisions at $\sqrt{s}=13$ TeV}},\ }\href@noop {} {\  (\bibinfo {year}
  {2021}{\natexlab{b}})}\BibitemShut {NoStop}%
\bibitem [{\citenamefont {Aad}\ \emph {et~al.}(2021{\natexlab{b}})\citenamefont
  {Aad} \emph {et~al.}}]{ATLAS:2020tlo}%
  \BibitemOpen
  \bibfield  {author} {\bibinfo {author} {\bibfnamefont {G.}~\bibnamefont
  {Aad}} \emph {et~al.} (\bibinfo {collaboration} {ATLAS}),\ }\bibfield
  {title} {\bibinfo {title} {{Search for heavy resonances decaying into a pair
  of Z bosons in the $\ell ^+\ell ^-\ell '^+\ell '^-$ and $\ell ^+\ell ^-\nu
  {{\bar{\nu }}}$ final states using 139 $\mathrm {fb}^{-1}$ of
  proton\textendash{}proton collisions at $\sqrt{s} = 13\,$TeV with the ATLAS
  detector}},\ }\href {https://doi.org/10.1140/epjc/s10052-021-09013-y}
  {\bibfield  {journal} {\bibinfo  {journal} {Eur. Phys. J. C}\ }\textbf
  {\bibinfo {volume} {81}},\ \bibinfo {pages} {332} (\bibinfo {year}
  {2021}{\natexlab{b}})},\ \Eprint {https://arxiv.org/abs/2009.14791}
  {arXiv:2009.14791 [hep-ex]} \BibitemShut {NoStop}%
\bibitem [{\citenamefont {Tumasyan}\ \emph
  {et~al.}(2021{\natexlab{b}})\citenamefont {Tumasyan} \emph
  {et~al.}}]{CMS:2021yci}%
  \BibitemOpen
  \bibfield  {author} {\bibinfo {author} {\bibfnamefont {A.}~\bibnamefont
  {Tumasyan}} \emph {et~al.} (\bibinfo {collaboration} {CMS}),\ }\bibfield
  {title} {\bibinfo {title} {{Search for a heavy Higgs boson decaying into two
  lighter Higgs bosons in the $\tau\tau$bb final state at 13 TeV}},\
  }\href@noop {} {\  (\bibinfo {year} {2021}{\natexlab{b}})},\ \Eprint
  {https://arxiv.org/abs/2106.10361} {arXiv:2106.10361 [hep-ex]} \BibitemShut
  {NoStop}%
\bibitem [{\citenamefont {Aad}\ \emph {et~al.}(2020)\citenamefont {Aad} \emph
  {et~al.}}]{ATLAS:2020pcy}%
  \BibitemOpen
  \bibfield  {author} {\bibinfo {author} {\bibfnamefont {G.}~\bibnamefont
  {Aad}} \emph {et~al.} (\bibinfo {collaboration} {ATLAS}),\ }\bibfield
  {title} {\bibinfo {title} {{Search for Higgs Boson Decays into a $Z$ Boson
  and a Light Hadronically Decaying Resonance Using 13 TeV $pp$ Collision Data
  from the ATLAS Detector}},\ }\href
  {https://doi.org/10.1103/PhysRevLett.125.221802} {\bibfield  {journal}
  {\bibinfo  {journal} {Phys. Rev. Lett.}\ }\textbf {\bibinfo {volume} {125}},\
  \bibinfo {pages} {221802} (\bibinfo {year} {2020})},\ \Eprint
  {https://arxiv.org/abs/2004.01678} {arXiv:2004.01678 [hep-ex]} \BibitemShut
  {NoStop}%
\bibitem [{\citenamefont {Aad}\ \emph {et~al.}(2021{\natexlab{c}})\citenamefont
  {Aad} \emph {et~al.}}]{ATLAS:2021jbf}%
  \BibitemOpen
  \bibfield  {author} {\bibinfo {author} {\bibfnamefont {G.}~\bibnamefont
  {Aad}} \emph {et~al.} (\bibinfo {collaboration} {ATLAS}),\ }\bibfield
  {title} {\bibinfo {title} {{Search for dark matter in events with missing
  transverse momentum and a Higgs boson decaying into two photons in $pp$
  collisions at $\sqrt{s} = 13$ TeV with the ATLAS detector}},\ }\href@noop {}
  {\  (\bibinfo {year} {2021}{\natexlab{c}})},\ \Eprint
  {https://arxiv.org/abs/2104.13240} {arXiv:2104.13240 [hep-ex]} \BibitemShut
  {NoStop}%
\bibitem [{\citenamefont {Chen}\ and\ \citenamefont
  {Guestrin}(2016)}]{Chen:2016:XST:2939672.2939785}%
  \BibitemOpen
  \bibfield  {author} {\bibinfo {author} {\bibfnamefont {T.}~\bibnamefont
  {Chen}}\ and\ \bibinfo {author} {\bibfnamefont {C.}~\bibnamefont
  {Guestrin}},\ }\bibfield  {title} {\bibinfo {title} {{XGBoost}: A scalable
  tree boosting system},\ }in\ \href {https://doi.org/10.1145/2939672.2939785}
  {\emph {\bibinfo {booktitle} {Proceedings of the 22nd ACM SIGKDD
  International Conference on Knowledge Discovery and Data Mining}}},\ \bibinfo
  {series and number} {KDD '16}\ (\bibinfo  {publisher} {ACM},\ \bibinfo
  {address} {New York, NY, USA},\ \bibinfo {year} {2016})\ pp.\ \bibinfo
  {pages} {785--794}\BibitemShut {NoStop}%
\bibitem [{\citenamefont {Bertacchi}\ \emph {et~al.}(2021)\citenamefont
  {Bertacchi} \emph {et~al.}}]{BelleIITrackingGroup:2020hpx}%
  \BibitemOpen
  \bibfield  {author} {\bibinfo {author} {\bibfnamefont {V.}~\bibnamefont
  {Bertacchi}} \emph {et~al.} (\bibinfo {collaboration} {Belle II Tracking
  Group}),\ }\bibfield  {title} {\bibinfo {title} {{Track finding at Belle
  II}},\ }\href {https://doi.org/10.1016/j.cpc.2020.107610} {\bibfield
  {journal} {\bibinfo  {journal} {Comput. Phys. Commun.}\ }\textbf {\bibinfo
  {volume} {259}},\ \bibinfo {pages} {107610} (\bibinfo {year} {2021})},\
  \Eprint {https://arxiv.org/abs/2003.12466} {arXiv:2003.12466
  [physics.ins-det]} \BibitemShut {NoStop}%
\bibitem [{\citenamefont {Abazajian}\ \emph {et~al.}(2012)\citenamefont
  {Abazajian} \emph {et~al.}}]{Abazajian:2012ys}%
  \BibitemOpen
  \bibfield  {author} {\bibinfo {author} {\bibfnamefont {K.~N.}\ \bibnamefont
  {Abazajian}} \emph {et~al.},\ }\bibfield  {title} {\bibinfo {title} {{Light
  Sterile Neutrinos: A White Paper}},\ }\href@noop {} {\  (\bibinfo {year}
  {2012})},\ \Eprint {https://arxiv.org/abs/1204.5379} {arXiv:1204.5379
  [hep-ph]} \BibitemShut {NoStop}%
\bibitem [{\citenamefont {Dentler}\ \emph {et~al.}(2018)\citenamefont
  {Dentler}, \citenamefont {Hern\'andez-Cabezudo}, \citenamefont {Kopp},
  \citenamefont {Machado}, \citenamefont {Maltoni}, \citenamefont
  {Martinez-Soler},\ and\ \citenamefont {Schwetz}}]{Dentler:2018sju}%
  \BibitemOpen
  \bibfield  {author} {\bibinfo {author} {\bibfnamefont {M.}~\bibnamefont
  {Dentler}}, \bibinfo {author} {\bibfnamefont {A.}~\bibnamefont
  {Hern\'andez-Cabezudo}}, \bibinfo {author} {\bibfnamefont {J.}~\bibnamefont
  {Kopp}}, \bibinfo {author} {\bibfnamefont {P.~A.~N.}\ \bibnamefont
  {Machado}}, \bibinfo {author} {\bibfnamefont {M.}~\bibnamefont {Maltoni}},
  \bibinfo {author} {\bibfnamefont {I.}~\bibnamefont {Martinez-Soler}},\ and\
  \bibinfo {author} {\bibfnamefont {T.}~\bibnamefont {Schwetz}},\ }\bibfield
  {title} {\bibinfo {title} {{Updated Global Analysis of Neutrino Oscillations
  in the Presence of eV-Scale Sterile Neutrinos}},\ }\href
  {https://doi.org/10.1007/JHEP08(2018)010} {\bibfield  {journal} {\bibinfo
  {journal} {JHEP}\ }\textbf {\bibinfo {volume} {08}},\ \bibinfo {pages}
  {010}},\ \Eprint {https://arxiv.org/abs/1803.10661} {arXiv:1803.10661
  [hep-ph]} \BibitemShut {NoStop}%
\bibitem [{\citenamefont {Bertuzzo}\ \emph {et~al.}(2018)\citenamefont
  {Bertuzzo}, \citenamefont {Jana}, \citenamefont {Machado},\ and\
  \citenamefont {Zukanovich~Funchal}}]{Bertuzzo:2018itn}%
  \BibitemOpen
  \bibfield  {author} {\bibinfo {author} {\bibfnamefont {E.}~\bibnamefont
  {Bertuzzo}}, \bibinfo {author} {\bibfnamefont {S.}~\bibnamefont {Jana}},
  \bibinfo {author} {\bibfnamefont {P.~A.~N.}\ \bibnamefont {Machado}},\ and\
  \bibinfo {author} {\bibfnamefont {R.}~\bibnamefont {Zukanovich~Funchal}},\
  }\bibfield  {title} {\bibinfo {title} {{Dark Neutrino Portal to Explain
  MiniBooNE excess}},\ }\href {https://doi.org/10.1103/PhysRevLett.121.241801}
  {\bibfield  {journal} {\bibinfo  {journal} {Phys. Rev. Lett.}\ }\textbf
  {\bibinfo {volume} {121}},\ \bibinfo {pages} {241801} (\bibinfo {year}
  {2018})},\ \Eprint {https://arxiv.org/abs/1807.09877} {arXiv:1807.09877
  [hep-ph]} \BibitemShut {NoStop}%
\bibitem [{\citenamefont {Ballett}\ \emph {et~al.}(2019)\citenamefont
  {Ballett}, \citenamefont {Pascoli},\ and\ \citenamefont
  {Ross-Lonergan}}]{Ballett:2018ynz}%
  \BibitemOpen
  \bibfield  {author} {\bibinfo {author} {\bibfnamefont {P.}~\bibnamefont
  {Ballett}}, \bibinfo {author} {\bibfnamefont {S.}~\bibnamefont {Pascoli}},\
  and\ \bibinfo {author} {\bibfnamefont {M.}~\bibnamefont {Ross-Lonergan}},\
  }\bibfield  {title} {\bibinfo {title} {{U(1)' mediated decays of heavy
  sterile neutrinos in MiniBooNE}},\ }\href
  {https://doi.org/10.1103/PhysRevD.99.071701} {\bibfield  {journal} {\bibinfo
  {journal} {Phys. Rev. D}\ }\textbf {\bibinfo {volume} {99}},\ \bibinfo
  {pages} {071701} (\bibinfo {year} {2019})},\ \Eprint
  {https://arxiv.org/abs/1808.02915} {arXiv:1808.02915 [hep-ph]} \BibitemShut
  {NoStop}%
\bibitem [{\citenamefont {Adamson}\ \emph {et~al.}(2016)\citenamefont {Adamson}
  \emph {et~al.}}]{MINOS:2016vvv}%
  \BibitemOpen
  \bibfield  {author} {\bibinfo {author} {\bibfnamefont {P.}~\bibnamefont
  {Adamson}} \emph {et~al.} (\bibinfo {collaboration} {MINOS}),\ }\bibfield
  {title} {\bibinfo {title} {{Constraints on Large Extra Dimensions from the
  MINOS Experiment}},\ }\href {https://doi.org/10.1103/PhysRevD.94.111101}
  {\bibfield  {journal} {\bibinfo  {journal} {Phys. Rev. D}\ }\textbf {\bibinfo
  {volume} {94}},\ \bibinfo {pages} {111101} (\bibinfo {year} {2016})},\
  \Eprint {https://arxiv.org/abs/1608.06964} {arXiv:1608.06964 [hep-ex]}
  \BibitemShut {NoStop}%
\bibitem [{\citenamefont {Kostelecky}\ and\ \citenamefont
  {Mewes}(2004)}]{Kostelecky:2003cr}%
  \BibitemOpen
  \bibfield  {author} {\bibinfo {author} {\bibfnamefont {V.~A.}\ \bibnamefont
  {Kostelecky}}\ and\ \bibinfo {author} {\bibfnamefont {M.}~\bibnamefont
  {Mewes}},\ }\bibfield  {title} {\bibinfo {title} {{Lorentz and CPT violation
  in neutrinos}},\ }\href {https://doi.org/10.1103/PhysRevD.69.016005}
  {\bibfield  {journal} {\bibinfo  {journal} {Phys. Rev. D}\ }\textbf {\bibinfo
  {volume} {69}},\ \bibinfo {pages} {016005} (\bibinfo {year} {2004})},\
  \Eprint {https://arxiv.org/abs/hep-ph/0309025} {arXiv:hep-ph/0309025}
  \BibitemShut {NoStop}%
\bibitem [{\citenamefont {Miranda}\ and\ \citenamefont
  {Nunokawa}(2015)}]{Miranda:2015dra}%
  \BibitemOpen
  \bibfield  {author} {\bibinfo {author} {\bibfnamefont {O.~G.}\ \bibnamefont
  {Miranda}}\ and\ \bibinfo {author} {\bibfnamefont {H.}~\bibnamefont
  {Nunokawa}},\ }\bibfield  {title} {\bibinfo {title} {{Non standard neutrino
  interactions: current status and future prospects}},\ }\href
  {https://doi.org/10.1088/1367-2630/17/9/095002} {\bibfield  {journal}
  {\bibinfo  {journal} {New J. Phys.}\ }\textbf {\bibinfo {volume} {17}},\
  \bibinfo {pages} {095002} (\bibinfo {year} {2015})},\ \Eprint
  {https://arxiv.org/abs/1505.06254} {arXiv:1505.06254 [hep-ph]} \BibitemShut
  {NoStop}%
\bibitem [{\citenamefont {de~Gouv\^ea}\ and\ \citenamefont
  {Kelly}(2016)}]{deGouvea:2015ndi}%
  \BibitemOpen
  \bibfield  {author} {\bibinfo {author} {\bibfnamefont {A.}~\bibnamefont
  {de~Gouv\^ea}}\ and\ \bibinfo {author} {\bibfnamefont {K.~J.}\ \bibnamefont
  {Kelly}},\ }\bibfield  {title} {\bibinfo {title} {{Non-standard Neutrino
  Interactions at DUNE}},\ }\href
  {https://doi.org/10.1016/j.nuclphysb.2016.03.013} {\bibfield  {journal}
  {\bibinfo  {journal} {Nucl. Phys. B}\ }\textbf {\bibinfo {volume} {908}},\
  \bibinfo {pages} {318} (\bibinfo {year} {2016})},\ \Eprint
  {https://arxiv.org/abs/1511.05562} {arXiv:1511.05562 [hep-ph]} \BibitemShut
  {NoStop}%
\bibitem [{\citenamefont {Jwa}\ \emph {et~al.}(2019)\citenamefont {Jwa},
  \citenamefont {Guglielmo}, \citenamefont {Carloni},\ and\ \citenamefont
  {Karagiorgi}}]{Jwa:2019zlh}%
  \BibitemOpen
  \bibfield  {author} {\bibinfo {author} {\bibfnamefont {Y.-J.}\ \bibnamefont
  {Jwa}}, \bibinfo {author} {\bibfnamefont {G.~D.}\ \bibnamefont {Guglielmo}},
  \bibinfo {author} {\bibfnamefont {L.~P.}\ \bibnamefont {Carloni}},\ and\
  \bibinfo {author} {\bibfnamefont {G.}~\bibnamefont {Karagiorgi}},\ }\bibfield
   {title} {\bibinfo {title} {{Accelerating Deep Neural Networks for Real-time
  Data Selection for High-resolution Imaging Particle Detectors}},\ }in\ \href
  {https://doi.org/10.1109/NYSDS.2019.8909784} {\emph {\bibinfo {booktitle}
  {{2019 New York Scientific Data Summit}: {Data-Driven Discovery in Science
  and Industry}}}}\ (\bibinfo {year} {2019})\BibitemShut {NoStop}%
\bibitem [{\citenamefont {Acciarri}\ \emph {et~al.}(2021)\citenamefont
  {Acciarri} \emph {et~al.}}]{ArgoNeuT:2021xtd}%
  \BibitemOpen
  \bibfield  {author} {\bibinfo {author} {\bibfnamefont {R.}~\bibnamefont
  {Acciarri}} \emph {et~al.} (\bibinfo {collaboration} {ArgoNeuT}),\ }\bibfield
   {title} {\bibinfo {title} {{A deep-learning based raw waveform
  region-of-interest finder for the liquid argon time projection chamber}},\
  }\href@noop {} {\  (\bibinfo {year} {2021})},\ \Eprint
  {https://arxiv.org/abs/2103.06391} {arXiv:2103.06391 [physics.ins-det]}
  \BibitemShut {NoStop}%
\bibitem [{\citenamefont {Uboldi}\ \emph {et~al.}(2021)\citenamefont {Uboldi},
  \citenamefont {Ruth}, \citenamefont {Andrews}, \citenamefont {Wang},
  \citenamefont {Wenzel}, \citenamefont {Wu},\ and\ \citenamefont
  {Yang}}]{Uboldi:2021jyj}%
  \BibitemOpen
  \bibfield  {author} {\bibinfo {author} {\bibfnamefont {L.}~\bibnamefont
  {Uboldi}}, \bibinfo {author} {\bibfnamefont {D.}~\bibnamefont {Ruth}},
  \bibinfo {author} {\bibfnamefont {M.}~\bibnamefont {Andrews}}, \bibinfo
  {author} {\bibfnamefont {M.~H. L.~S.}\ \bibnamefont {Wang}}, \bibinfo
  {author} {\bibfnamefont {H.-J.}\ \bibnamefont {Wenzel}}, \bibinfo {author}
  {\bibfnamefont {W.}~\bibnamefont {Wu}},\ and\ \bibinfo {author}
  {\bibfnamefont {T.}~\bibnamefont {Yang}},\ }\bibfield  {title} {\bibinfo
  {title} {{Extracting low energy signals from raw LArTPC waveforms using deep
  learning techniques -- A proof of concept}},\ }\href@noop {} {\  (\bibinfo
  {year} {2021})},\ \Eprint {https://arxiv.org/abs/2106.09911}
  {arXiv:2106.09911 [physics.ins-det]} \BibitemShut {NoStop}%
\bibitem [{\citenamefont {Anker}\ \emph {et~al.}(2021)\citenamefont {Anker}
  \emph {et~al.}}]{ARIANNA:2021inb}%
  \BibitemOpen
  \bibfield  {author} {\bibinfo {author} {\bibfnamefont {A.}~\bibnamefont
  {Anker}} \emph {et~al.} (\bibinfo {collaboration} {ARIANNA}),\ }\bibfield
  {title} {\bibinfo {title} {{A novel trigger based on neural networks for
  radio neutrino detectors}},\ }\href {https://doi.org/10.22323/1.395.1074}
  {\bibfield  {journal} {\bibinfo  {journal} {PoS}\ }\textbf {\bibinfo {volume}
  {ICRC2021}},\ \bibinfo {pages} {1074} (\bibinfo {year} {2021})}\BibitemShut
  {NoStop}%
\bibitem [{\citenamefont {Acero}\ \emph {et~al.}(2021)\citenamefont {Acero}
  \emph {et~al.}}]{NOvA:2021smv}%
  \BibitemOpen
  \bibfield  {author} {\bibinfo {author} {\bibfnamefont {M.~A.}\ \bibnamefont
  {Acero}} \emph {et~al.} (\bibinfo {collaboration} {NOvA}),\ }\bibfield
  {title} {\bibinfo {title} {{Search for Active-Sterile Antineutrino Mixing
  Using Neutral-Current Interactions with the NOvA Experiment}},\ }\href
  {https://doi.org/10.1103/PhysRevLett.127.201801} {\bibfield  {journal}
  {\bibinfo  {journal} {Phys. Rev. Lett.}\ }\textbf {\bibinfo {volume} {127}},\
  \bibinfo {pages} {201801} (\bibinfo {year} {2021})},\ \Eprint
  {https://arxiv.org/abs/2106.04673} {arXiv:2106.04673 [hep-ex]} \BibitemShut
  {NoStop}%
\bibitem [{\citenamefont {Abratenko}\ \emph
  {et~al.}(2021{\natexlab{b}})\citenamefont {Abratenko} \emph
  {et~al.}}]{MicroBooNE:2021jwr}%
  \BibitemOpen
  \bibfield  {author} {\bibinfo {author} {\bibfnamefont {P.}~\bibnamefont
  {Abratenko}} \emph {et~al.} (\bibinfo {collaboration} {MicroBooNE}),\
  }\bibfield  {title} {\bibinfo {title} {{Search for an anomalous excess of
  charged-current quasi-elastic $\nu_e$ interactions with the MicroBooNE
  experiment using Deep-Learning-based reconstruction}},\ }\href@noop {} {\
  (\bibinfo {year} {2021}{\natexlab{b}})},\ \Eprint
  {https://arxiv.org/abs/2110.14080} {arXiv:2110.14080 [hep-ex]} \BibitemShut
  {NoStop}%
\bibitem [{\citenamefont {Baldi}\ \emph {et~al.}(2019)\citenamefont {Baldi},
  \citenamefont {Bian}, \citenamefont {Hertel},\ and\ \citenamefont
  {Li}}]{Baldi:2018qhe}%
  \BibitemOpen
  \bibfield  {author} {\bibinfo {author} {\bibfnamefont {P.}~\bibnamefont
  {Baldi}}, \bibinfo {author} {\bibfnamefont {J.}~\bibnamefont {Bian}},
  \bibinfo {author} {\bibfnamefont {L.}~\bibnamefont {Hertel}},\ and\ \bibinfo
  {author} {\bibfnamefont {L.}~\bibnamefont {Li}},\ }\bibfield  {title}
  {\bibinfo {title} {{Improved Energy Reconstruction in NOvA with Regression
  Convolutional Neural Networks}},\ }\href
  {https://doi.org/10.1103/PhysRevD.99.012011} {\bibfield  {journal} {\bibinfo
  {journal} {Phys. Rev. D}\ }\textbf {\bibinfo {volume} {99}},\ \bibinfo
  {pages} {012011} (\bibinfo {year} {2019})},\ \Eprint
  {https://arxiv.org/abs/1811.04557} {arXiv:1811.04557 [physics.ins-det]}
  \BibitemShut {NoStop}%
\bibitem [{\citenamefont {Abratenko}\ \emph
  {et~al.}(2021{\natexlab{c}})\citenamefont {Abratenko} \emph
  {et~al.}}]{MicroBooNE:2021ojx}%
  \BibitemOpen
  \bibfield  {author} {\bibinfo {author} {\bibfnamefont {P.}~\bibnamefont
  {Abratenko}} \emph {et~al.} (\bibinfo {collaboration} {MicroBooNE}),\
  }\bibfield  {title} {\bibinfo {title} {{Wire-Cell 3D Pattern Recognition
  Techniques for Neutrino Event Reconstruction in Large LArTPCs: Algorithm
  Description and Quantitative Evaluation with MicroBooNE Simulation}},\
  }\href@noop {} {\  (\bibinfo {year} {2021}{\natexlab{c}})},\ \Eprint
  {https://arxiv.org/abs/2110.13961} {arXiv:2110.13961 [physics.ins-det]}
  \BibitemShut {NoStop}%
\bibitem [{\citenamefont {Aiello}\ \emph {et~al.}(2020)\citenamefont {Aiello}
  \emph {et~al.}}]{KM3NeT:2020zod}%
  \BibitemOpen
  \bibfield  {author} {\bibinfo {author} {\bibfnamefont {S.}~\bibnamefont
  {Aiello}} \emph {et~al.} (\bibinfo {collaboration} {KM3NeT}),\ }\bibfield
  {title} {\bibinfo {title} {{Event reconstruction for KM3NeT/ORCA using
  convolutional neural networks}},\ }\href
  {https://doi.org/10.1088/1748-0221/15/10/P10005} {\bibfield  {journal}
  {\bibinfo  {journal} {JINST}\ }\textbf {\bibinfo {volume} {15}}\bibfield
  {number} {\bibinfo  {number} { (10)},\ \bibinfo {pages} {P10005}},\ }\Eprint
  {https://arxiv.org/abs/2004.08254} {arXiv:2004.08254 [astro-ph.IM]}
  \BibitemShut {NoStop}%
\bibitem [{\citenamefont {Ayres}\ \emph {et~al.}(2007)\citenamefont {Ayres}
  \emph {et~al.}}]{NOvA:2007rmc}%
  \BibitemOpen
  \bibfield  {author} {\bibinfo {author} {\bibfnamefont {D.~S.}\ \bibnamefont
  {Ayres}} \emph {et~al.} (\bibinfo {collaboration} {NOvA}),\ }\bibfield
  {title} {\bibinfo {title} {{The NOvA Technical Design Report}}\ }\href
  {https://doi.org/10.2172/935497} {10.2172/935497} (\bibinfo {year}
  {2007})\BibitemShut {NoStop}%
\bibitem [{\citenamefont {Psihas}\ \emph {et~al.}(2020)\citenamefont {Psihas},
  \citenamefont {Groh}, \citenamefont {Tunnell},\ and\ \citenamefont
  {Warburton}}]{Psihas:2020pby}%
  \BibitemOpen
  \bibfield  {author} {\bibinfo {author} {\bibfnamefont {F.}~\bibnamefont
  {Psihas}}, \bibinfo {author} {\bibfnamefont {M.}~\bibnamefont {Groh}},
  \bibinfo {author} {\bibfnamefont {C.}~\bibnamefont {Tunnell}},\ and\ \bibinfo
  {author} {\bibfnamefont {K.}~\bibnamefont {Warburton}},\ }\bibfield  {title}
  {\bibinfo {title} {{A Review on Machine Learning for Neutrino Experiments}}\
  }\href {https://doi.org/10.1142/S0217751X20430058}
  {10.1142/S0217751X20430058} (\bibinfo {year} {2020}),\ \Eprint
  {https://arxiv.org/abs/2008.01242} {arXiv:2008.01242 [physics.comp-ph]}
  \BibitemShut {NoStop}%
\bibitem [{\citenamefont {Abi}\ \emph {et~al.}(2020{\natexlab{a}})\citenamefont
  {Abi} \emph {et~al.}}]{DUNE:2020lwj}%
  \BibitemOpen
  \bibfield  {author} {\bibinfo {author} {\bibfnamefont {B.}~\bibnamefont
  {Abi}} \emph {et~al.} (\bibinfo {collaboration} {DUNE}),\ }\bibfield  {title}
  {\bibinfo {title} {{Deep Underground Neutrino Experiment (DUNE), Far Detector
  Technical Design Report, Volume I Introduction to DUNE}},\ }\href
  {https://doi.org/10.1088/1748-0221/15/08/T08008} {\bibfield  {journal}
  {\bibinfo  {journal} {JINST}\ }\textbf {\bibinfo {volume} {15}}\bibfield
  {number} {\bibinfo  {number} { (08)},\ \bibinfo {pages} {T08008}},\ }\Eprint
  {https://arxiv.org/abs/2002.02967} {arXiv:2002.02967 [physics.ins-det]}
  \BibitemShut {NoStop}%
\bibitem [{\citenamefont {Acciarri}\ \emph
  {et~al.}(2017{\natexlab{a}})\citenamefont {Acciarri} \emph
  {et~al.}}]{MicroBooNE:2016pwy}%
  \BibitemOpen
  \bibfield  {author} {\bibinfo {author} {\bibfnamefont {R.}~\bibnamefont
  {Acciarri}} \emph {et~al.} (\bibinfo {collaboration} {MicroBooNE}),\
  }\bibfield  {title} {\bibinfo {title} {{Design and Construction of the
  MicroBooNE Detector}},\ }\href
  {https://doi.org/10.1088/1748-0221/12/02/P02017} {\bibfield  {journal}
  {\bibinfo  {journal} {JINST}\ }\textbf {\bibinfo {volume} {12}}\bibfield
  {number} {\bibinfo  {number} { (02)},\ \bibinfo {pages} {P02017}},\ }\Eprint
  {https://arxiv.org/abs/1612.05824} {arXiv:1612.05824 [physics.ins-det]}
  \BibitemShut {NoStop}%
\bibitem [{\citenamefont {Antonello}\ \emph {et~al.}(2015)\citenamefont
  {Antonello} \emph {et~al.}}]{MicroBooNE:2015bmn}%
  \BibitemOpen
  \bibfield  {author} {\bibinfo {author} {\bibfnamefont {M.}~\bibnamefont
  {Antonello}} \emph {et~al.} (\bibinfo {collaboration} {MicroBooNE, LAr1-ND,
  ICARUS-WA104}),\ }\bibfield  {title} {\bibinfo {title} {{A Proposal for a
  Three Detector Short-Baseline Neutrino Oscillation Program in the Fermilab
  Booster Neutrino Beam}},\ }\href@noop {} {\  (\bibinfo {year} {2015})},\
  \Eprint {https://arxiv.org/abs/1503.01520} {arXiv:1503.01520
  [physics.ins-det]} \BibitemShut {NoStop}%
\bibitem [{\citenamefont {Abi}\ \emph {et~al.}(2020{\natexlab{b}})\citenamefont
  {Abi} \emph {et~al.}}]{DUNE:2020gpm}%
  \BibitemOpen
  \bibfield  {author} {\bibinfo {author} {\bibfnamefont {B.}~\bibnamefont
  {Abi}} \emph {et~al.} (\bibinfo {collaboration} {DUNE}),\ }\bibfield  {title}
  {\bibinfo {title} {{Neutrino interaction classification with a convolutional
  neural network in the DUNE far detector}},\ }\href
  {https://doi.org/10.1103/PhysRevD.102.092003} {\bibfield  {journal} {\bibinfo
   {journal} {Phys. Rev. D}\ }\textbf {\bibinfo {volume} {102}},\ \bibinfo
  {pages} {092003} (\bibinfo {year} {2020}{\natexlab{b}})},\ \Eprint
  {https://arxiv.org/abs/2006.15052} {arXiv:2006.15052 [physics.ins-det]}
  \BibitemShut {NoStop}%
\bibitem [{\citenamefont {Liu}\ \emph {et~al.}(2020)\citenamefont {Liu},
  \citenamefont {Ott}, \citenamefont {Collado}, \citenamefont {Jargowsky},
  \citenamefont {Wu}, \citenamefont {Bian},\ and\ \citenamefont
  {Baldi}}]{Liu:2020pzv}%
  \BibitemOpen
  \bibfield  {author} {\bibinfo {author} {\bibfnamefont {J.}~\bibnamefont
  {Liu}}, \bibinfo {author} {\bibfnamefont {J.}~\bibnamefont {Ott}}, \bibinfo
  {author} {\bibfnamefont {J.}~\bibnamefont {Collado}}, \bibinfo {author}
  {\bibfnamefont {B.}~\bibnamefont {Jargowsky}}, \bibinfo {author}
  {\bibfnamefont {W.}~\bibnamefont {Wu}}, \bibinfo {author} {\bibfnamefont
  {J.}~\bibnamefont {Bian}},\ and\ \bibinfo {author} {\bibfnamefont
  {P.}~\bibnamefont {Baldi}},\ }\bibfield  {title} {\bibinfo {title}
  {{Deep-Learning-Based Kinematic Reconstruction for DUNE}},\ }\href@noop {} {\
   (\bibinfo {year} {2020})},\ \Eprint {https://arxiv.org/abs/2012.06181}
  {arXiv:2012.06181 [physics.ins-det]} \BibitemShut {NoStop}%
\bibitem [{\citenamefont {Acciarri}\ \emph
  {et~al.}(2017{\natexlab{b}})\citenamefont {Acciarri} \emph
  {et~al.}}]{MicroBooNE:2016dpb}%
  \BibitemOpen
  \bibfield  {author} {\bibinfo {author} {\bibfnamefont {R.}~\bibnamefont
  {Acciarri}} \emph {et~al.} (\bibinfo {collaboration} {MicroBooNE}),\
  }\bibfield  {title} {\bibinfo {title} {{Convolutional Neural Networks Applied
  to Neutrino Events in a Liquid Argon Time Projection Chamber}},\ }\href
  {https://doi.org/10.1088/1748-0221/12/03/P03011} {\bibfield  {journal}
  {\bibinfo  {journal} {JINST}\ }\textbf {\bibinfo {volume} {12}}\bibfield
  {number} {\bibinfo  {number} { (03)},\ \bibinfo {pages} {P03011}},\ }\Eprint
  {https://arxiv.org/abs/1611.05531} {arXiv:1611.05531 [physics.ins-det]}
  \BibitemShut {NoStop}%
\bibitem [{\citenamefont {Adams}\ \emph {et~al.}(2019)\citenamefont {Adams}
  \emph {et~al.}}]{MicroBooNE:2018kka}%
  \BibitemOpen
  \bibfield  {author} {\bibinfo {author} {\bibfnamefont {C.}~\bibnamefont
  {Adams}} \emph {et~al.} (\bibinfo {collaboration} {MicroBooNE}),\ }\bibfield
  {title} {\bibinfo {title} {{Deep neural network for pixel-level
  electromagnetic particle identification in the MicroBooNE liquid argon time
  projection chamber}},\ }\href {https://doi.org/10.1103/PhysRevD.99.092001}
  {\bibfield  {journal} {\bibinfo  {journal} {Phys. Rev. D}\ }\textbf {\bibinfo
  {volume} {99}},\ \bibinfo {pages} {092001} (\bibinfo {year} {2019})},\
  \Eprint {https://arxiv.org/abs/1808.07269} {arXiv:1808.07269 [hep-ex]}
  \BibitemShut {NoStop}%
\bibitem [{\citenamefont {Abratenko}\ \emph
  {et~al.}(2021{\natexlab{d}})\citenamefont {Abratenko} \emph
  {et~al.}}]{MicroBooNE:2020yze}%
  \BibitemOpen
  \bibfield  {author} {\bibinfo {author} {\bibfnamefont {P.}~\bibnamefont
  {Abratenko}} \emph {et~al.} (\bibinfo {collaboration} {MicroBooNE}),\
  }\bibfield  {title} {\bibinfo {title} {{Semantic segmentation with a sparse
  convolutional neural network for event reconstruction in MicroBooNE}},\
  }\href {https://doi.org/10.1103/PhysRevD.103.052012} {\bibfield  {journal}
  {\bibinfo  {journal} {Phys. Rev. D}\ }\textbf {\bibinfo {volume} {103}},\
  \bibinfo {pages} {052012} (\bibinfo {year} {2021}{\natexlab{d}})},\ \Eprint
  {https://arxiv.org/abs/2012.08513} {arXiv:2012.08513 [physics.ins-det]}
  \BibitemShut {NoStop}%
\bibitem [{\citenamefont {Domin\'e}\ and\ \citenamefont
  {Terao}(2020)}]{Domine:2019zhm}%
  \BibitemOpen
  \bibfield  {author} {\bibinfo {author} {\bibfnamefont {L.}~\bibnamefont
  {Domin\'e}}\ and\ \bibinfo {author} {\bibfnamefont {K.}~\bibnamefont
  {Terao}},\ }\bibfield  {title} {\bibinfo {title} {{Scalable deep
  convolutional neural networks for sparse, locally dense liquid argon time
  projection chamber data}},\ }\href
  {https://doi.org/10.1103/PhysRevD.102.012005} {\bibfield  {journal} {\bibinfo
   {journal} {Phys. Rev. D}\ }\textbf {\bibinfo {volume} {102}},\ \bibinfo
  {pages} {012005} (\bibinfo {year} {2020})},\ \Eprint
  {https://arxiv.org/abs/1903.05663} {arXiv:1903.05663 [hep-ex]} \BibitemShut
  {NoStop}%
\bibitem [{\citenamefont {Acciarri}\ \emph {et~al.}(2020)\citenamefont
  {Acciarri} \emph {et~al.}}]{SBND:2020eho}%
  \BibitemOpen
  \bibfield  {author} {\bibinfo {author} {\bibfnamefont {R.}~\bibnamefont
  {Acciarri}} \emph {et~al.} (\bibinfo {collaboration} {SBND}),\ }\bibfield
  {title} {\bibinfo {title} {{Cosmic Background Removal with Deep Neural
  Networks in SBND}},\ }\href@noop {} {\  (\bibinfo {year} {2020})},\ \Eprint
  {https://arxiv.org/abs/2012.01301} {arXiv:2012.01301 [physics.data-an]}
  \BibitemShut {NoStop}%
\bibitem [{\citenamefont {Drielsma}\ \emph
  {et~al.}(2021{\natexlab{a}})\citenamefont {Drielsma}, \citenamefont {Terao},
  \citenamefont {Domin\'e},\ and\ \citenamefont {Koh}}]{Drielsma:2021jdv}%
  \BibitemOpen
  \bibfield  {author} {\bibinfo {author} {\bibfnamefont {F.}~\bibnamefont
  {Drielsma}}, \bibinfo {author} {\bibfnamefont {K.}~\bibnamefont {Terao}},
  \bibinfo {author} {\bibfnamefont {L.}~\bibnamefont {Domin\'e}},\ and\
  \bibinfo {author} {\bibfnamefont {D.~H.}\ \bibnamefont {Koh}},\ }\bibfield
  {title} {\bibinfo {title} {{Scalable, End-to-End, Deep-Learning-Based Data
  Reconstruction Chain for Particle Imaging Detectors}},\ }in\ \href@noop {}
  {\emph {\bibinfo {booktitle} {{34th Conference on Neural Information
  Processing Systems}}}}\ (\bibinfo {year} {2021})\ \Eprint
  {https://arxiv.org/abs/2102.01033} {arXiv:2102.01033 [hep-ex]} \BibitemShut
  {NoStop}%
\bibitem [{\citenamefont {Domin{\'e}}\ and\ \citenamefont
  {Terao}(2020)}]{Domine:2020tlx}%
  \BibitemOpen
  \bibfield  {author} {\bibinfo {author} {\bibfnamefont {L.}~\bibnamefont
  {Domin{\'e}}}\ and\ \bibinfo {author} {\bibfnamefont {K.}~\bibnamefont
  {Terao}},\ }\bibfield  {title} {\bibinfo {title} {{Point Proposal Network for
  Reconstructing 3D Particle Positions with Sub-Pixel Precision in Liquid Argon
  Time Projection Chambers}},\ }\href@noop {} {\  (\bibinfo {year} {2020})},\
  \Eprint {https://arxiv.org/abs/2006.14745} {arXiv:2006.14745 [hep-ex]}
  \BibitemShut {NoStop}%
\bibitem [{\citenamefont {Koh}\ \emph {et~al.}(2020)\citenamefont {Koh},
  \citenamefont {de~Soux}, \citenamefont {Domine}, \citenamefont {Drielsma},
  \citenamefont {Itay}, \citenamefont {Lin}, \citenamefont {Terao},
  \citenamefont {Tsang},\ and\ \citenamefont {Usher}}]{Koh:2020snv}%
  \BibitemOpen
  \bibfield  {author} {\bibinfo {author} {\bibfnamefont {D.~H.}\ \bibnamefont
  {Koh}}, \bibinfo {author} {\bibfnamefont {P.~C.}\ \bibnamefont {de~Soux}},
  \bibinfo {author} {\bibfnamefont {L.}~\bibnamefont {Domine}}, \bibinfo
  {author} {\bibfnamefont {F.}~\bibnamefont {Drielsma}}, \bibinfo {author}
  {\bibfnamefont {R.}~\bibnamefont {Itay}}, \bibinfo {author} {\bibfnamefont
  {Q.}~\bibnamefont {Lin}}, \bibinfo {author} {\bibfnamefont {K.}~\bibnamefont
  {Terao}}, \bibinfo {author} {\bibfnamefont {K.~V.}\ \bibnamefont {Tsang}},\
  and\ \bibinfo {author} {\bibfnamefont {T.}~\bibnamefont {Usher}} (\bibinfo
  {collaboration} {DeepLearnPhysics}),\ }\bibfield  {title} {\bibinfo {title}
  {Scalable, proposal-free instance segmentation network for 3d pixel
  clustering and particle trajectory reconstruction in liquid argon time
  projection chambers},\ }\href@noop {} {\  (\bibinfo {year} {2020})},\ \Eprint
  {https://arxiv.org/abs/2007.03083} {arXiv:2007.03083 [physics]} \BibitemShut
  {NoStop}%
\bibitem [{\citenamefont {Drielsma}\ \emph
  {et~al.}(2021{\natexlab{b}})\citenamefont {Drielsma}, \citenamefont {Lin},
  \citenamefont {de~Soux}, \citenamefont {Domin\'e}, \citenamefont {Itay},
  \citenamefont {Koh}, \citenamefont {Nelson}, \citenamefont {Terao},
  \citenamefont {Tsang},\ and\ \citenamefont
  {Usher}}]{DeepLearnPhysics:2020hut}%
  \BibitemOpen
  \bibfield  {author} {\bibinfo {author} {\bibfnamefont {F.}~\bibnamefont
  {Drielsma}}, \bibinfo {author} {\bibfnamefont {Q.}~\bibnamefont {Lin}},
  \bibinfo {author} {\bibfnamefont {P.~C.}\ \bibnamefont {de~Soux}}, \bibinfo
  {author} {\bibfnamefont {L.}~\bibnamefont {Domin\'e}}, \bibinfo {author}
  {\bibfnamefont {R.}~\bibnamefont {Itay}}, \bibinfo {author} {\bibfnamefont
  {D.~H.}\ \bibnamefont {Koh}}, \bibinfo {author} {\bibfnamefont {B.~J.}\
  \bibnamefont {Nelson}}, \bibinfo {author} {\bibfnamefont {K.}~\bibnamefont
  {Terao}}, \bibinfo {author} {\bibfnamefont {K.~V.}\ \bibnamefont {Tsang}},\
  and\ \bibinfo {author} {\bibfnamefont {T.~L.}\ \bibnamefont {Usher}}
  (\bibinfo {collaboration} {DeepLearnPhysics}),\ }\bibfield  {title} {\bibinfo
  {title} {{Clustering of electromagnetic showers and particle interactions
  with graph neural networks in liquid argon time projection chambers}},\
  }\href {https://doi.org/10.1103/PhysRevD.104.072004} {\bibfield  {journal}
  {\bibinfo  {journal} {Phys. Rev. D}\ }\textbf {\bibinfo {volume} {104}},\
  \bibinfo {pages} {072004} (\bibinfo {year} {2021}{\natexlab{b}})},\ \Eprint
  {https://arxiv.org/abs/2007.01335} {arXiv:2007.01335 [physics.ins-det]}
  \BibitemShut {NoStop}%
\bibitem [{\citenamefont {Adams}\ \emph {et~al.}(2020)\citenamefont {Adams},
  \citenamefont {Terao},\ and\ \citenamefont {Wongjirad}}]{Adams:2020vlj}%
  \BibitemOpen
  \bibfield  {author} {\bibinfo {author} {\bibfnamefont {C.}~\bibnamefont
  {Adams}}, \bibinfo {author} {\bibfnamefont {K.}~\bibnamefont {Terao}},\ and\
  \bibinfo {author} {\bibfnamefont {T.}~\bibnamefont {Wongjirad}},\ }\bibfield
  {title} {\bibinfo {title} {{PILArNet: Public Dataset for Particle Imaging
  Liquid Argon Detectors in High Energy Physics}},\ }\href@noop {} {\
  (\bibinfo {year} {2020})},\ \Eprint {https://arxiv.org/abs/2006.01993}
  {arXiv:2006.01993 [physics.ins-det]} \BibitemShut {NoStop}%
\bibitem [{\citenamefont {Psihas}(2017)}]{Psihas:2017yuc}%
  \BibitemOpen
  \bibfield  {author} {\bibinfo {author} {\bibfnamefont {F.}~\bibnamefont
  {Psihas}} (\bibinfo {collaboration} {NOvA}),\ }\bibfield  {title} {\bibinfo
  {title} {{The Convolutional Visual Network for Identification and
  Reconstruction of NOvA Events}},\ }\href
  {https://doi.org/10.1088/1742-6596/898/7/072053} {\bibfield  {journal}
  {\bibinfo  {journal} {J. Phys. Conf. Ser.}\ }\textbf {\bibinfo {volume}
  {898}},\ \bibinfo {pages} {072053} (\bibinfo {year} {2017})}\BibitemShut
  {NoStop}%
\bibitem [{\citenamefont {Gando}\ \emph {et~al.}(2016)\citenamefont {Gando}
  \emph {et~al.}}]{KamLAND-Zen:2016pfg}%
  \BibitemOpen
  \bibfield  {author} {\bibinfo {author} {\bibfnamefont {A.}~\bibnamefont
  {Gando}} \emph {et~al.} (\bibinfo {collaboration} {KamLAND-Zen}),\ }\bibfield
   {title} {\bibinfo {title} {{Search for Majorana Neutrinos near the Inverted
  Mass Hierarchy Region with KamLAND-Zen}},\ }\href
  {https://doi.org/10.1103/PhysRevLett.117.082503} {\bibfield  {journal}
  {\bibinfo  {journal} {Phys. Rev. Lett.}\ }\textbf {\bibinfo {volume} {117}},\
  \bibinfo {pages} {082503} (\bibinfo {year} {2016})},\ \bibinfo {note}
  {[Addendum: Phys.Rev.Lett. 117, 109903 (2016)]},\ \Eprint
  {https://arxiv.org/abs/1605.02889} {arXiv:1605.02889 [hep-ex]} \BibitemShut
  {NoStop}%
\bibitem [{\citenamefont {Racah}\ \emph {et~al.}(2016)\citenamefont {Racah},
  \citenamefont {Ko}, \citenamefont {Sadowski}, \citenamefont {Bhimji},
  \citenamefont {Tull}, \citenamefont {Oh}, \citenamefont {Baldi},\ and\
  \citenamefont {Prabhat}}]{Racah:2016gnm}%
  \BibitemOpen
  \bibfield  {author} {\bibinfo {author} {\bibfnamefont {E.}~\bibnamefont
  {Racah}}, \bibinfo {author} {\bibfnamefont {S.}~\bibnamefont {Ko}}, \bibinfo
  {author} {\bibfnamefont {P.}~\bibnamefont {Sadowski}}, \bibinfo {author}
  {\bibfnamefont {W.}~\bibnamefont {Bhimji}}, \bibinfo {author} {\bibfnamefont
  {C.}~\bibnamefont {Tull}}, \bibinfo {author} {\bibfnamefont {S.-Y.}\
  \bibnamefont {Oh}}, \bibinfo {author} {\bibfnamefont {P.}~\bibnamefont
  {Baldi}},\ and\ \bibinfo {author} {\bibnamefont {Prabhat}},\ }\bibfield
  {title} {\bibinfo {title} {{Revealing Fundamental Physics from the Daya Bay
  Neutrino Experiment using Deep Neural Networks}},\ }\href@noop {} {\
  (\bibinfo {year} {2016})},\ \Eprint {https://arxiv.org/abs/1601.07621}
  {arXiv:1601.07621 [stat.ML]} \BibitemShut {NoStop}%
\bibitem [{\citenamefont {Choma}\ \emph
  {et~al.}(2018{\natexlab{b}})\citenamefont {Choma}, \citenamefont {Monti},
  \citenamefont {Gerhardt}, \citenamefont {Palczewski}, \citenamefont
  {Ronaghi}, \citenamefont {Prabhat}, \citenamefont {Bhimji}, \citenamefont
  {Bronstein}, \citenamefont {Klein},\ and\ \citenamefont
  {Bruna}}]{IceCube:2018gms}%
  \BibitemOpen
  \bibfield  {author} {\bibinfo {author} {\bibfnamefont {N.}~\bibnamefont
  {Choma}}, \bibinfo {author} {\bibfnamefont {F.}~\bibnamefont {Monti}},
  \bibinfo {author} {\bibfnamefont {L.}~\bibnamefont {Gerhardt}}, \bibinfo
  {author} {\bibfnamefont {T.}~\bibnamefont {Palczewski}}, \bibinfo {author}
  {\bibfnamefont {Z.}~\bibnamefont {Ronaghi}}, \bibinfo {author} {\bibnamefont
  {Prabhat}}, \bibinfo {author} {\bibfnamefont {W.}~\bibnamefont {Bhimji}},
  \bibinfo {author} {\bibfnamefont {M.~M.}\ \bibnamefont {Bronstein}}, \bibinfo
  {author} {\bibfnamefont {S.~R.}\ \bibnamefont {Klein}},\ and\ \bibinfo
  {author} {\bibfnamefont {J.}~\bibnamefont {Bruna}} (\bibinfo {collaboration}
  {IceCube}),\ }\bibfield  {title} {\bibinfo {title} {{Graph Neural Networks
  for IceCube Signal Classification}},\ }\href@noop {} {\  (\bibinfo {year}
  {2018}{\natexlab{b}})},\ \Eprint {https://arxiv.org/abs/1809.06166}
  {arXiv:1809.06166 [cs.LG]} \BibitemShut {NoStop}%
\bibitem [{\citenamefont {Abbasi}\ \emph {et~al.}(2021)\citenamefont {Abbasi}
  \emph {et~al.}}]{IceCube:2021dvc}%
  \BibitemOpen
  \bibfield  {author} {\bibinfo {author} {\bibfnamefont {R.}~\bibnamefont
  {Abbasi}} \emph {et~al.} (\bibinfo {collaboration} {IceCube}),\ }\bibfield
  {title} {\bibinfo {title} {{Reconstruction of Neutrino Events in IceCube
  using Graph Neural Networks}},\ }\href {https://doi.org/10.22323/1.395.1044}
  {\bibfield  {journal} {\bibinfo  {journal} {PoS}\ }\textbf {\bibinfo {volume}
  {ICRC2021}},\ \bibinfo {pages} {1044} (\bibinfo {year} {2021})},\ \Eprint
  {https://arxiv.org/abs/2107.12187} {arXiv:2107.12187 [astro-ph.IM]}
  \BibitemShut {NoStop}%
\bibitem [{\citenamefont {Abi}\ \emph {et~al.}(2020{\natexlab{c}})\citenamefont
  {Abi} \emph {et~al.}}]{DUNE:2020mra}%
  \BibitemOpen
  \bibfield  {author} {\bibinfo {author} {\bibfnamefont {B.}~\bibnamefont
  {Abi}} \emph {et~al.} (\bibinfo {collaboration} {DUNE}),\ }\bibfield  {title}
  {\bibinfo {title} {{Deep Underground Neutrino Experiment (DUNE), Far Detector
  Technical Design Report, Volume III: DUNE Far Detector Technical
  Coordination}},\ }\href {https://doi.org/10.1088/1748-0221/15/08/T08009}
  {\bibfield  {journal} {\bibinfo  {journal} {JINST}\ }\textbf {\bibinfo
  {volume} {15}}\bibfield  {number} {\bibinfo  {number} { (08)},\ \bibinfo
  {pages} {T08009}},\ }\Eprint {https://arxiv.org/abs/2002.03008}
  {arXiv:2002.03008 [physics.ins-det]} \BibitemShut {NoStop}%
\bibitem [{\citenamefont {Wang}\ \emph {et~al.}(2020)\citenamefont {Wang},
  \citenamefont {Yang}, \citenamefont {Acosta~Flechas}, \citenamefont {Harris},
  \citenamefont {Hawks}, \citenamefont {Holzman}, \citenamefont {Knoepfel},
  \citenamefont {Krupa}, \citenamefont {Pedro},\ and\ \citenamefont
  {Tran}}]{Wang:2020fjr}%
  \BibitemOpen
  \bibfield  {author} {\bibinfo {author} {\bibfnamefont {M.}~\bibnamefont
  {Wang}}, \bibinfo {author} {\bibfnamefont {T.}~\bibnamefont {Yang}}, \bibinfo
  {author} {\bibfnamefont {M.}~\bibnamefont {Acosta~Flechas}}, \bibinfo
  {author} {\bibfnamefont {P.}~\bibnamefont {Harris}}, \bibinfo {author}
  {\bibfnamefont {B.}~\bibnamefont {Hawks}}, \bibinfo {author} {\bibfnamefont
  {B.}~\bibnamefont {Holzman}}, \bibinfo {author} {\bibfnamefont
  {K.}~\bibnamefont {Knoepfel}}, \bibinfo {author} {\bibfnamefont
  {J.}~\bibnamefont {Krupa}}, \bibinfo {author} {\bibfnamefont
  {K.}~\bibnamefont {Pedro}},\ and\ \bibinfo {author} {\bibfnamefont
  {N.}~\bibnamefont {Tran}},\ }\bibfield  {title} {\bibinfo {title}
  {{GPU-accelerated machine learning inference as a service for computing in
  neutrino experiments}}\ }\href {https://doi.org/10.3389/fdata.2020.604083}
  {10.3389/fdata.2020.604083} (\bibinfo {year} {2020}),\ \Eprint
  {https://arxiv.org/abs/2009.04509} {arXiv:2009.04509 [physics.comp-ph]}
  \BibitemShut {NoStop}%
\bibitem [{\citenamefont {Akerib}\ \emph
  {et~al.}(2021{\natexlab{a}})\citenamefont {Akerib} \emph {et~al.}}]{LUX-S2}%
  \BibitemOpen
  \bibfield  {author} {\bibinfo {author} {\bibfnamefont {D.~S.}\ \bibnamefont
  {Akerib}} \emph {et~al.} (\bibinfo {collaboration} {LUX}),\ }\bibfield
  {title} {\bibinfo {title} {{Improving sensitivity to low-mass dark matter in
  LUX using a novel electrode background mitigation technique}},\ }\href
  {https://doi.org/10.1103/PhysRevD.104.012011} {\bibfield  {journal} {\bibinfo
   {journal} {Phys. Rev. D}\ }\textbf {\bibinfo {volume} {104}},\ \bibinfo
  {pages} {012011} (\bibinfo {year} {2021}{\natexlab{a}})},\ \Eprint
  {https://arxiv.org/abs/2011.09602} {arXiv:2011.09602 [hep-ex]} \BibitemShut
  {NoStop}%
\bibitem [{\citenamefont {Akerib}\ \emph
  {et~al.}(2021{\natexlab{b}})\citenamefont {Akerib} \emph
  {et~al.}}]{LUX-EFT-Run4}%
  \BibitemOpen
  \bibfield  {author} {\bibinfo {author} {\bibfnamefont {D.~S.}\ \bibnamefont
  {Akerib}} \emph {et~al.} (\bibinfo {collaboration} {LUX}),\ }\bibfield
  {title} {\bibinfo {title} {{Constraints on Effective Field Theory Couplings
  Using 311.2 days of LUX Data}},\ }\href@noop {} {\bibfield  {journal}
  {\bibinfo  {journal} {arXiv}\ } (\bibinfo {year} {2021}{\natexlab{b}})},\
  \Eprint {https://arxiv.org/abs/2102.06998} {arXiv:2102.06998 [astro-ph.CO]}
  \BibitemShut {NoStop}%
\bibitem [{\citenamefont {Rossiter}(2021)}]{Rossiter-thesis}%
  \BibitemOpen
  \bibfield  {author} {\bibinfo {author} {\bibfnamefont {P.}~\bibnamefont
  {Rossiter}},\ }\emph {\bibinfo {title} {{Background Mitigation in Dual Phase
  Xenon Time Projection Chambers}}},\ \href@noop {} {Ph.D. thesis},\ \bibinfo
  {school} {Sheffield U.} (\bibinfo {year} {2021})\BibitemShut {NoStop}%
\bibitem [{\citenamefont {Aprile}\ \emph {et~al.}(2021)\citenamefont {Aprile}
  \emph {et~al.}}]{XENON1T-nu}%
  \BibitemOpen
  \bibfield  {author} {\bibinfo {author} {\bibfnamefont {E.}~\bibnamefont
  {Aprile}} \emph {et~al.} (\bibinfo {collaboration} {XENON}),\ }\bibfield
  {title} {\bibinfo {title} {{Search for Coherent Elastic Scattering of Solar
  $^8$B Neutrinos in the XENON1T Dark Matter Experiment}},\ }\href
  {https://doi.org/10.1103/PhysRevLett.126.091301} {\bibfield  {journal}
  {\bibinfo  {journal} {Phys. Rev. Lett.}\ }\textbf {\bibinfo {volume} {126}},\
  \bibinfo {pages} {091301} (\bibinfo {year} {2021})},\ \Eprint
  {https://arxiv.org/abs/2012.02846} {arXiv:2012.02846 [hep-ex]} \BibitemShut
  {NoStop}%
\bibitem [{\citenamefont {Agnese}\ \emph {et~al.}(2019)\citenamefont {Agnese}
  \emph {et~al.}}]{SuperCDMS-2018}%
  \BibitemOpen
  \bibfield  {author} {\bibinfo {author} {\bibfnamefont {R.}~\bibnamefont
  {Agnese}} \emph {et~al.} (\bibinfo {collaboration} {SuperCDMS}),\ }\bibfield
  {title} {\bibinfo {title} {{Search for Low-Mass Dark Matter with CDMSlite
  Using a Profile Likelihood Fit}},\ }\href
  {https://doi.org/10.1103/PhysRevD.99.062001} {\bibfield  {journal} {\bibinfo
  {journal} {Phys. Rev. D}\ }\textbf {\bibinfo {volume} {99}},\ \bibinfo
  {pages} {062001} (\bibinfo {year} {2019})},\ \Eprint
  {https://arxiv.org/abs/1808.09098} {arXiv:1808.09098 [astro-ph.CO]}
  \BibitemShut {NoStop}%
\bibitem [{\citenamefont {Adhikari}\ \emph
  {et~al.}(2021{\natexlab{a}})\citenamefont {Adhikari} \emph
  {et~al.}}]{COSINE-100-threshold}%
  \BibitemOpen
  \bibfield  {author} {\bibinfo {author} {\bibfnamefont {G.}~\bibnamefont
  {Adhikari}} \emph {et~al.} (\bibinfo {collaboration} {COSINE-100}),\
  }\bibfield  {title} {\bibinfo {title} {{Lowering the energy threshold in
  COSINE-100 dark matter searches}},\ }\href
  {https://doi.org/10.1016/j.astropartphys.2021.102581} {\bibfield  {journal}
  {\bibinfo  {journal} {Astropart. Phys.}\ }\textbf {\bibinfo {volume} {130}},\
  \bibinfo {pages} {102581} (\bibinfo {year} {2021}{\natexlab{a}})},\ \Eprint
  {https://arxiv.org/abs/2005.13784} {arXiv:2005.13784 [physics.ins-det]}
  \BibitemShut {NoStop}%
\bibitem [{\citenamefont {Albert}\ \emph {et~al.}(2016)\citenamefont {Albert}
  \emph {et~al.}}]{EXO-excited}%
  \BibitemOpen
  \bibfield  {author} {\bibinfo {author} {\bibfnamefont {J.~B.}\ \bibnamefont
  {Albert}} \emph {et~al.} (\bibinfo {collaboration} {EXO-200}),\ }\bibfield
  {title} {\bibinfo {title} {{Search for $2\nu\beta\beta$ decay of $^{136}$Xe
  to the 0$_1^+$ excited state of $^{136}$Ba with EXO-200}},\ }\href
  {https://doi.org/10.1103/PhysRevC.93.035501} {\bibfield  {journal} {\bibinfo
  {journal} {Phys. Rev. C}\ }\textbf {\bibinfo {volume} {93}},\ \bibinfo
  {pages} {035501} (\bibinfo {year} {2016})},\ \Eprint
  {https://arxiv.org/abs/1511.04770} {arXiv:1511.04770 [nucl-ex]} \BibitemShut
  {NoStop}%
\bibitem [{\citenamefont {Albert}\ \emph {et~al.}(2018)\citenamefont {Albert}
  \emph {et~al.}}]{EXO-upgraded}%
  \BibitemOpen
  \bibfield  {author} {\bibinfo {author} {\bibfnamefont {J.~B.}\ \bibnamefont
  {Albert}} \emph {et~al.} (\bibinfo {collaboration} {EXO}),\ }\bibfield
  {title} {\bibinfo {title} {{Search for Neutrinoless Double-Beta Decay with
  the Upgraded EXO-200 Detector}},\ }\href
  {https://doi.org/10.1103/PhysRevLett.120.072701} {\bibfield  {journal}
  {\bibinfo  {journal} {Phys. Rev. Lett.}\ }\textbf {\bibinfo {volume} {120}},\
  \bibinfo {pages} {072701} (\bibinfo {year} {2018})},\ \Eprint
  {https://arxiv.org/abs/1707.08707} {arXiv:1707.08707 [hep-ex]} \BibitemShut
  {NoStop}%
\bibitem [{\citenamefont {Anton}\ \emph {et~al.}(2019)\citenamefont {Anton}
  \emph {et~al.}}]{EXO-complete}%
  \BibitemOpen
  \bibfield  {author} {\bibinfo {author} {\bibfnamefont {G.}~\bibnamefont
  {Anton}} \emph {et~al.} (\bibinfo {collaboration} {EXO-200}),\ }\bibfield
  {title} {\bibinfo {title} {{Search for Neutrinoless Double-$\beta$ Decay with
  the Complete EXO-200 Dataset}},\ }\href
  {https://doi.org/10.1103/PhysRevLett.123.161802} {\bibfield  {journal}
  {\bibinfo  {journal} {Phys. Rev. Lett.}\ }\textbf {\bibinfo {volume} {123}},\
  \bibinfo {pages} {161802} (\bibinfo {year} {2019})},\ \Eprint
  {https://arxiv.org/abs/1906.02723} {arXiv:1906.02723 [hep-ex]} \BibitemShut
  {NoStop}%
\bibitem [{\citenamefont {Yu}(2019)}]{Yu-PCA}%
  \BibitemOpen
  \bibfield  {author} {\bibinfo {author} {\bibfnamefont {T.~C.}\ \bibnamefont
  {Yu}},\ }\bibfield  {title} {\bibinfo {title} {{Template-free Pulse Height
  Estimation of Microcalorimeter Responses with PCA}},\ }\href@noop {}
  {\bibfield  {journal} {\bibinfo  {journal} {arXiV}\ } (\bibinfo {year}
  {2019})},\ \Eprint {https://arxiv.org/abs/1910.14261} {arXiv:1910.14261
  [physics.data-an]} \BibitemShut {NoStop}%
\bibitem [{\citenamefont {Wagner}(2020)}]{wagner2020-thesis}%
  \BibitemOpen
  \bibfield  {author} {\bibinfo {author} {\bibfnamefont {F.}~\bibnamefont
  {Wagner}},\ }\emph {\bibinfo {title} {Machine learning methods for the raw
  data analysis of cryogenic dark matter experiments}},\ \href
  {https://doi.org/10.34726/hss.2020.77322} {Ph.D. thesis},\ \bibinfo  {school}
  {Wien} (\bibinfo {year} {2020})\BibitemShut {NoStop}%
\bibitem [{\citenamefont {Holl}\ \emph {et~al.}(2019)\citenamefont {Holl},
  \citenamefont {Hauertmann}, \citenamefont {Majorovits}, \citenamefont
  {Schulz}, \citenamefont {Schuster},\ and\ \citenamefont
  {Zsigmond}}]{Ge-pulse-shape}%
  \BibitemOpen
  \bibfield  {author} {\bibinfo {author} {\bibfnamefont {P.}~\bibnamefont
  {Holl}}, \bibinfo {author} {\bibfnamefont {L.}~\bibnamefont {Hauertmann}},
  \bibinfo {author} {\bibfnamefont {B.}~\bibnamefont {Majorovits}}, \bibinfo
  {author} {\bibfnamefont {O.}~\bibnamefont {Schulz}}, \bibinfo {author}
  {\bibfnamefont {M.}~\bibnamefont {Schuster}},\ and\ \bibinfo {author}
  {\bibfnamefont {A.~J.}\ \bibnamefont {Zsigmond}},\ }\bibfield  {title}
  {\bibinfo {title} {{Deep learning based pulse shape discrimination for
  germanium detectors}},\ }\href
  {https://doi.org/10.1140/epjc/s10052-019-6869-2} {\bibfield  {journal}
  {\bibinfo  {journal} {Eur. Phys. J. C}\ }\textbf {\bibinfo {volume} {79}},\
  \bibinfo {pages} {450} (\bibinfo {year} {2019})},\ \Eprint
  {https://arxiv.org/abs/1903.01462} {arXiv:1903.01462 [physics.ins-det]}
  \BibitemShut {NoStop}%
\bibitem [{\citenamefont {Matusch}\ \emph {et~al.}(2018)\citenamefont {Matusch}
  \emph {et~al.}}]{PICO:2018fbf}%
  \BibitemOpen
  \bibfield  {author} {\bibinfo {author} {\bibfnamefont {B.}~\bibnamefont
  {Matusch}} \emph {et~al.} (\bibinfo {collaboration} {PICO}),\ }\bibfield
  {title} {\bibinfo {title} {{Developing a Bubble Chamber Particle
  Discriminator Using Semi-Supervised Learning}},\ }\href@noop {} {\  (\bibinfo
  {year} {2018})},\ \Eprint {https://arxiv.org/abs/1811.11308}
  {arXiv:1811.11308 [physics.comp-ph]} \BibitemShut {NoStop}%
\bibitem [{\citenamefont {Matusch}\ and\ \citenamefont
  {Cao}(2020)}]{PICO-semi-supervised}%
  \BibitemOpen
  \bibfield  {author} {\bibinfo {author} {\bibfnamefont {B.}~\bibnamefont
  {Matusch}}\ and\ \bibinfo {author} {\bibfnamefont {G.}~\bibnamefont {Cao}},\
  }\bibfield  {title} {\bibinfo {title} {{Particle identification using
  semi-supervised learning in the PICO-60 dark matter detector}},\ }\href
  {https://doi.org/10.1088/1742-6596/1525/1/012085} {\bibfield  {journal}
  {\bibinfo  {journal} {J. Phys. Conf. Ser.}\ }\textbf {\bibinfo {volume}
  {1525}},\ \bibinfo {pages} {012085} (\bibinfo {year} {2020})}\BibitemShut
  {NoStop}%
\bibitem [{\citenamefont {M{\"u}hlmann}(2019)}]{muhlmann2019-thesis}%
  \BibitemOpen
  \bibfield  {author} {\bibinfo {author} {\bibfnamefont {C.}~\bibnamefont
  {M{\"u}hlmann}},\ }\emph {\bibinfo {title} {Pulse-shape dicrimination with
  deep learning in CRESST}},\ \href
  {https://resolver.obvsg.at/urn:nbn:at:at-ubtuw:1-122102} {Ph.D. thesis},\
  \bibinfo  {school} {Wien} (\bibinfo {year} {2019})\BibitemShut {NoStop}%
\bibitem [{\citenamefont {Ai}\ \emph {et~al.}(2018)\citenamefont {Ai},
  \citenamefont {Wang}, \citenamefont {Huang},\ and\ \citenamefont
  {Sun}}]{Ai-gas-TPC-2018}%
  \BibitemOpen
  \bibfield  {author} {\bibinfo {author} {\bibfnamefont {P.}~\bibnamefont
  {Ai}}, \bibinfo {author} {\bibfnamefont {D.}~\bibnamefont {Wang}}, \bibinfo
  {author} {\bibfnamefont {G.}~\bibnamefont {Huang}},\ and\ \bibinfo {author}
  {\bibfnamefont {X.}~\bibnamefont {Sun}},\ }\bibfield  {title} {\bibinfo
  {title} {{Three-dimensional convolutional neural networks for neutrinoless
  double-beta decay signal/background discrimination in high-pressure gaseous
  Time Projection Chamber}},\ }\href
  {https://doi.org/10.1088/1748-0221/13/08/P08015} {\bibfield  {journal}
  {\bibinfo  {journal} {JINST}\ }\textbf {\bibinfo {volume} {13}}\bibfield
  {number} {\bibinfo  {number} { (08)},\ \bibinfo {pages} {P08015}},\ }\Eprint
  {https://arxiv.org/abs/1803.01482} {arXiv:1803.01482 [physics.data-an]}
  \BibitemShut {NoStop}%
\bibitem [{\citenamefont {Renner}\ \emph {et~al.}(2017)\citenamefont {Renner}
  \emph {et~al.}}]{NEXT-2017}%
  \BibitemOpen
  \bibfield  {author} {\bibinfo {author} {\bibfnamefont {J.}~\bibnamefont
  {Renner}} \emph {et~al.} (\bibinfo {collaboration} {NEXT}),\ }\bibfield
  {title} {\bibinfo {title} {{Background rejection in NEXT using deep neural
  networks}},\ }\href {https://doi.org/10.1088/1748-0221/12/01/T01004}
  {\bibfield  {journal} {\bibinfo  {journal} {JINST}\ }\textbf {\bibinfo
  {volume} {12}}\bibfield  {number} {\bibinfo  {number} { (01)},\ \bibinfo
  {pages} {T01004}},\ }\Eprint {https://arxiv.org/abs/1609.06202}
  {arXiv:1609.06202 [physics.ins-det]} \BibitemShut {NoStop}%
\bibitem [{\citenamefont {Kekic}\ \emph {et~al.}(2021)\citenamefont {Kekic}
  \emph {et~al.}}]{NEXT-2021}%
  \BibitemOpen
  \bibfield  {author} {\bibinfo {author} {\bibfnamefont {M.}~\bibnamefont
  {Kekic}} \emph {et~al.} (\bibinfo {collaboration} {NEXT}),\ }\bibfield
  {title} {\bibinfo {title} {{Demonstration of background rejection using deep
  convolutional neural networks in the NEXT experiment}},\ }\href
  {https://doi.org/10.1007/JHEP01(2021)189} {\bibfield  {journal} {\bibinfo
  {journal} {JHEP}\ }\textbf {\bibinfo {volume} {01}},\ \bibinfo {pages}
  {189}},\ \Eprint {https://arxiv.org/abs/2009.10783} {arXiv:2009.10783
  [physics.ins-det]} \BibitemShut {NoStop}%
\bibitem [{\citenamefont {Li}\ \emph {et~al.}(2019{\natexlab{a}})\citenamefont
  {Li} \emph {et~al.}}]{nEXO-charge}%
  \BibitemOpen
  \bibfield  {author} {\bibinfo {author} {\bibfnamefont {Z.}~\bibnamefont {Li}}
  \emph {et~al.} (\bibinfo {collaboration} {nEXO}),\ }\bibfield  {title}
  {\bibinfo {title} {{Simulation of charge readout with segmented tiles in
  nEXO}},\ }\href {https://doi.org/10.1088/1748-0221/14/09/P09020} {\bibfield
  {journal} {\bibinfo  {journal} {JINST}\ }\textbf {\bibinfo {volume}
  {14}}\bibfield  {number} {\bibinfo  {number} { (09)},\ \bibinfo {pages}
  {P09020}},\ }\Eprint {https://arxiv.org/abs/1907.07512} {arXiv:1907.07512
  [physics.ins-det]} \BibitemShut {NoStop}%
\bibitem [{\citenamefont {Adhikari}\ \emph
  {et~al.}(2021{\natexlab{b}})\citenamefont {Adhikari} \emph
  {et~al.}}]{nEXO-design}%
  \BibitemOpen
  \bibfield  {author} {\bibinfo {author} {\bibfnamefont {G.}~\bibnamefont
  {Adhikari}} \emph {et~al.} (\bibinfo {collaboration} {nEXO}),\ }\bibfield
  {title} {\bibinfo {title} {{nEXO: Neutrinoless double beta decay search
  beyond $10^{28}$ year half-life sensitivity}},\ }\href@noop {} {\bibfield
  {journal} {\bibinfo  {journal} {arXiv}\ } (\bibinfo {year}
  {2021}{\natexlab{b}})},\ \Eprint {https://arxiv.org/abs/2106.16243}
  {arXiv:2106.16243 [nucl-ex]} \BibitemShut {NoStop}%
\bibitem [{\citenamefont {Qiao}\ \emph {et~al.}(2018)\citenamefont {Qiao},
  \citenamefont {Lu}, \citenamefont {Chen}, \citenamefont {Han}, \citenamefont
  {Ji},\ and\ \citenamefont {Wang}}]{PandaX-III-MC}%
  \BibitemOpen
  \bibfield  {author} {\bibinfo {author} {\bibfnamefont {H.}~\bibnamefont
  {Qiao}}, \bibinfo {author} {\bibfnamefont {C.}~\bibnamefont {Lu}}, \bibinfo
  {author} {\bibfnamefont {X.}~\bibnamefont {Chen}}, \bibinfo {author}
  {\bibfnamefont {K.}~\bibnamefont {Han}}, \bibinfo {author} {\bibfnamefont
  {X.}~\bibnamefont {Ji}},\ and\ \bibinfo {author} {\bibfnamefont
  {S.}~\bibnamefont {Wang}},\ }\bibfield  {title} {\bibinfo {title}
  {{Signal-background discrimination with convolutional neural networks in the
  PandaX-III experiment using MC simulation}},\ }\href
  {https://doi.org/10.1007/s11433-018-9233-5} {\bibfield  {journal} {\bibinfo
  {journal} {Sci. China Phys. Mech. Astron.}\ }\textbf {\bibinfo {volume}
  {61}},\ \bibinfo {pages} {101007} (\bibinfo {year} {2018})},\ \Eprint
  {https://arxiv.org/abs/1802.03489} {arXiv:1802.03489 [physics.ins-det]}
  \BibitemShut {NoStop}%
\bibitem [{\citenamefont {Li}\ \emph {et~al.}(2019{\natexlab{b}})\citenamefont
  {Li}, \citenamefont {Elagin}, \citenamefont {Fraker}, \citenamefont {Grant},\
  and\ \citenamefont {Winslow}}]{Kamland-2019}%
  \BibitemOpen
  \bibfield  {author} {\bibinfo {author} {\bibfnamefont {A.}~\bibnamefont
  {Li}}, \bibinfo {author} {\bibfnamefont {A.}~\bibnamefont {Elagin}}, \bibinfo
  {author} {\bibfnamefont {S.}~\bibnamefont {Fraker}}, \bibinfo {author}
  {\bibfnamefont {C.}~\bibnamefont {Grant}},\ and\ \bibinfo {author}
  {\bibfnamefont {L.}~\bibnamefont {Winslow}},\ }\bibfield  {title} {\bibinfo
  {title} {{Suppression of Cosmic Muon Spallation Backgrounds in Liquid
  Scintillator Detectors Using Convolutional Neural Networks}},\ }\href
  {https://doi.org/10.1016/j.nima.2019.162604} {\bibfield  {journal} {\bibinfo
  {journal} {Nucl. Instrum. Meth. A}\ }\textbf {\bibinfo {volume} {947}},\
  \bibinfo {pages} {162604} (\bibinfo {year} {2019}{\natexlab{b}})},\ \Eprint
  {https://arxiv.org/abs/1812.02906} {arXiv:1812.02906 [physics.ins-det]}
  \BibitemShut {NoStop}%
\bibitem [{\citenamefont {Golovatiuk}\ \emph {et~al.}(2021)\citenamefont
  {Golovatiuk}, \citenamefont {Ustyuzhanin}, \citenamefont {Alexandrov},\ and\
  \citenamefont {De~Lellis}}]{emulsion-dm}%
  \BibitemOpen
  \bibfield  {author} {\bibinfo {author} {\bibfnamefont {A.}~\bibnamefont
  {Golovatiuk}}, \bibinfo {author} {\bibfnamefont {A.}~\bibnamefont
  {Ustyuzhanin}}, \bibinfo {author} {\bibfnamefont {A.}~\bibnamefont
  {Alexandrov}},\ and\ \bibinfo {author} {\bibfnamefont {G.}~\bibnamefont
  {De~Lellis}},\ }\bibfield  {title} {\bibinfo {title} {{Deep Learning for
  direct Dark Matter search with nuclear emulsions}},\ }\href@noop {} {\
  (\bibinfo {year} {2021})},\ \Eprint {https://arxiv.org/abs/2106.11995}
  {arXiv:2106.11995 [hep-ex]} \BibitemShut {NoStop}%
\bibitem [{\citenamefont {Simola}\ \emph {et~al.}(2019)\citenamefont {Simola},
  \citenamefont {Pelssers}, \citenamefont {Barge}, \citenamefont {Conrad},\
  and\ \citenamefont {Corander}}]{bolfi-dm}%
  \BibitemOpen
  \bibfield  {author} {\bibinfo {author} {\bibfnamefont {U.}~\bibnamefont
  {Simola}}, \bibinfo {author} {\bibfnamefont {B.}~\bibnamefont {Pelssers}},
  \bibinfo {author} {\bibfnamefont {D.}~\bibnamefont {Barge}}, \bibinfo
  {author} {\bibfnamefont {J.}~\bibnamefont {Conrad}},\ and\ \bibinfo {author}
  {\bibfnamefont {J.}~\bibnamefont {Corander}},\ }\bibfield  {title} {\bibinfo
  {title} {{Machine Learning Accelerated Likelihood-Free Event Reconstruction
  in Dark Matter Direct Detection}},\ }\href
  {https://doi.org/10.1088/1748-0221/14/03/P03004} {\bibfield  {journal}
  {\bibinfo  {journal} {JINST}\ }\textbf {\bibinfo {volume} {14}}\bibfield
  {number} {\bibinfo  {number} { (03)},\ \bibinfo {pages} {P03004}},\ }\Eprint
  {https://arxiv.org/abs/1810.09930} {arXiv:1810.09930 [astro-ph.IM]}
  \BibitemShut {NoStop}%
\bibitem [{\citenamefont {Delaquis}\ \emph {et~al.}(2018)\citenamefont
  {Delaquis} \emph {et~al.}}]{EXO-deep}%
  \BibitemOpen
  \bibfield  {author} {\bibinfo {author} {\bibfnamefont {S.}~\bibnamefont
  {Delaquis}} \emph {et~al.} (\bibinfo {collaboration} {EXO-200
  Collaboration}),\ }\bibfield  {title} {\bibinfo {title} {{Deep Neural
  Networks for Energy and Position Reconstruction in EXO-200}},\ }\href
  {https://doi.org/10.1088/1748-0221/13/08/P08023} {\bibfield  {journal}
  {\bibinfo  {journal} {JINST}\ }\textbf {\bibinfo {volume} {13}}\bibinfo
  {number} { (08)},\ \bibinfo {pages} {P08023}}\BibitemShut {NoStop}%
\bibitem [{\citenamefont {Aprile}\ \emph {et~al.}(2019)\citenamefont {Aprile}
  \emph {et~al.}}]{XENON1T-recon}%
  \BibitemOpen
\bibfield  {number} {  }\bibfield  {author} {\bibinfo {author} {\bibfnamefont
  {E.}~\bibnamefont {Aprile}} \emph {et~al.} (\bibinfo {collaboration}
  {XENON}),\ }\bibfield  {title} {\bibinfo {title} {{XENON1T Dark Matter Data
  Analysis: Signal Reconstruction, Calibration and Event Selection}},\ }\href
  {https://doi.org/10.1103/PhysRevD.100.052014} {\bibfield  {journal} {\bibinfo
   {journal} {Phys. Rev. D}\ }\textbf {\bibinfo {volume} {100}},\ \bibinfo
  {pages} {052014} (\bibinfo {year} {2019})},\ \Eprint
  {https://arxiv.org/abs/1906.04717} {arXiv:1906.04717 [physics.ins-det]}
  \BibitemShut {NoStop}%
\bibitem [{\citenamefont {Goicoechea-Casanueva}\ \emph
  {et~al.}(2021)\citenamefont {Goicoechea-Casanueva}, \citenamefont {Kish},\
  and\ \citenamefont {Maricic}}]{DarkSide-recon}%
  \BibitemOpen
  \bibfield  {author} {\bibinfo {author} {\bibfnamefont {V.}~\bibnamefont
  {Goicoechea-Casanueva}}, \bibinfo {author} {\bibfnamefont {A.}~\bibnamefont
  {Kish}},\ and\ \bibinfo {author} {\bibfnamefont {J.}~\bibnamefont {Maricic}}
  (\bibinfo {collaboration} {DarkSide}),\ }\bibfield  {title} {\bibinfo {title}
  {{Event vertex reconstruction with deep neural networks for the DarkSide-20k
  experiment}},\ }\href {https://doi.org/10.1051/epjconf/202125103029}
  {\bibfield  {journal} {\bibinfo  {journal} {EPJ Web Conf.}\ }\textbf
  {\bibinfo {volume} {251}},\ \bibinfo {pages} {03029} (\bibinfo {year}
  {2021})}\BibitemShut {NoStop}%
\bibitem [{\citenamefont {Grobov}\ and\ \citenamefont
  {Ilyasov}(2020)}]{DarkSide-CNN}%
  \BibitemOpen
  \bibfield  {author} {\bibinfo {author} {\bibfnamefont {A.}~\bibnamefont
  {Grobov}}\ and\ \bibinfo {author} {\bibfnamefont {A.}~\bibnamefont {Ilyasov}}
  (\bibinfo {collaboration} {DarkSide}),\ }\bibfield  {title} {\bibinfo {title}
  {{Convolutional neural network approach to event position reconstruction in
  DarkSide-50 experiment}},\ }\href
  {https://doi.org/10.1088/1742-6596/1690/1/012013} {\bibfield  {journal}
  {\bibinfo  {journal} {J. Phys. Conf. Ser.}\ }\textbf {\bibinfo {volume}
  {1690}},\ \bibinfo {pages} {012013} (\bibinfo {year} {2020})}\BibitemShut
  {NoStop}%
\bibitem [{\citenamefont {Abadi}\ \emph {et~al.}(2015)\citenamefont {Abadi},
  \citenamefont {Agarwal}, \citenamefont {Barham}, \citenamefont {Brevdo},
  \citenamefont {Chen}, \citenamefont {Citro}, \citenamefont {Corrado},
  \citenamefont {Davis}, \citenamefont {Dean}, \citenamefont {Devin},
  \citenamefont {Ghemawat}, \citenamefont {Goodfellow}, \citenamefont {Harp},
  \citenamefont {Irving}, \citenamefont {Isard}, \citenamefont {Jia},
  \citenamefont {Jozefowicz}, \citenamefont {Kaiser}, \citenamefont {Kudlur},
  \citenamefont {Levenberg}, \citenamefont {Man\'{e}}, \citenamefont {Monga},
  \citenamefont {Moore}, \citenamefont {Murray}, \citenamefont {Olah},
  \citenamefont {Schuster}, \citenamefont {Shlens}, \citenamefont {Steiner},
  \citenamefont {Sutskever}, \citenamefont {Talwar}, \citenamefont {Tucker},
  \citenamefont {Vanhoucke}, \citenamefont {Vasudevan}, \citenamefont
  {Vi\'{e}gas}, \citenamefont {Vinyals}, \citenamefont {Warden}, \citenamefont
  {Wattenberg}, \citenamefont {Wicke}, \citenamefont {Yu},\ and\ \citenamefont
  {Zheng}}]{tensorflow2015-whitepaper}%
  \BibitemOpen
  \bibfield  {author} {\bibinfo {author} {\bibfnamefont {M.}~\bibnamefont
  {Abadi}}, \bibinfo {author} {\bibfnamefont {A.}~\bibnamefont {Agarwal}},
  \bibinfo {author} {\bibfnamefont {P.}~\bibnamefont {Barham}}, \bibinfo
  {author} {\bibfnamefont {E.}~\bibnamefont {Brevdo}}, \bibinfo {author}
  {\bibfnamefont {Z.}~\bibnamefont {Chen}}, \bibinfo {author} {\bibfnamefont
  {C.}~\bibnamefont {Citro}}, \bibinfo {author} {\bibfnamefont {G.~S.}\
  \bibnamefont {Corrado}}, \bibinfo {author} {\bibfnamefont {A.}~\bibnamefont
  {Davis}}, \bibinfo {author} {\bibfnamefont {J.}~\bibnamefont {Dean}},
  \bibinfo {author} {\bibfnamefont {M.}~\bibnamefont {Devin}}, \bibinfo
  {author} {\bibfnamefont {S.}~\bibnamefont {Ghemawat}}, \bibinfo {author}
  {\bibfnamefont {I.}~\bibnamefont {Goodfellow}}, \bibinfo {author}
  {\bibfnamefont {A.}~\bibnamefont {Harp}}, \bibinfo {author} {\bibfnamefont
  {G.}~\bibnamefont {Irving}}, \bibinfo {author} {\bibfnamefont
  {M.}~\bibnamefont {Isard}}, \bibinfo {author} {\bibfnamefont
  {Y.}~\bibnamefont {Jia}}, \bibinfo {author} {\bibfnamefont {R.}~\bibnamefont
  {Jozefowicz}}, \bibinfo {author} {\bibfnamefont {L.}~\bibnamefont {Kaiser}},
  \bibinfo {author} {\bibfnamefont {M.}~\bibnamefont {Kudlur}}, \bibinfo
  {author} {\bibfnamefont {J.}~\bibnamefont {Levenberg}}, \bibinfo {author}
  {\bibfnamefont {D.}~\bibnamefont {Man\'{e}}}, \bibinfo {author}
  {\bibfnamefont {R.}~\bibnamefont {Monga}}, \bibinfo {author} {\bibfnamefont
  {S.}~\bibnamefont {Moore}}, \bibinfo {author} {\bibfnamefont
  {D.}~\bibnamefont {Murray}}, \bibinfo {author} {\bibfnamefont
  {C.}~\bibnamefont {Olah}}, \bibinfo {author} {\bibfnamefont {M.}~\bibnamefont
  {Schuster}}, \bibinfo {author} {\bibfnamefont {J.}~\bibnamefont {Shlens}},
  \bibinfo {author} {\bibfnamefont {B.}~\bibnamefont {Steiner}}, \bibinfo
  {author} {\bibfnamefont {I.}~\bibnamefont {Sutskever}}, \bibinfo {author}
  {\bibfnamefont {K.}~\bibnamefont {Talwar}}, \bibinfo {author} {\bibfnamefont
  {P.}~\bibnamefont {Tucker}}, \bibinfo {author} {\bibfnamefont
  {V.}~\bibnamefont {Vanhoucke}}, \bibinfo {author} {\bibfnamefont
  {V.}~\bibnamefont {Vasudevan}}, \bibinfo {author} {\bibfnamefont
  {F.}~\bibnamefont {Vi\'{e}gas}}, \bibinfo {author} {\bibfnamefont
  {O.}~\bibnamefont {Vinyals}}, \bibinfo {author} {\bibfnamefont
  {P.}~\bibnamefont {Warden}}, \bibinfo {author} {\bibfnamefont
  {M.}~\bibnamefont {Wattenberg}}, \bibinfo {author} {\bibfnamefont
  {M.}~\bibnamefont {Wicke}}, \bibinfo {author} {\bibfnamefont
  {Y.}~\bibnamefont {Yu}},\ and\ \bibinfo {author} {\bibfnamefont
  {X.}~\bibnamefont {Zheng}},\ }\href {https://www.tensorflow.org/} {\bibinfo
  {title} {{TensorFlow}: Large-scale machine learning on heterogeneous
  systems}} (\bibinfo {year} {2015}),\ \bibinfo {note} {software available from
  tensorflow.org}\BibitemShut {NoStop}%
\bibitem [{\citenamefont {Paszke}\ \emph {et~al.}(2019)\citenamefont {Paszke},
  \citenamefont {Gross}, \citenamefont {Massa}, \citenamefont {Lerer},
  \citenamefont {Bradbury}, \citenamefont {Chanan}, \citenamefont {Killeen},
  \citenamefont {Lin}, \citenamefont {Gimelshein}, \citenamefont {Antiga},
  \citenamefont {Desmaison}, \citenamefont {Kopf}, \citenamefont {Yang},
  \citenamefont {DeVito}, \citenamefont {Raison}, \citenamefont {Tejani},
  \citenamefont {Chilamkurthy}, \citenamefont {Steiner}, \citenamefont {Fang},
  \citenamefont {Bai},\ and\ \citenamefont {Chintala}}]{NEURIPS2019_9015}%
  \BibitemOpen
  \bibfield  {author} {\bibinfo {author} {\bibfnamefont {A.}~\bibnamefont
  {Paszke}}, \bibinfo {author} {\bibfnamefont {S.}~\bibnamefont {Gross}},
  \bibinfo {author} {\bibfnamefont {F.}~\bibnamefont {Massa}}, \bibinfo
  {author} {\bibfnamefont {A.}~\bibnamefont {Lerer}}, \bibinfo {author}
  {\bibfnamefont {J.}~\bibnamefont {Bradbury}}, \bibinfo {author}
  {\bibfnamefont {G.}~\bibnamefont {Chanan}}, \bibinfo {author} {\bibfnamefont
  {T.}~\bibnamefont {Killeen}}, \bibinfo {author} {\bibfnamefont
  {Z.}~\bibnamefont {Lin}}, \bibinfo {author} {\bibfnamefont {N.}~\bibnamefont
  {Gimelshein}}, \bibinfo {author} {\bibfnamefont {L.}~\bibnamefont {Antiga}},
  \bibinfo {author} {\bibfnamefont {A.}~\bibnamefont {Desmaison}}, \bibinfo
  {author} {\bibfnamefont {A.}~\bibnamefont {Kopf}}, \bibinfo {author}
  {\bibfnamefont {E.}~\bibnamefont {Yang}}, \bibinfo {author} {\bibfnamefont
  {Z.}~\bibnamefont {DeVito}}, \bibinfo {author} {\bibfnamefont
  {M.}~\bibnamefont {Raison}}, \bibinfo {author} {\bibfnamefont
  {A.}~\bibnamefont {Tejani}}, \bibinfo {author} {\bibfnamefont
  {S.}~\bibnamefont {Chilamkurthy}}, \bibinfo {author} {\bibfnamefont
  {B.}~\bibnamefont {Steiner}}, \bibinfo {author} {\bibfnamefont
  {L.}~\bibnamefont {Fang}}, \bibinfo {author} {\bibfnamefont {J.}~\bibnamefont
  {Bai}},\ and\ \bibinfo {author} {\bibfnamefont {S.}~\bibnamefont
  {Chintala}},\ }\bibfield  {title} {\bibinfo {title} {Pytorch: An imperative
  style, high-performance deep learning library},\ }\href@noop {} {\bibfield
  {journal} {\bibinfo  {journal} {Advances in Neural Information Processing
  Systems 32}\ ,\ \bibinfo {pages} {8024}} (\bibinfo {year}
  {2019})}\BibitemShut {NoStop}%
\bibitem [{\citenamefont {Bellagente}\ \emph {et~al.}(2020)\citenamefont
  {Bellagente}, \citenamefont {Butter}, \citenamefont {Kasieczka},
  \citenamefont {Plehn},\ and\ \citenamefont
  {Winterhalder}}]{Bellagente:2019uyp}%
  \BibitemOpen
  \bibfield  {author} {\bibinfo {author} {\bibfnamefont {M.}~\bibnamefont
  {Bellagente}}, \bibinfo {author} {\bibfnamefont {A.}~\bibnamefont {Butter}},
  \bibinfo {author} {\bibfnamefont {G.}~\bibnamefont {Kasieczka}}, \bibinfo
  {author} {\bibfnamefont {T.}~\bibnamefont {Plehn}},\ and\ \bibinfo {author}
  {\bibfnamefont {R.}~\bibnamefont {Winterhalder}},\ }\bibfield  {title}
  {\bibinfo {title} {{How to GAN away Detector Effects}},\ }\href
  {https://doi.org/10.21468/SciPostPhys.8.4.070} {\bibfield  {journal}
  {\bibinfo  {journal} {SciPost Phys.}\ }\textbf {\bibinfo {volume} {8}},\
  \bibinfo {pages} {070} (\bibinfo {year} {2020})},\ \Eprint
  {https://arxiv.org/abs/1912.00477} {arXiv:1912.00477 [hep-ph]} \BibitemShut
  {NoStop}%
\bibitem [{\citenamefont {Komiske}\ \emph {et~al.}(2021)\citenamefont
  {Komiske}, \citenamefont {McCormack},\ and\ \citenamefont
  {Nachman}}]{Komiske:2021vym}%
  \BibitemOpen
  \bibfield  {author} {\bibinfo {author} {\bibfnamefont {P.}~\bibnamefont
  {Komiske}}, \bibinfo {author} {\bibfnamefont {W.~P.}\ \bibnamefont
  {McCormack}},\ and\ \bibinfo {author} {\bibfnamefont {B.}~\bibnamefont
  {Nachman}},\ }\bibfield  {title} {\bibinfo {title} {{Preserving New Physics
  while Simultaneously Unfolding All Observables}},\ }\href@noop {} {\
  (\bibinfo {year} {2021})},\ \Eprint {https://arxiv.org/abs/2105.09923}
  {arXiv:2105.09923 [hep-ph]} \BibitemShut {NoStop}%
\bibitem [{\citenamefont {Chatterjee}\ \emph {et~al.}(2021)\citenamefont
  {Chatterjee}, \citenamefont {Frohner}, \citenamefont {Lechner}, \citenamefont
  {Sch\"ofbeck},\ and\ \citenamefont {Schwarz}}]{Chatterjee:2021nms}%
  \BibitemOpen
  \bibfield  {author} {\bibinfo {author} {\bibfnamefont {S.}~\bibnamefont
  {Chatterjee}}, \bibinfo {author} {\bibfnamefont {N.}~\bibnamefont {Frohner}},
  \bibinfo {author} {\bibfnamefont {L.}~\bibnamefont {Lechner}}, \bibinfo
  {author} {\bibfnamefont {R.}~\bibnamefont {Sch\"ofbeck}},\ and\ \bibinfo
  {author} {\bibfnamefont {D.}~\bibnamefont {Schwarz}},\ }\bibfield  {title}
  {\bibinfo {title} {{Tree boosting for learning EFT parameters}},\ }\href@noop
  {} {\  (\bibinfo {year} {2021})},\ \Eprint {https://arxiv.org/abs/2107.10859}
  {arXiv:2107.10859 [hep-ph]} \BibitemShut {NoStop}%
\bibitem [{\citenamefont {Chen}\ \emph {et~al.}(2021)\citenamefont {Chen},
  \citenamefont {Glioti}, \citenamefont {Panico},\ and\ \citenamefont
  {Wulzer}}]{Chen:2020mev}%
  \BibitemOpen
  \bibfield  {author} {\bibinfo {author} {\bibfnamefont {S.}~\bibnamefont
  {Chen}}, \bibinfo {author} {\bibfnamefont {A.}~\bibnamefont {Glioti}},
  \bibinfo {author} {\bibfnamefont {G.}~\bibnamefont {Panico}},\ and\ \bibinfo
  {author} {\bibfnamefont {A.}~\bibnamefont {Wulzer}},\ }\bibfield  {title}
  {\bibinfo {title} {{Parametrized classifiers for optimal EFT sensitivity}},\
  }\href {https://doi.org/10.1007/JHEP05(2021)247} {\bibfield  {journal}
  {\bibinfo  {journal} {JHEP}\ }\textbf {\bibinfo {volume} {05}},\ \bibinfo
  {pages} {247}},\ \Eprint {https://arxiv.org/abs/2007.10356} {arXiv:2007.10356
  [hep-ph]} \BibitemShut {NoStop}%
\bibitem [{\citenamefont {Erbin}\ and\ \citenamefont
  {Krippendorf}(2020)}]{Erbin:2018csv}%
  \BibitemOpen
  \bibfield  {author} {\bibinfo {author} {\bibfnamefont {H.}~\bibnamefont
  {Erbin}}\ and\ \bibinfo {author} {\bibfnamefont {S.}~\bibnamefont
  {Krippendorf}},\ }\bibfield  {title} {\bibinfo {title} {{GANs for generating
  EFT models}},\ }\href {https://doi.org/10.1016/j.physletb.2020.135798}
  {\bibfield  {journal} {\bibinfo  {journal} {Phys. Lett. B}\ }\textbf
  {\bibinfo {volume} {810}},\ \bibinfo {pages} {135798} (\bibinfo {year}
  {2020})},\ \Eprint {https://arxiv.org/abs/1809.02612} {arXiv:1809.02612
  [cs.LG]} \BibitemShut {NoStop}%
\bibitem [{\citenamefont {Caron}\ \emph {et~al.}(2017)\citenamefont {Caron},
  \citenamefont {Kim}, \citenamefont {Rolbiecki}, \citenamefont {Ruiz~de
  Austri},\ and\ \citenamefont {Stienen}}]{Caron:2016hib}%
  \BibitemOpen
  \bibfield  {author} {\bibinfo {author} {\bibfnamefont {S.}~\bibnamefont
  {Caron}}, \bibinfo {author} {\bibfnamefont {J.~S.}\ \bibnamefont {Kim}},
  \bibinfo {author} {\bibfnamefont {K.}~\bibnamefont {Rolbiecki}}, \bibinfo
  {author} {\bibfnamefont {R.}~\bibnamefont {Ruiz~de Austri}},\ and\ \bibinfo
  {author} {\bibfnamefont {B.}~\bibnamefont {Stienen}},\ }\bibfield  {title}
  {\bibinfo {title} {{The BSM-AI project: SUSY-AI\textendash{}generalizing LHC
  limits on supersymmetry with machine learning}},\ }\href
  {https://doi.org/10.1140/epjc/s10052-017-4814-9} {\bibfield  {journal}
  {\bibinfo  {journal} {Eur. Phys. J. C}\ }\textbf {\bibinfo {volume} {77}},\
  \bibinfo {pages} {257} (\bibinfo {year} {2017})},\ \Eprint
  {https://arxiv.org/abs/1605.02797} {arXiv:1605.02797 [hep-ph]} \BibitemShut
  {NoStop}%
\bibitem [{\citenamefont {Bertone}\ \emph {et~al.}(2019)\citenamefont
  {Bertone}, \citenamefont {Deisenroth}, \citenamefont {Kim}, \citenamefont
  {Liem}, \citenamefont {Ruiz~de Austri},\ and\ \citenamefont
  {Welling}}]{Bertone:2016mdy}%
  \BibitemOpen
  \bibfield  {author} {\bibinfo {author} {\bibfnamefont {G.}~\bibnamefont
  {Bertone}}, \bibinfo {author} {\bibfnamefont {M.~P.}\ \bibnamefont
  {Deisenroth}}, \bibinfo {author} {\bibfnamefont {J.~S.}\ \bibnamefont {Kim}},
  \bibinfo {author} {\bibfnamefont {S.}~\bibnamefont {Liem}}, \bibinfo {author}
  {\bibfnamefont {R.}~\bibnamefont {Ruiz~de Austri}},\ and\ \bibinfo {author}
  {\bibfnamefont {M.}~\bibnamefont {Welling}},\ }\bibfield  {title} {\bibinfo
  {title} {{Accelerating the BSM interpretation of LHC data with machine
  learning}},\ }\href {https://doi.org/10.1016/j.dark.2019.100293} {\bibfield
  {journal} {\bibinfo  {journal} {Phys. Dark Univ.}\ }\textbf {\bibinfo
  {volume} {24}},\ \bibinfo {pages} {100293} (\bibinfo {year} {2019})},\
  \Eprint {https://arxiv.org/abs/1611.02704} {arXiv:1611.02704 [hep-ph]}
  \BibitemShut {NoStop}%
\bibitem [{\citenamefont {Kronheim}\ \emph {et~al.}(2021)\citenamefont
  {Kronheim}, \citenamefont {Kuchera}, \citenamefont {Prosper},\ and\
  \citenamefont {Karbo}}]{Kronheim:2020vct}%
  \BibitemOpen
  \bibfield  {author} {\bibinfo {author} {\bibfnamefont {B.~S.}\ \bibnamefont
  {Kronheim}}, \bibinfo {author} {\bibfnamefont {M.~P.}\ \bibnamefont
  {Kuchera}}, \bibinfo {author} {\bibfnamefont {H.~B.}\ \bibnamefont
  {Prosper}},\ and\ \bibinfo {author} {\bibfnamefont {A.}~\bibnamefont
  {Karbo}},\ }\bibfield  {title} {\bibinfo {title} {{Bayesian Neural Networks
  for Fast SUSY Predictions}},\ }\href
  {https://doi.org/10.1016/j.physletb.2020.136041} {\bibfield  {journal}
  {\bibinfo  {journal} {Phys. Lett. B}\ }\textbf {\bibinfo {volume} {813}},\
  \bibinfo {pages} {136041} (\bibinfo {year} {2021})},\ \Eprint
  {https://arxiv.org/abs/2007.04506} {arXiv:2007.04506 [hep-ph]} \BibitemShut
  {NoStop}%
\bibitem [{\citenamefont {{Lindegren}}\ \emph {et~al.}(2018)\citenamefont
  {{Lindegren}}, \citenamefont {{Hern{\'a}ndez}}, \citenamefont {{Bombrun}},
  \citenamefont {{Klioner}}, \citenamefont {{Bastian}}, \citenamefont
  {{Ramos-Lerate}}, \citenamefont {{de Torres}}, \citenamefont
  {{Steidelm{\"u}ller}}, \citenamefont {{Stephenson}}, \citenamefont {{Hobbs}},
  \citenamefont {{Lammers}}, \citenamefont {{Biermann}}, \citenamefont
  {{Geyer}}, \citenamefont {{Hilger}}, \citenamefont {{Michalik}},
  \citenamefont {{Stampa}}, \citenamefont {{McMillan}}, \citenamefont
  {{Casta{\~n}eda}}, \citenamefont {{Clotet}}, \citenamefont {{Comoretto}},
  \citenamefont {{Davidson}}, \citenamefont {{Fabricius}}, \citenamefont
  {{Gracia}}, \citenamefont {{Hambly}}, \citenamefont {{Hutton}}, \citenamefont
  {{Mora}}, \citenamefont {{Portell}}, \citenamefont {{van Leeuwen}},
  \citenamefont {{Abbas}}, \citenamefont {{Abreu}}, \citenamefont {{Altmann}},
  \citenamefont {{Andrei}}, \citenamefont {{Anglada}}, \citenamefont
  {{Balaguer-N{\'u}{\~n}ez}}, \citenamefont {{Barache}}, \citenamefont
  {{Becciani}}, \citenamefont {{Bertone}}, \citenamefont {{Bianchi}},
  \citenamefont {{Bouquillon}}, \citenamefont {{Bourda}}, \citenamefont
  {{Br{\"u}semeister}}, \citenamefont {{Bucciarelli}}, \citenamefont
  {{Busonero}}, \citenamefont {{Buzzi}}, \citenamefont {{Cancelliere}},
  \citenamefont {{Carlucci}}, \citenamefont {{Charlot}}, \citenamefont
  {{Cheek}}, \citenamefont {{Crosta}}, \citenamefont {{Crowley}}, \citenamefont
  {{de Bruijne}}, \citenamefont {{de Felice}}, \citenamefont {{Drimmel}},
  \citenamefont {{Esquej}}, \citenamefont {{Fienga}}, \citenamefont {{Fraile}},
  \citenamefont {{Gai}}, \citenamefont {{Garralda}}, \citenamefont
  {{Gonz{\'a}lez-Vidal}}, \citenamefont {{Guerra}}, \citenamefont {{Hauser}},
  \citenamefont {{Hofmann}}, \citenamefont {{Holl}}, \citenamefont {{Jordan}},
  \citenamefont {{Lattanzi}}, \citenamefont {{Lenhardt}}, \citenamefont
  {{Liao}}, \citenamefont {{Licata}}, \citenamefont {{Lister}}, \citenamefont
  {{L{\"o}ffler}}, \citenamefont {{Marchant}}, \citenamefont
  {{Martin-Fleitas}}, \citenamefont {{Messineo}}, \citenamefont {{Mignard}},
  \citenamefont {{Morbidelli}}, \citenamefont {{Poggio}}, \citenamefont
  {{Riva}}, \citenamefont {{Rowell}}, \citenamefont {{Salguero}}, \citenamefont
  {{Sarasso}}, \citenamefont {{Sciacca}}, \citenamefont {{Siddiqui}},
  \citenamefont {{Smart}}, \citenamefont {{Spagna}}, \citenamefont {{Steele}},
  \citenamefont {{Taris}}, \citenamefont {{Torra}}, \citenamefont {{van
  Elteren}}, \citenamefont {{van Reeven}},\ and\ \citenamefont
  {{Vecchiato}}}]{2018A&A...616A...2L}%
  \BibitemOpen
  \bibfield  {author} {\bibinfo {author} {\bibfnamefont {L.}~\bibnamefont
  {{Lindegren}}}, \bibinfo {author} {\bibfnamefont {J.}~\bibnamefont
  {{Hern{\'a}ndez}}}, \bibinfo {author} {\bibfnamefont {A.}~\bibnamefont
  {{Bombrun}}}, \bibinfo {author} {\bibfnamefont {S.}~\bibnamefont
  {{Klioner}}}, \bibinfo {author} {\bibfnamefont {U.}~\bibnamefont
  {{Bastian}}}, \bibinfo {author} {\bibfnamefont {M.}~\bibnamefont
  {{Ramos-Lerate}}}, \bibinfo {author} {\bibfnamefont {A.}~\bibnamefont {{de
  Torres}}}, \bibinfo {author} {\bibfnamefont {H.}~\bibnamefont
  {{Steidelm{\"u}ller}}}, \bibinfo {author} {\bibfnamefont {C.}~\bibnamefont
  {{Stephenson}}}, \bibinfo {author} {\bibfnamefont {D.}~\bibnamefont
  {{Hobbs}}}, \bibinfo {author} {\bibfnamefont {U.}~\bibnamefont {{Lammers}}},
  \bibinfo {author} {\bibfnamefont {M.}~\bibnamefont {{Biermann}}}, \bibinfo
  {author} {\bibfnamefont {R.}~\bibnamefont {{Geyer}}}, \bibinfo {author}
  {\bibfnamefont {T.}~\bibnamefont {{Hilger}}}, \bibinfo {author}
  {\bibfnamefont {D.}~\bibnamefont {{Michalik}}}, \bibinfo {author}
  {\bibfnamefont {U.}~\bibnamefont {{Stampa}}}, \bibinfo {author}
  {\bibfnamefont {P.~J.}\ \bibnamefont {{McMillan}}}, \bibinfo {author}
  {\bibfnamefont {J.}~\bibnamefont {{Casta{\~n}eda}}}, \bibinfo {author}
  {\bibfnamefont {M.}~\bibnamefont {{Clotet}}}, \bibinfo {author}
  {\bibfnamefont {G.}~\bibnamefont {{Comoretto}}}, \bibinfo {author}
  {\bibfnamefont {M.}~\bibnamefont {{Davidson}}}, \bibinfo {author}
  {\bibfnamefont {C.}~\bibnamefont {{Fabricius}}}, \bibinfo {author}
  {\bibfnamefont {G.}~\bibnamefont {{Gracia}}}, \bibinfo {author}
  {\bibfnamefont {N.~C.}\ \bibnamefont {{Hambly}}}, \bibinfo {author}
  {\bibfnamefont {A.}~\bibnamefont {{Hutton}}}, \bibinfo {author}
  {\bibfnamefont {A.}~\bibnamefont {{Mora}}}, \bibinfo {author} {\bibfnamefont
  {J.}~\bibnamefont {{Portell}}}, \bibinfo {author} {\bibfnamefont
  {F.}~\bibnamefont {{van Leeuwen}}}, \bibinfo {author} {\bibfnamefont
  {U.}~\bibnamefont {{Abbas}}}, \bibinfo {author} {\bibfnamefont
  {A.}~\bibnamefont {{Abreu}}}, \bibinfo {author} {\bibfnamefont
  {M.}~\bibnamefont {{Altmann}}}, \bibinfo {author} {\bibfnamefont
  {A.}~\bibnamefont {{Andrei}}}, \bibinfo {author} {\bibfnamefont
  {E.}~\bibnamefont {{Anglada}}}, \bibinfo {author} {\bibfnamefont
  {L.}~\bibnamefont {{Balaguer-N{\'u}{\~n}ez}}}, \bibinfo {author}
  {\bibfnamefont {C.}~\bibnamefont {{Barache}}}, \bibinfo {author}
  {\bibfnamefont {U.}~\bibnamefont {{Becciani}}}, \bibinfo {author}
  {\bibfnamefont {S.}~\bibnamefont {{Bertone}}}, \bibinfo {author}
  {\bibfnamefont {L.}~\bibnamefont {{Bianchi}}}, \bibinfo {author}
  {\bibfnamefont {S.}~\bibnamefont {{Bouquillon}}}, \bibinfo {author}
  {\bibfnamefont {G.}~\bibnamefont {{Bourda}}}, \bibinfo {author}
  {\bibfnamefont {T.}~\bibnamefont {{Br{\"u}semeister}}}, \bibinfo {author}
  {\bibfnamefont {B.}~\bibnamefont {{Bucciarelli}}}, \bibinfo {author}
  {\bibfnamefont {D.}~\bibnamefont {{Busonero}}}, \bibinfo {author}
  {\bibfnamefont {R.}~\bibnamefont {{Buzzi}}}, \bibinfo {author} {\bibfnamefont
  {R.}~\bibnamefont {{Cancelliere}}}, \bibinfo {author} {\bibfnamefont
  {T.}~\bibnamefont {{Carlucci}}}, \bibinfo {author} {\bibfnamefont
  {P.}~\bibnamefont {{Charlot}}}, \bibinfo {author} {\bibfnamefont
  {N.}~\bibnamefont {{Cheek}}}, \bibinfo {author} {\bibfnamefont
  {M.}~\bibnamefont {{Crosta}}}, \bibinfo {author} {\bibfnamefont
  {C.}~\bibnamefont {{Crowley}}}, \bibinfo {author} {\bibfnamefont
  {J.}~\bibnamefont {{de Bruijne}}}, \bibinfo {author} {\bibfnamefont
  {F.}~\bibnamefont {{de Felice}}}, \bibinfo {author} {\bibfnamefont
  {R.}~\bibnamefont {{Drimmel}}}, \bibinfo {author} {\bibfnamefont
  {P.}~\bibnamefont {{Esquej}}}, \bibinfo {author} {\bibfnamefont
  {A.}~\bibnamefont {{Fienga}}}, \bibinfo {author} {\bibfnamefont
  {E.}~\bibnamefont {{Fraile}}}, \bibinfo {author} {\bibfnamefont
  {M.}~\bibnamefont {{Gai}}}, \bibinfo {author} {\bibfnamefont
  {N.}~\bibnamefont {{Garralda}}}, \bibinfo {author} {\bibfnamefont {J.~J.}\
  \bibnamefont {{Gonz{\'a}lez-Vidal}}}, \bibinfo {author} {\bibfnamefont
  {R.}~\bibnamefont {{Guerra}}}, \bibinfo {author} {\bibfnamefont
  {M.}~\bibnamefont {{Hauser}}}, \bibinfo {author} {\bibfnamefont
  {W.}~\bibnamefont {{Hofmann}}}, \bibinfo {author} {\bibfnamefont
  {B.}~\bibnamefont {{Holl}}}, \bibinfo {author} {\bibfnamefont
  {S.}~\bibnamefont {{Jordan}}}, \bibinfo {author} {\bibfnamefont {M.~G.}\
  \bibnamefont {{Lattanzi}}}, \bibinfo {author} {\bibfnamefont
  {H.}~\bibnamefont {{Lenhardt}}}, \bibinfo {author} {\bibfnamefont
  {S.}~\bibnamefont {{Liao}}}, \bibinfo {author} {\bibfnamefont
  {E.}~\bibnamefont {{Licata}}}, \bibinfo {author} {\bibfnamefont
  {T.}~\bibnamefont {{Lister}}}, \bibinfo {author} {\bibfnamefont
  {W.}~\bibnamefont {{L{\"o}ffler}}}, \bibinfo {author} {\bibfnamefont
  {J.}~\bibnamefont {{Marchant}}}, \bibinfo {author} {\bibfnamefont {J.~M.}\
  \bibnamefont {{Martin-Fleitas}}}, \bibinfo {author} {\bibfnamefont
  {R.}~\bibnamefont {{Messineo}}}, \bibinfo {author} {\bibfnamefont
  {F.}~\bibnamefont {{Mignard}}}, \bibinfo {author} {\bibfnamefont
  {R.}~\bibnamefont {{Morbidelli}}}, \bibinfo {author} {\bibfnamefont
  {E.}~\bibnamefont {{Poggio}}}, \bibinfo {author} {\bibfnamefont
  {A.}~\bibnamefont {{Riva}}}, \bibinfo {author} {\bibfnamefont
  {N.}~\bibnamefont {{Rowell}}}, \bibinfo {author} {\bibfnamefont
  {E.}~\bibnamefont {{Salguero}}}, \bibinfo {author} {\bibfnamefont
  {M.}~\bibnamefont {{Sarasso}}}, \bibinfo {author} {\bibfnamefont
  {E.}~\bibnamefont {{Sciacca}}}, \bibinfo {author} {\bibfnamefont
  {H.}~\bibnamefont {{Siddiqui}}}, \bibinfo {author} {\bibfnamefont {R.~L.}\
  \bibnamefont {{Smart}}}, \bibinfo {author} {\bibfnamefont {A.}~\bibnamefont
  {{Spagna}}}, \bibinfo {author} {\bibfnamefont {I.}~\bibnamefont {{Steele}}},
  \bibinfo {author} {\bibfnamefont {F.}~\bibnamefont {{Taris}}}, \bibinfo
  {author} {\bibfnamefont {J.}~\bibnamefont {{Torra}}}, \bibinfo {author}
  {\bibfnamefont {A.}~\bibnamefont {{van Elteren}}}, \bibinfo {author}
  {\bibfnamefont {W.}~\bibnamefont {{van Reeven}}},\ and\ \bibinfo {author}
  {\bibfnamefont {A.}~\bibnamefont {{Vecchiato}}},\ }\bibfield  {title}
  {\bibinfo {title} {{Gaia Data Release 2. The astrometric solution}},\ }\href
  {https://doi.org/10.1051/0004-6361/201832727} {\bibfield  {journal} {\bibinfo
   {journal} {Astron. \& Astrophys.}\ }\textbf {\bibinfo {volume} {616}},\
  \bibinfo {eid} {A2} (\bibinfo {year} {2018})},\ \Eprint
  {https://arxiv.org/abs/1804.09366} {arXiv:1804.09366 [astro-ph.IM]}
  \BibitemShut {NoStop}%
\bibitem [{\citenamefont {{Ivezi{\'c}}}\ \emph {et~al.}(2019)\citenamefont
  {{Ivezi{\'c}}}, \citenamefont {{Kahn}}, \citenamefont {{Tyson}},
  \citenamefont {{Abel}}, \citenamefont {{Acosta}}, \citenamefont {{Allsman}},
  \citenamefont {{Alonso}}, \citenamefont {{AlSayyad}}, \citenamefont
  {{Anderson}}, \citenamefont {{Andrew}},\ and\ \citenamefont
  {et~al.}}]{2019ApJ...873..111I}%
  \BibitemOpen
  \bibfield  {author} {\bibinfo {author} {\bibfnamefont {{\v Z}.}~\bibnamefont
  {{Ivezi{\'c}}}}, \bibinfo {author} {\bibfnamefont {S.~M.}\ \bibnamefont
  {{Kahn}}}, \bibinfo {author} {\bibfnamefont {J.~A.}\ \bibnamefont {{Tyson}}},
  \bibinfo {author} {\bibfnamefont {B.}~\bibnamefont {{Abel}}}, \bibinfo
  {author} {\bibfnamefont {E.}~\bibnamefont {{Acosta}}}, \bibinfo {author}
  {\bibfnamefont {R.}~\bibnamefont {{Allsman}}}, \bibinfo {author}
  {\bibfnamefont {D.}~\bibnamefont {{Alonso}}}, \bibinfo {author}
  {\bibfnamefont {Y.}~\bibnamefont {{AlSayyad}}}, \bibinfo {author}
  {\bibfnamefont {S.~F.}\ \bibnamefont {{Anderson}}}, \bibinfo {author}
  {\bibfnamefont {J.}~\bibnamefont {{Andrew}}},\ and\ \bibinfo {author}
  {\bibnamefont {et~al.}},\ }\bibfield  {title} {\bibinfo {title} {{LSST: From
  Science Drivers to Reference Design and Anticipated Data Products}},\ }\href
  {https://doi.org/10.3847/1538-4357/ab042c} {\bibfield  {journal} {\bibinfo
  {journal} {\apj}\ }\textbf {\bibinfo {volume} {873}},\ \bibinfo {eid} {111}
  (\bibinfo {year} {2019})},\ \Eprint {https://arxiv.org/abs/0805.2366}
  {arXiv:0805.2366} \BibitemShut {NoStop}%
\bibitem [{\citenamefont {Abbott}\ \emph {et~al.}(2009)\citenamefont {Abbott}
  \emph {et~al.}}]{LIGOScientific:2007fwp}%
  \BibitemOpen
  \bibfield  {author} {\bibinfo {author} {\bibfnamefont {B.~P.}\ \bibnamefont
  {Abbott}} \emph {et~al.} (\bibinfo {collaboration} {LIGO Scientific}),\
  }\bibfield  {title} {\bibinfo {title} {{LIGO: The Laser interferometer
  gravitational-wave observatory}},\ }\href
  {https://doi.org/10.1088/0034-4885/72/7/076901} {\bibfield  {journal}
  {\bibinfo  {journal} {Rept. Prog. Phys.}\ }\textbf {\bibinfo {volume} {72}},\
  \bibinfo {pages} {076901} (\bibinfo {year} {2009})},\ \Eprint
  {https://arxiv.org/abs/0711.3041} {arXiv:0711.3041 [gr-qc]} \BibitemShut
  {NoStop}%
\bibitem [{\citenamefont {{Weltman}}\ \emph {et~al.}(2020)\citenamefont
  {{Weltman}}, \citenamefont {{Bull}}, \citenamefont {{Camera}}, \citenamefont
  {{Kelley}}, \citenamefont {{Padmanabhan}}, \citenamefont {{Pritchard}},
  \citenamefont {{Raccanelli}}, \citenamefont {{Riemer-S{\o}rensen}},
  \citenamefont {{Shao}}, \citenamefont {{Andrianomena}}, \citenamefont
  {{Athanassoula}}, \citenamefont {{Bacon}}, \citenamefont {{Barkana}},
  \citenamefont {{Bertone}}, \citenamefont {{B{\oe}hm}}, \citenamefont
  {{Bonvin}}, \citenamefont {{Bosma}}, \citenamefont {{Br{\"u}ggen}},
  \citenamefont {{Burigana}}, \citenamefont {{Calore}}, \citenamefont
  {{Cembranos}}, \citenamefont {{Clarkson}}, \citenamefont {{Connors}},
  \citenamefont {{Cruz-Dombriz}}, \citenamefont {{Dunsby}}, \citenamefont
  {{Fonseca}}, \citenamefont {{Fornengo}}, \citenamefont {{Gaggero}},
  \citenamefont {{Harrison}}, \citenamefont {{Larena}}, \citenamefont {{Ma}},
  \citenamefont {{Maartens}}, \citenamefont {{M{\'e}ndez-Isla}}, \citenamefont
  {{Mohanty}}, \citenamefont {{Murray}}, \citenamefont {{Parkinson}},
  \citenamefont {{Pourtsidou}}, \citenamefont {{Quinn}}, \citenamefont
  {{Regis}}, \citenamefont {{Saha}}, \citenamefont {{Sahl{\'e}n}},
  \citenamefont {{Sakellariadou}}, \citenamefont {{Silk}}, \citenamefont
  {{Trombetti}}, \citenamefont {{Vazza}}, \citenamefont {{Venumadhav}},
  \citenamefont {{Vidotto}}, \citenamefont {{Villaescusa-Navarro}},
  \citenamefont {{Wang}}, \citenamefont {{Weniger}}, \citenamefont {{Wolz}},
  \citenamefont {{Zhang}},\ and\ \citenamefont
  {{Gaensler}}}]{2020PASA...37....2W}%
  \BibitemOpen
  \bibfield  {author} {\bibinfo {author} {\bibfnamefont {A.}~\bibnamefont
  {{Weltman}}}, \bibinfo {author} {\bibfnamefont {P.}~\bibnamefont {{Bull}}},
  \bibinfo {author} {\bibfnamefont {S.}~\bibnamefont {{Camera}}}, \bibinfo
  {author} {\bibfnamefont {K.}~\bibnamefont {{Kelley}}}, \bibinfo {author}
  {\bibfnamefont {H.}~\bibnamefont {{Padmanabhan}}}, \bibinfo {author}
  {\bibfnamefont {J.}~\bibnamefont {{Pritchard}}}, \bibinfo {author}
  {\bibfnamefont {A.}~\bibnamefont {{Raccanelli}}}, \bibinfo {author}
  {\bibfnamefont {S.}~\bibnamefont {{Riemer-S{\o}rensen}}}, \bibinfo {author}
  {\bibfnamefont {L.}~\bibnamefont {{Shao}}}, \bibinfo {author} {\bibfnamefont
  {S.}~\bibnamefont {{Andrianomena}}}, \bibinfo {author} {\bibfnamefont
  {E.}~\bibnamefont {{Athanassoula}}}, \bibinfo {author} {\bibfnamefont
  {D.}~\bibnamefont {{Bacon}}}, \bibinfo {author} {\bibfnamefont
  {R.}~\bibnamefont {{Barkana}}}, \bibinfo {author} {\bibfnamefont
  {G.}~\bibnamefont {{Bertone}}}, \bibinfo {author} {\bibfnamefont
  {C.}~\bibnamefont {{B{\oe}hm}}}, \bibinfo {author} {\bibfnamefont
  {C.}~\bibnamefont {{Bonvin}}}, \bibinfo {author} {\bibfnamefont
  {A.}~\bibnamefont {{Bosma}}}, \bibinfo {author} {\bibfnamefont
  {M.}~\bibnamefont {{Br{\"u}ggen}}}, \bibinfo {author} {\bibfnamefont
  {C.}~\bibnamefont {{Burigana}}}, \bibinfo {author} {\bibfnamefont
  {F.}~\bibnamefont {{Calore}}}, \bibinfo {author} {\bibfnamefont {J.~A.~R.}\
  \bibnamefont {{Cembranos}}}, \bibinfo {author} {\bibfnamefont
  {C.}~\bibnamefont {{Clarkson}}}, \bibinfo {author} {\bibfnamefont {R.~M.~T.}\
  \bibnamefont {{Connors}}}, \bibinfo {author} {\bibfnamefont {{\'A}.~d.~l.}\
  \bibnamefont {{Cruz-Dombriz}}}, \bibinfo {author} {\bibfnamefont {P.~K.~S.}\
  \bibnamefont {{Dunsby}}}, \bibinfo {author} {\bibfnamefont {J.}~\bibnamefont
  {{Fonseca}}}, \bibinfo {author} {\bibfnamefont {N.}~\bibnamefont
  {{Fornengo}}}, \bibinfo {author} {\bibfnamefont {D.}~\bibnamefont
  {{Gaggero}}}, \bibinfo {author} {\bibfnamefont {I.}~\bibnamefont
  {{Harrison}}}, \bibinfo {author} {\bibfnamefont {J.}~\bibnamefont
  {{Larena}}}, \bibinfo {author} {\bibfnamefont {Y.~Z.}\ \bibnamefont {{Ma}}},
  \bibinfo {author} {\bibfnamefont {R.}~\bibnamefont {{Maartens}}}, \bibinfo
  {author} {\bibfnamefont {M.}~\bibnamefont {{M{\'e}ndez-Isla}}}, \bibinfo
  {author} {\bibfnamefont {S.~D.}\ \bibnamefont {{Mohanty}}}, \bibinfo {author}
  {\bibfnamefont {S.}~\bibnamefont {{Murray}}}, \bibinfo {author}
  {\bibfnamefont {D.}~\bibnamefont {{Parkinson}}}, \bibinfo {author}
  {\bibfnamefont {A.}~\bibnamefont {{Pourtsidou}}}, \bibinfo {author}
  {\bibfnamefont {P.~J.}\ \bibnamefont {{Quinn}}}, \bibinfo {author}
  {\bibfnamefont {M.}~\bibnamefont {{Regis}}}, \bibinfo {author} {\bibfnamefont
  {P.}~\bibnamefont {{Saha}}}, \bibinfo {author} {\bibfnamefont
  {M.}~\bibnamefont {{Sahl{\'e}n}}}, \bibinfo {author} {\bibfnamefont
  {M.}~\bibnamefont {{Sakellariadou}}}, \bibinfo {author} {\bibfnamefont
  {J.}~\bibnamefont {{Silk}}}, \bibinfo {author} {\bibfnamefont
  {T.}~\bibnamefont {{Trombetti}}}, \bibinfo {author} {\bibfnamefont
  {F.}~\bibnamefont {{Vazza}}}, \bibinfo {author} {\bibfnamefont
  {T.}~\bibnamefont {{Venumadhav}}}, \bibinfo {author} {\bibfnamefont
  {F.}~\bibnamefont {{Vidotto}}}, \bibinfo {author} {\bibfnamefont
  {F.}~\bibnamefont {{Villaescusa-Navarro}}}, \bibinfo {author} {\bibfnamefont
  {Y.}~\bibnamefont {{Wang}}}, \bibinfo {author} {\bibfnamefont
  {C.}~\bibnamefont {{Weniger}}}, \bibinfo {author} {\bibfnamefont
  {L.}~\bibnamefont {{Wolz}}}, \bibinfo {author} {\bibfnamefont
  {F.}~\bibnamefont {{Zhang}}},\ and\ \bibinfo {author} {\bibfnamefont {B.~M.}\
  \bibnamefont {{Gaensler}}},\ }\bibfield  {title} {\bibinfo {title}
  {{Fundamental physics with the Square Kilometre Array}},\ }\href
  {https://doi.org/10.1017/pasa.2019.42} {\bibfield  {journal} {\bibinfo
  {journal} {Publ. Astron. Soc. Aust.}\ }\textbf {\bibinfo {volume} {37}},\
  \bibinfo {eid} {e002} (\bibinfo {year} {2020})},\ \Eprint
  {https://arxiv.org/abs/1810.02680} {arXiv:1810.02680 [astro-ph.CO]}
  \BibitemShut {NoStop}%
\bibitem [{\citenamefont {Stein}(2020)}]{george_stein_2020_4024768}%
  \BibitemOpen
  \bibfield  {author} {\bibinfo {author} {\bibfnamefont {G.}~\bibnamefont
  {Stein}},\ }\bibfield  {title} {\bibinfo {title}
  {{georgestein/ml-in-cosmology: Machine learning in cosmology}},\ }\href
  {https://doi.org/10.5281/zenodo.4024768} {10.5281/zenodo.4024768} (\bibinfo
  {year} {2020})\BibitemShut {NoStop}%
\end{thebibliography}%

\clearpage
\appendix

\end{document}